\documentclass[authoryear,preprint,dvipsnames,letterpaper,12pt]{elsarticle}
\RequirePackage[dvipsnames]{xcolor}
\usepackage[a4paper, top=1in, bottom=1in, left=1in, right=1in]{geometry} 
\usepackage{amssymb}
\usepackage{graphicx}
\usepackage{amsmath}

\usepackage{placeins}
\usepackage{orcidlink} 
\usepackage{float}
\usepackage[dvipsnames]{xcolor}
\usepackage{silence}
\usepackage{amssymb}
\usepackage{float} 

\journal{}

\begin{document}

\begin{frontmatter}


\title{Mechanics Informatics: A paradigm for efficiently learning constitutive models}


\author[inst1]{Royal C. Ihuaenyi}
\author[inst1]{Wei Li}
\author[inst2,inst3]{Martin Z. Bazant}
\author[inst1]{Juner Zhu\corref{cor1}}
\ead{j.zhu@northeastern.edu}
\cortext[cor1]{Corresponding author.}

\affiliation[inst1]{organization={Department of Mechanical and Industrial Engineering, Northeastern University},
            addressline={360 Huntington Ave}, 
            city={Boston},
            postcode={02115}, 
            state={MA},
            country={USA}}

\affiliation[inst2]{organization={Department of Chemical Engineering, Massachusetts Institute of Technology},
            addressline={77 Massachusetts Ave}, 
            city={Cambridge},
            postcode={02139}, 
            state={MA},
            country={USA}}
            
\affiliation[inst3]{organization={Department of Mathematics, Massachusetts Institute of Technology},
            addressline={77 Massachusetts Ave}, 
            city={Cambridge},
            postcode={02139}, 
            state={MA},
            country={USA}}

\begin{abstract}
Efficient and accurate learning of constitutive laws is crucial for accurately predicting the mechanical behavior of materials under complex loading conditions. Accurate model calibration hinges on a delicate interplay between the information embedded in experimental data and the parameters that define our constitutive models.
The information encoded in the parameters of the constitutive model must be complemented by the information in the data used for calibration. This interplay raises fundamental questions: How can we quantify the information content of test data? How much information does a single test convey? Also, how much information is required to accurately learn a constitutive model? To address these questions, we introduce \textit{mechanics informatics}, a paradigm for efficient and accurate constitutive model learning. At its core is the \textit{stress state entropy}, a metric for quantifying the information content of experimental data. Using this framework, we analyzed specimen geometries with varying information content for learning an anisotropic inelastic law. Specimens with limited information enabled accurate identification of a few parameters sensitive to the information in the data. Furthermore, we optimized specimen design by incorporating stress state entropy into a Bayesian optimization scheme. This led to the design of cruciform specimens with maximized entropy for accurate parameter identification. Conversely, minimizing entropy in Peirs shear specimens yielded a uniform shear stress state, showcasing the framework's flexibility in tailoring designs for specific experimental goals. Finally, we addressed experimental uncertainties, demonstrated the potential of transfer learning for replacing challenging testing protocols with simpler alternatives, and extension of the framework to different material laws.

\end{abstract}


\end{frontmatter}

\section{Introduction}
\label{sec:intro}
How many material parameters can be learned from a given experimental dataset?
This has been a longstanding question, central to the field of mechanics for
generations, with some of the greatest minds working to address it. In 1638,
Galileo inferred a single parameter, the ratio of gravitational force to
deflection from a bar bending experiment~\citep{Galilei1638}, laying the
foundation for elasticity theory. Between 1865 and 1925,
Maxwell,~\citep{huber1904},~\citep{mises1913mechanik},
and~\citep{hencky1924theorie} noted differences in material strength under
tension and shear, leading to the J2 yield criterion, effectively learning two
parameters from two test types. Hill later extended this to anisotropic
materials by introducing a six-parameter yield surface~\citep{hill1948theory}.
In deterministic settings, $n$ stress-state sensitive parameters require
$n$ independent experiments. Consequently, advances in experimental
techniques have significantly enabled modern mechanicians to develop more
sophisticated constitutive models.

With a wide range of constitutive models available, parameter identification
has become a central challenge. These models capture complex behaviors governed
by microstructures, as in sheet metal
forming~\citep{Cao2020Opportunities,cao2024artificial}. While some parameters
can be computed from first
principles~\citep{qi2014lithium,dreyer2015brittle,chang2017lithiation}, most
are obtained through mechanical
testing~\citep{rossi2022testing,guery2016identification,zhu2014influence}. This
human-centered process is difficult to automate. Simple tests such as uniaxial
tension, equibiaxial tension, and simple shear are traditionally used to
isolate parameters, each targeting a distinct stress state. This approach has
proven effective across disciplines like metal forming and automotive
design~\citep{rossi2022testing,fu2016identification,fu2020method,anghileri2005inverse},
valued for its robustness and interpretability. Nonetheless, three major
limitations of this paradigm are becoming increasingly apparent as experimental
mechanics advances.

The first limitation is that simple mechanical tests are often complicated by
structural effects such as rotation, necking, and friction. Ideal stress states
(uniaxial, biaxial, or shear) are rarely achieved. Uniaxial tension (UT) tests
suffer from strain localization, necessitating inverse methods to extract
post-localization
behavior~\citep{dunand2010hybrid,KAMAYA2011243,ZHANG19993497}. Uniaxial
compression (UC) tests are affected by surface friction, effectively becoming
upsetting tests~\citep{BAO2004,WIERZBICKI2005Seven,Vilotic2003}. Simple shear
(SS) tests are preferred over pure shear tests because the former has lower
controlling force requirement in the direction perpendicular to the shear
load~\citep{SEGAL2002331}. The design of SS specimen also deals with
eliminating the effect of rigid-body rotation and the bending moment on the
critical section. Equibiaxial (EB) tension tests demand specialized cruciform
specimens and precise loading to achieve uniform
deformation~\citep{KUWABARA1998,KUWABARA2007,Meng2016}. Consequently,
extracting intrinsic material properties often relies on advanced analysis
aided by computational methods.

The second limitation encompasses the limited information content of simple
tests, which are typically designed so that most material points undergo nearly
identical loading paths. Calibrating complex constitutive models requires
diverse deformation states pertinent to the model parameters, necessitating
multiple tests and increasing experimental cost. This is especially critical
when using advanced imaging techniques such as scanning electron microscopy
(SEM), transmission electron microscopy (TEM), or X-ray-based computer
tomography (XCT) for in situ experiments~\citep{GORJI2017,NI2021,ZHU2018APEN}.
For such testing protocols, cost constraints limit the number of feasible
tests. Therefore, maximizing information from each experiment is essential.

The third limitation is the reliance on expert knowledge within the
human-centered paradigm. Designing effective specimens and analyzing results
require extensive training and experience. For example, optimizing simple shear
tests to achieve ideal stress states often involves intensive
design~\citep{dunand2010hybrid,dunand2011optimized,morin2017prediction,GORJI2017,tancogne2021ductile}.
As engineering moves into the Industry 4.0 era, the demand for skills in
automation, digital twins, and data analytics
grows~\citep{hernandez2020engineering}. This calls for a shift from traditional
human-centered paradigms toward more automated and scalable approaches.

Overcoming these limitations requires a paradigm shift involving the use of
geometrically complex specimens to maximize information per test, and
leveraging algorithms for automated parameter identification. Ortiz and
co-workers established a comprehensive framework for data-driven computational
solid mechanics in a series of
publications~\citep{KIRCHDOERFER2016CMAME,KIRCHDOERFER2017CMAME,KIRCHDOERFER2018NME,stainier2019model,EGGERSMANN2019,CARRARA2020,EGGERSMANN2021}.
To enrich stress states, they used specimens with multiple holes under biaxial
loading~\citep{stainier2019model}. Other researchers adopted similar
strategies,~\citep{jones2018parameter} designed a ``D-shaped'' specimen for
viscoplastic parameter calibration,~\citep{kim2014determination} introduced a
``$\Sigma$-shaped'' specimen for identifying Hill48 parameters through
uniaxial tests, and~\citep{Barroqueiro2020Design} used topology optimization to
design test specimen with a complex geometry.

As the complexity of test specimens and boundary conditions increases, it becomes impractical to analytically solve the stress and strain fields to determine the constitutive relationship directly. Consequently, learning the constitutive models of materials becomes an inverse problem. In deformable solids, the inverse problem is inherently over-determined, meaning there are more equations than unknowns. Typically, it is addressed by optimizing a set of constitutive parameters for a chosen model, minimizing the difference between model predictions and experimental data. Success depends on a sufficiently rich dataset and an effective optimization algorithm, both of which have advanced significantly over recent decades, driving two major waves of inverse learning. 

The first wave emerged in the late 1990s--early 2000s, driven by digital image
correlation (DIC), which enabled low-cost, full-field deformation
measurements~\citep{Sutton2009}. This paved the way for large experimental
datasets and inverse methods such as finite element model updating (FEMU) and
the virtual fields method (VFM)~\citep{avril2008overview}. FEMU iteratively
adjusts model parameters via FE simulations to minimize the objective
function~\citep[e.g.][]{dunand2010hybrid,ZHANG19993497,zhu2014influence}. FEMU
has a low data requirement in the sense that it does not necessarily require
full-field data. VFM, in contrast, relies on the principle of virtual work with
carefully constructed virtual
fields~\citep{avril2008identification,kramer2014implementation,LATTANZI2020}.
While powerful, VFM requires full-field measurements or sufficient surface data
to infer volume displacements. A major challenge of the VFM is how to
effectively construct virtual fields. This is mainly because for a constitutive
relation with $m$ unknown parameters, $m$ virtual fields are
required to generate $m$ independent
equations~\citep{avril2008overview}.

We are now amid a second wave of inverse learning, fueled by rapid advances in
computational capability and the blooming of artificial intelligence (AI)
algorithms, particularly deep learning. Modern AI techniques have proven
successful across scientific domains, including recognizing and predicting
phase separation patterns~\citep{ZHAO2020PRL,ZHAO2021JComP,zhao2023learning},
identifying material parameters~\citep{YAMANAKA2020}, visualizing safety
envelopes~\citep{LI2019Joule}, and optimizing engineered
systems~\citep{attia2020Nature,Aykol2021}. These developments offer promising
strategies for solving inverse problems in mechanics with increased efficiency.
For example, deep neural networks have been employed to infer constitutive
parameters from indirect observations, where stress--strain pairs are not
directly accessible~\citep{HUANG2020JComP,XU2021JComP}. Similarly, unsupervised
and automated learning methods have been proposed to extract material laws
without requiring labeled
data~\citep{marino2023automated,flaschel2021unsupervised,flaschel2023automated}.
Other work has demonstrated the learning of chemo-mechanical constitutive laws
and the detection of plasticity using atomic-scale imaging of 3D strain and
concentration fields~\citep{deng2022correlative}. A recent review
by~\citep{fuhg2024review} surveys this growing field of data-driven
constitutive modeling.

Despite this progress, several fundamental questions remain that challenge the transition from traditional, human-centered, multi-test paradigms to AI-assisted, single-test inverse learning. Specifically, how much information does a single test convey?, how much information is required to accurately learn a model?, and how can we optimally design test specimens to meet a target information content?. 

To address these challenges, we introduce the concept of
\textbf{\emph{mechanics informatics}}, inspired by analogous developments in
materials informatics~\citep{ramprasad2017machine} and medical
informatics~\citep{haux2010medical}. This emerging field seeks to apply
principles from information theory and data science to optimize experimental
design, improve model calibration, and enhance the predictive fidelity of
mechanical models. In this paper, we elaborate on the theoretical foundation
and practical implementation of mechanics informatics as a new lens for
accurately learning constitutive models.

\section{Problem formulation of learning constitutive models}
\label{Section:Formulation} 
\subsection{Governing equations}
\label{Subsection:GE} 
For a deformable domain $\Omega \subset \mathbb{R}^d$, the strain-displacement kinematic relationship is defined by:
\begin{equation}
    \label{eq:2-1}
    \boldsymbol{\varepsilon} \left( \mathbf{u} \right) = \frac{1}{2} \left( \nabla \mathbf{u} + \nabla \mathbf{u}^\intercal \right), \quad \text{in } \Omega,
\end{equation}
where $\boldsymbol{\varepsilon}$ denotes the strain tensor and $\mathbf{u}$ is the displacement field. The Dirichlet boundary condition prescribed on $\Omega_u$ is given by:
\begin{equation}
    \label{eq:2-2}
    \mathbf{u} = \hat{\mathbf{u}}, \quad \text{on } \Omega_u,
\end{equation}
where $\hat{\mathbf{u}}$ represents the prescribed displacement field along the boundary.

The equilibrium condition for forces within the domain is expressed as:
\begin{equation}
    \label{eq:2-3}
    \mathrm{div} \, \boldsymbol{\sigma} + \mathbf{b} = \mathbf{0}, \quad \text{in } \Omega,
\end{equation}
where $\boldsymbol{\sigma}$ is the stress tensor and $\mathbf{b}$ is the body force vector. Additionally, the Neumann boundary condition applied on $\Omega_t$ is given by:
\begin{equation}
    \label{eq:2-4}
    \boldsymbol{\sigma} \mathbf{n} = \hat{\mathbf{t}}, \quad \text{on } \Omega_t,
\end{equation}
where $\mathbf{n}$ is the unit normal to the boundary and $\hat{\mathbf{t}}$ represents the prescribed traction. Together, these equations govern the behavior of the deformable domain $\Omega$ under applied displacements and forces.

\subsection{Constitutive relations}
\label{Subsection:CR} 
A general form of the constitutive relation is typically expressed as, $\boldsymbol{\sigma} = \boldsymbol{\sigma}\left(\boldsymbol{\varepsilon}, \dot{\boldsymbol{\varepsilon}}, T ,\ldots \right)$,  
where $\dot{\boldsymbol{\varepsilon}}$ and $T$ are the strain rate and temperature, respectively. In this study, we focus on material models that are independent of strain rate and temperature, simplifying the constitutive relation to the form, $\boldsymbol{\sigma} = \boldsymbol{\sigma}\left(\boldsymbol{\varepsilon}\right)$. This approach follows small-deformation plasticity theory, a widely used framework for modeling sheet metals, based on the additive decomposition of the strain tensor:
\begin{equation}
    \label{eq:2-6}
    \boldsymbol{\varepsilon} = \boldsymbol{\varepsilon}^e + \boldsymbol{\varepsilon}^p,
\end{equation}
where $\boldsymbol{\varepsilon}^e$ and $\boldsymbol{\varepsilon}^p$ are the elastic and plastic components of the strain tensor, respectively.

The Cauchy stress is determined by the elastic response as:
\begin{equation}
    \label{eq:2-7}
    \boldsymbol{\sigma} = \mathbb{C} : \boldsymbol{\varepsilon}^e,
\end{equation}
where $\mathbb{C}$ is the fourth-order elastic modulus tensor.

The yielding behavior of the material is governed by a yield function, defined as:
\begin{equation}
    \label{eq:2-8}
    f\left(\boldsymbol{\sigma}, \boldsymbol{\varepsilon}^p\right) = 0.
\end{equation}

Furthermore, the plastic flow is characterized by the relation:
\begin{equation}
    \label{eq:2-9}
    \dot{\boldsymbol{\varepsilon}}^p = |\dot{\boldsymbol{\varepsilon}}^p| \frac{\partial g\left(\boldsymbol{\sigma}, \boldsymbol{\varepsilon}^p\right)}{\partial \boldsymbol{\sigma}},
\end{equation}
where $g$ is the flow potential function. When $g$ is identical to the yield function $f$, the plastic flow is termed \textit{associative}; otherwise, it is \textit{non-associative} \citep{simo2006computational, gurtin2010mechanics}.

Additionally, for characterizing how stress states evolve relative to the yield surface, the Kuhn-Tucker loading-unloading conditions are expressed as:
\begin{equation}
    \label{eq:2-10}
    |\dot{\boldsymbol{\varepsilon}}^p| \geq 0, \quad f \leq 0, \quad |\dot{\boldsymbol{\varepsilon}}^p| f = 0,
\end{equation}
along with the consistency condition:
\begin{equation}
    \label{eq:2-11}
    |\dot{\boldsymbol{\varepsilon}}^p| \dot{f} = 0.
\end{equation}

\subsection{Forward and inverse problems}
\label{Subsection:F&IP} 
Equations \ref{eq:2-1} through \ref{eq:2-11} fully describe a boundary value problem (BVP) that can be solved for a unique solution when appropriate boundary conditions are specified, and the parameters of the constitutive relations are known. This formulation defines the forward problem as follows:
\begin{equation}
    \label{eq:2-12}
    \textrm{Forward problem: } \{ \hat{\mathbf{u}}, \hat{\mathbf{t}}, \boldsymbol{\theta} \} \longrightarrow \{ \mathbf{u}(\boldsymbol{x}), \boldsymbol{\varepsilon}(\boldsymbol{x}), \boldsymbol{\sigma}(\boldsymbol{x}) \},
\end{equation}
where $\boldsymbol{\theta}$ represents the parameter vector or parameter space associated with the constitutive framework.

Solving the forward problem is essential when material properties are fully characterized, and the parameter space is known, as it facilitates the prediction of stress fields. The forward problem is well-posed, ensuring both the uniqueness and existence of a solution given the defined parameters. However, in many practical scenarios, the parameter space $\boldsymbol{\theta}$ that defines the constitutive model is totally or partially unknown. This uncertainty often arises from the complex nature of material responses, which necessitates constitutive models involving parameters that are challenging to identify experimentally with high reliability. In such cases, the problem must be approached inversely:
\begin{equation}
    \label{eq:2-13}
    \textrm{Type-I Inverse problem: } \{\mathbf{u}\left(\boldsymbol{x}\right) \to \boldsymbol{\varepsilon}\left(\boldsymbol{x}\right), \hat{\mathbf{t}}\} \longrightarrow \{\boldsymbol{\sigma}\left(\boldsymbol{x}\right), \boldsymbol{\theta}\}.
\end{equation}

In this work, we assume that the continuum model is homogeneous, i.e., the parameter space $\boldsymbol{\theta}$ comprises material parameters that are invariant with respect to the material coordinates $\boldsymbol{x}$. However, some advanced models incorporate material heterogeneity and anisotropy, making the parameter space coordinate-dependent \citep{ihuaenyi2021orthotropic,ihuaenyi2023orthotropic,ihuaenyi2023coupled, iqbal2023probabilistic}. Theoretically, kinematic, equilibrium and constitutive equations hold at every material point, resulting in an over-determined system for the inverse problem. This over-determined nature of the problem often requires identifying the parameter space from noisy experimental observations, inherently rendering the inverse problem ill-posed due to potential issues with non-uniqueness or even non-existence of a solution. Additionally, the accuracy of parameter identification relies heavily on the robustness of the inverse learning procedure, which must not only capture the bulk material response effectively through informative test data but also demonstrate robustness to measurement noise.

It is noteworthy that several studies have explored the direct inference of stress fields from full-field displacement measurements without explicitly specifying a constitutive model, such as the works by Cameron and Tasan~\citep{cameron2021full} and Réthoré et al.~\citep{Rethore2018}. This class of problems can be expressed as:

\begin{equation}
    \label{eq:2-14}
    \textrm{Type-II inverse problem: } \{\mathbf{u}(\boldsymbol{x}) \to \boldsymbol{\varepsilon}(\boldsymbol{x}), \hat{\mathbf{t}}\} \xrightarrow{\text{Assumptions}} \{\boldsymbol{\sigma}(\boldsymbol{x})\}.
\end{equation}

Such problems are inherently under-determined and require additional assumptions to render them solvable. For instance, Cameron and Tasan~\citep{cameron2021full} introduced an ``alignment assumption,'' which postulates that the principal directions of the Cauchy stress tensor align with those of either the strain or strain-rate tensors. Similarly, Réthoré et al.~\citep{Rethore2018} proposed the mechanical image correlation (MIC) method, which reconstructs displacement, strain, and stress fields using a parametric representation based on an assumed anelastic strain field. In MIC, displacement parameters are extracted via digital image correlation (DIC), while stress parameters are inferred through thermodynamic considerations, enabling the identification of admissible material states.

The feasibility of directly estimating stress fields from displacement thus hinges on the validity of these simplifying assumptions, which in effect serve as surrogate constitutive constraints without the need to specify material parameters. Accordingly, the inverse formulation in Eq.~\ref{eq:2-14} may be viewed as a special case of the more general formulation presented in Eq.~\ref{eq:2-13}. In this work, we focus on type-I inverse problems, as defined by Eq.~\ref{eq:2-13}.

\subsection{A special case of anisotropic plasticity}
\label{Subsection:Anisotropic Plasticity} 
In this study, we focus on a special case of the outlined constitutive framework, particularly examining the behavior of anisotropic sheet metal under a planar stress state. Within the small deformation range, the assumption of isotropic elasticity is typically employed due to the nearly isotropic behavior of sheet metals and for simplicity and computational efficiency \citep{Martins2019,zhu2014influence,Pottier2011,avril2008identification,Martins2019}. The isotropic linear elastic response of the material is described by its elastic modulus ($E$) and Poisson's ratio ($\nu$).

As the material deforms plastically, its response becomes anisotropic and can be well represented by a quadratic yield function. This yield function is coupled with an associated flow rule and an isotropic hardening law to capture the progression of the plastic deformation. The onset of plastic deformation is governed by the yield function, expressed as:
\begin{equation}
    f(\boldsymbol{\sigma}, \boldsymbol{\varepsilon}^p) = \sqrt{\boldsymbol{\sigma} : \mathbf{A} : \boldsymbol{\sigma}} - \sigma_Y(\bar{\varepsilon}^p) = 0,
\end{equation}
where \( \mathbf{A} \) is a symmetric tensor characterizing material anisotropy. In the Hill48 \citep{hill1948theory} formulation, this tensor is defined as:
\begin{equation}
    \mathbf{A} = 
    \begin{bmatrix}
        1 & -H & -G & 0 & 0 & 0 \\
        -H & H + F & -F & 0 & 0 & 0  \\
        -G & -F & F + G & 0 & 0 & 0  \\
        0 & 0 & 0 & 2L & 0 & 0 \\
        0 & 0 & 0 & 0 & 2M & 0 \\
        0 & 0 & 0 & 0 & 0 & 2N 
    \end{bmatrix}.
\end{equation}

Here, \( H \), \( G \), \( F \), \( L \), \( M \), and \( N \) are yield constants, typically determined experimentally. Under the plane stress assumption, the stress components $\sigma_{33} = \tau_{23} = \tau_{13} = 0$ , rendering the coefficients $M$ and $N$ irrelevant to the formulation. Notably, when the anisotropic parameters take values $F = G = H = 0.5$ and $N = 1.5$, the anisotropic yield surface simplifies to the isotropic von Mises yield surface.
Furthermore, the isotropic hardening law $\sigma_Y(\bar{\varepsilon}^p)$ is modeled using the Swift formulation, expressed as:
\begin{equation}
    \sigma_{\text{Y}} (\bar{\varepsilon}^p) = A(\bar{\varepsilon}^p + \varepsilon_0)^n,
\end{equation}
where $\varepsilon_0 = \left( \frac{\sigma_0}{A} \right)^{1/n}$ defines the reference strain. Here, $A$, $\sigma_0$, and $n$ are material parameters, with $\sigma_0$ representing the initial yield stress. In the inverse learning procedure, six material parameters ($A$, $\sigma_0$, $n$, $F$, $G$, and $N$) that characterize the material response are to be identified. This reduced parameter set is sufficient because the relationship $H + G = 1$ is enforced by using the initial yield stress in the rolling direction (RD) as the reference yield stress.

\section{Mechanics informatics}
\label{Section:Informatics} 
Mechanics informatics is introduced in this work as a framework aimed at addressing key challenges in quantifying the information content of mechanical tests and enabling the efficient and accurate learning of material laws. By evaluating the information content and richness of test data, this framework seeks to address the critical question: \textit{``How much information
does a single test convey?"}. Additionally, we consider the complementary question: \textit{``How much information is required to accurately learn a constitutive model?"}. These questions are fundamental in mechanics since each test carries a unique amount and type of information. Furthermore, a material law represents the material's response under specific stress states, requiring a defined level of information from tests for learning. 
   
\subsection{Information entropy}
\label{Subsection:Information entropy} 
Information theory, a groundbreaking framework pioneered by Shannon in 1948 \citep{Shannon1948mathematical}, provides the essential tools to quantify, encode, and transmit information across varied systems. At its core lies the concept of \textit{entropy}, a fundamental measure of uncertainty or information content within a dataset, signal, or system. In essence, systems with greater complexity or variability reflect higher entropy, while

To formally define information entropy, consider a discrete random variable \( \text{X} \) that takes on values \( \{x_1, x_2, \ldots, x_n\} \), where \( n \) represents the total number of possible outcomes. The probability of observing a specific outcome \( x \) is given by \( p(x) \). The entropy, \( \text{H}(\text{X}) \) of this variable, representing the average level of ``surprise" or uncertainty across its outcomes, is calculated by the formula:
\begin{equation}
\text{H}(\text{X}) = -\sum_{x \in \text{X}} p(x) \ln p(x).
\end{equation}
Since $p(x) >0 $ and $\sum\limits_{x \in \text{X}} p(x)=1$, the minimum entropy $\text{H}_{\text{min}}=0$ is achieved if a single outcome $x_0$ occurs with probability $p(x_0)=1$.

For continuous cases, where \( \text{X} \) is a random variable with a probability density function \( f(x) \), the entropy generalizes to the form:
\begin{equation}
\text{H}(X) = -\int f(x) \ln f(x) \, dx,
\end{equation}
which is minimized ($\text{H}=0$) when the continuous probability density is localized on one point outcome as a Dirac delta function, $f(x)=\delta(x-x_0)$. 

By extension, if we consider two related discrete-valued variables \( \text{X} \) and \(\text{Y} \), with \( p(y) \) representing the probability of any \( y \in Y \), where \( \text{Y} \) is the set of all possible values \(\{y_1, y_2, \ldots, y_n\}\), the joint entropy associated with these variables is defined as:
\begin{equation}
\text{H}(\text{X},\text{Y}) = -\sum_{x \in \text{X}} \sum_{y \in \text{Y}} p(x,y) \ln p(x,y).
\end{equation}

Consequently, if the variables \( \text{X} \) and \( \text{Y} \) are continuous with a joint probability density function \( f(x,y) \), their joint entropy is given by:
\begin{equation}
\text{H}(\text{X},\text{Y}) = -\iint f(x,y) \ln f(x,y) \, dx \, dy.
\end{equation}

When an independent relationship exists between the probability distributions of the two variables, their joint entropy can be expressed as the sum of their individual entropy values. 

Additionally, the concept of maximum entropy is a fundamental aspect of information theory. A system reaches maximum entropy when the probabilities of all possible values of the discrete variable span a uniform distribution, which is expressed as:
\begin{equation}
\text{H}_{\text{max}}(\text{X}) = -\ln \left( \frac{1}{n} \right).
\end{equation}

\subsection{Stress state analysis}
\label{Subsection:stres sstate} 
The stress state of a material encapsulates the stress distribution arising from its composition, loading conditions, and boundary constraints. Different stress states can be induced in a test depending on the specific mechanical response or features targeted for extraction by a testing protocol. For instance, to characterize a material's uniaxial tensile behavior, a standard dog-bone tensile specimen is typically employed. The gage section of this specimen is subjected to a uniaxial tensile stress as the tensile specimen is loaded primarily along one principal direction, with only one non-zero nominal stress, $\boldsymbol{\sigma} = (\sigma_{11}, 0, 0, 0, 0, 0)^\text{T}$.

Hence, to precisely capture material behavior under pure stress states, specialized testing methods have been developed, ensuring a robust and accurate representation of the bulk material properties. For uniaxial compression response, dog-bone-shaped specimens with a short gage length and stack compression tests are widely utilized \citep{alves2011revisiting,astm2000compression}. The torsion test is a standard approach applied for characterizing shear response using cylindrical or tubular specimens \citep{papasidero2015ductile}. Additionally, test specimens have been designed in literature to induce a pure shear stress state within a given gage section by normal stresses on the test specimen \citep{GORJI2017, BAO2004}. Furthermore, testing methods such as the cruciform specimen test \citep{deng2015cruciform}, hydraulic bulge test \citep{mulder2015accurate}, and mini-punch test \citep{ROTH2016} have been effectively implemented in representing the equibiaxial tensile response of materials.

These testing configurations ensure that the resulting strain field or force response accurately reflects the material's bulk response under each specific stress state. While these traditional tests are robust, they are designed to provide detailed information about the material behavior under a single stress state in each test. However, heterogeneous stress states can be induced through the design of test specimens with complex geometries, thereby enriching the information content of the data obtained from a single testing protocol \citep{kim2014determination, Barroqueiro2020Design, jones2018parameter}. The variation in stress states within a single test reflects the diverse range of mechanical responses that can be captured. Hence, test protocols that generate heterogeneous stress states within a single experiment are valuable for reducing experimental costs when learning complex material models. 

Analyzing the stress state within a given testing protocol is critical for selecting the appropriate experimental procedures to generate data for learning constitutive models. To accurately quantify the stress state(s) in a given test, it is essential to define metrics that characterize the nature of the applied loading conditions. Such metrics can include the ratios of the principal stresses and their respective signs. For example, in the case of an isotropic material subjected to uniaxial tension in a planar stress state, only one principal stress is non-zero and positive ($\sigma_{\text{11}} > 0$, $\sigma_{\text{22}} = 0$). In contrast, in shear, the principal stress ratio is equal to $-1$, and the principal stresses have opposite signs ($\sigma_{\text{11}}/\sigma_{\text{22}} = -1$, $\sigma_{\text{11}} > 0$, $\sigma_{\text{22}} < 0$). In the context of anisotropic plasticity, as considered in this study, the stress state of a material undergoing plastic deformation can be described using two non-dimensionalized parameters, the stress triaxiality ($\eta$) and the Lode angle parameter ($\bar{\theta}$). The stress triaxiality, \(\eta\), serves as a measure of the relative contribution of hydrostatic stress within the overall stress state, defined as the ratio of hydrostatic stress to the equivalent stress:
\begin{equation}
\eta = \frac{\sigma_h}{\sigma_{\text{eqv}}},
\end{equation}
where \(\sigma_h = \frac{\sigma_{kk}}{3}\), with \(\sigma_{kk}\) representing the first invariant or trace of the stress tensor, and \(\sigma_{\text{eqv}}\) is the equivalent stress. For a two-dimensional stress state, the stress triaxiality is constrained within the range \(\eta \in \left[-\frac{2}{3}, \frac{2}{3}\right]\). In a general multiaxial stress state, the values of \(\eta\) can range from \(-\infty\) to \(\infty\).

The Lode angle, which quantifies the deviatoric nature of the stress state, is expressed in terms of the normalized third invariant of the deviatoric stress tensor, \(J_3\), as:
\begin{equation}
\cos(\theta) = \frac{27}{2} \frac{J_3}{\sigma_{\text{eq}}}.
\end{equation}
The Lode angle is often normalized within the range \(\bar{\theta} \in [-1, 1]\) as the Lode angle parameter:
\begin{equation}
\bar{\theta} = 1 - \frac{2}{\pi} \arccos\left(\frac{27}{2} \frac{J_3}{\sigma_{\text{eq}}}\right).
\end{equation}

\begin{figure}[H]
    \centering
    \includegraphics[width=0.7\columnwidth]{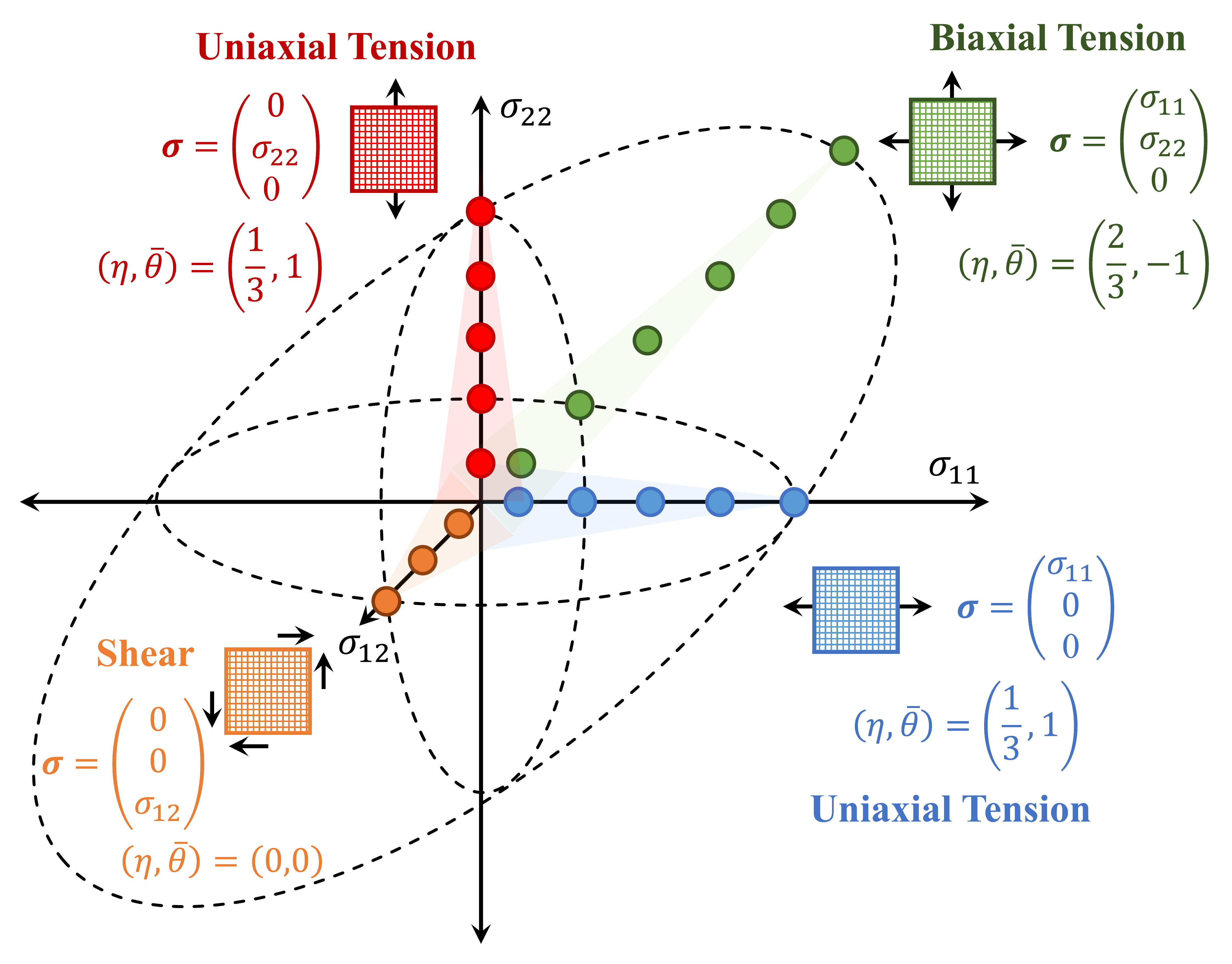} 
    \vspace{-1em}
    \caption{Two-dimensional yield surface demonstrating tension--compression symmetry, annotated with key stress states and their corresponding stress triaxiality and Lode angle parameter values.}
    \label{fig: Figure 1}
\end{figure}

Employing the stress triaxiality and Lode angle parameter as metrics, it is possible to precisely identify the stress state corresponding to a specific mechanical response during plastic deformation. Figure \ref{fig: Figure 1} illustrates the major stress states on a 2D yield surface of a material exhibiting tension--compression symmetry. The stress triaxiality and Lode angle values are shown for each stress state.

\subsection{Stress state entropy}
\label{Subsection:stressentropy}  
In a previous work \citep{ihuaenyi2024seeking}, we proposed a theoretical framework that integrates concepts from information theory to quantify the stress state information content of test specimens subjected to specific loading conditions. Here, we summarize the mathematical framework for the \textit{stress state entropy}, which serves as a measure of the information content in a mechanical test, incorporating stress metrics and information-theoretic principles.

Given \(\Theta\), the set of all considered stress states required for feature extraction that can be generated from a test specimen \(S\), and \(p(\sigma)\), the probability of any stress state \(\sigma \in \Theta\), the stress state information content or stress state entropy of the test specimen considering \(\Theta\) as a discrete-valued variable is defined as:
\begin{equation}
\label{eq: Stress State Entropy}
\text{H}(\Theta) = - \sum_{\sigma \in \Theta} p(\sigma) \ln p(\sigma).
\end{equation}

To address the fundamental question, \textit{How much information does a single test convey?} The amount of information conveyed by a single test is its stress state entropy. By spatially discretizing the specimen geometry, we can analyze the local stress state in each discrete region, and can compute the stress state entropy (Eq. \ref{eq: Stress State Entropy}) of the entire specimen. This provides a clear metric for evaluating how informative a specimen’s response is under a given testing protocol. 

To address the complementary question, \textit{How much information is required to learn a constitutive model?}, we highlight the importance of analyzing both the model parameters and the stress states to which they are sensitive, as these factors influence their accurate identification. For example, a simple 1D isotropic linear elastic model with a parameter space comprising of just the elastic modulus $\boldsymbol{\theta} = \{\text{E}\} $ can be accurately learned from a single uniaxial tensile test. This is because the material response under the uniaxial stress state provides sufficient information for parameter estimation. Thus, the information requirement for this model corresponds to the stress state entropy for a stress state space associated with only uniaxial tension, $\Theta = \{\sigma_{\text{UT}}\}$. However, for more complex models, where parameters depend on responses across multiple stress states, accurate identification necessitates test data covering a broader range of the relevant stress states.

To effectively learn the anisotropic inelastic material model within the current formulation, which assumes a state of planar stress and tension--compression symmetry, the test database must encompass stress states to which the material parameters are sensitive. Specifically, the hardening law parameters ($A$, $\varepsilon_0$, $n$) and the yield constant $G$ are primarily influenced by the uniaxial tensile response in the RD. The yield constants $F$ and $N$ are sensitive to the material response in the transverse direction (TD) and shear, respectively. Therefore, accurately identifying the material law requires test data capturing sufficient information on uniaxial tension in both the RD and TD, as well as shear ($\Theta = \{ \sigma_{\text{UT}}^{\text{RD}}, \sigma_{\text{UT}}^{\text{TD}}, \sigma_{\text{S}} \}$). Hence, the specific composition of $\Theta$ depends on the stress states required to accurately infer the material behavior. Thus, the information required to accurately learn the material law is the \textit{stress state entropy}, which considers all stress states required to learn the material law.

Furthermore, maximizing the stress state entropy of a test specimen requires a uniform probability distribution of stress states within the relevant stress state space. A specimen that achieves the maximum stress state entropy is defined as the \textit{theoretically optimal specimen}. The stress state entropy of such a specimen is defined as:
\begin{equation}
\text{H}_{\text{max}}(\Theta \mid=n) = -\ln \left(\frac{1}{n}\right).
\end{equation}

Designing test specimens to achieve the maximally achievable stress state entropy within a given stress state space necessitates advanced topological optimization techniques. However, the practical realization of such specimens can be constrained by uncertainties arising from impure stress states at specimen boundaries and experimental uncertainties. To address these limitations, in \citep{ihuaenyi2024seeking} we proposed an optimal stress state entropy hypothesis, which bounds the entropy of a test specimen designed to extract information about a specific stress state space between two limits, determined by the cardinality of the stress state space. This is defined as:
\begin{equation}
    \text{H}_{\text{max}}(\Theta \mid=n-1) < \text{H}(\Theta \mid=n) \leq \text{H}_{\text{max}}(\Theta|=n).
    \label{eq:entropy_criterion}
\end{equation}

In essence, Equation \ref{eq:entropy_criterion} establishes that for a stress state space \(\Theta\) with cardinality \(n\), the stress state entropy of the designed or selected specimen should exceed the maximum entropy associated with a stress state space of cardinality \(n-1\) while remaining less than or equal to the maximum entropy for the complete stress state space. This bound ensures that all required stress states for learning the considered material law are represented in the test specimen response. This criterion provides a practical guideline for experimentalists to optimize the information extracted from test specimens, ultimately enhancing the accuracy of inverse learning for constitutive models that rely on material responses across multiple stress states.

\section{Inverse learning}
\label{Section:Inverse learning}
In this study, we infer the parameters of the anisotropic inelastic material law from strain fields. The strain fields are generated from test specimens of various geometries, subjected to different boundary conditions, with each specimen contributing data that offers distinct information content. Solving the inverse problem is realized through the penalization of the discrepancy between synthetic experimental strain fields and numerical strain fields. The synthetic experimental strain fields are generated from FE simulations using the ground truth parameter values as input and applying a typical DIC resolution limit to the strain fields. The measurement accuracy and strain resolution limit of DIC are highly dependent on system setup, speckle pattern quality, and environmental conditions \citep{Sutton2009,pan2009review}. In large deformation testing, distortion the speckle pattern is a commonality, making it harder to track small displacements accurately. In this study, we apply a strain resolution limit of 0.05\% to FE strain fields obtained from simulations of selected geometries. At strain levels below this threshold, the mismatch between experimental and numerical strain fields becomes pronounced. This mismatch level between FE strain fields and experimental strain fields is typical in carefully carried out large deformation tests \citep{Lava2020}. The penalized mismatch is realized through an objective function defined as:
\begin{equation}
\mathcal{L}_{\boldsymbol{\theta}} = \frac{1}{n_p} \sum_{i=1}^{n_s} \sum_{j=1}^{n_p} \left\| \hat{\boldsymbol{\varepsilon}}_j - \boldsymbol{\varepsilon}_j(\boldsymbol{\theta}) \right\|^2_i,
\end{equation}

where $\hat{\boldsymbol{\varepsilon}}_j = [\hat{\varepsilon}_{11}, \hat{\varepsilon}_{22}, \hat{\varepsilon}_{12}]_j$ and $\boldsymbol{\varepsilon}_j(\boldsymbol{\theta}) = [\varepsilon_{11}(\boldsymbol{\theta}), \varepsilon_{22}(\boldsymbol{\theta}), \varepsilon_{12}(\boldsymbol{\theta})]_j$ are the experimental and numerical strain tensors for each measurement point, with $\boldsymbol{\theta}$ representing the parameter vector. $n_p$ is the number of measurement points, and $n_s$ is the number of time steps considered. In this study, ten uniformly spaced time steps are utilized within the learning framework ($n_s = 10$).

\subsection{Parameter optimization}
\label{subsection: Parameter opt}
A deterministic optimization approach is employed to learn the model parameters by minimizing the objective function, $\mathcal{L}_{\boldsymbol{\theta}}$, using the Nelder--Mead simplex method \citep{lagarias1998convergence, singer2009nelder}. The Nelder--Mead algorithm was chosen for its simplicity, versatility, and the distinct advantage of not requiring gradient information. This makes it especially well-suited for optimization problems where derivatives are either unavailable or computationally expensive. Additionally, the Nelder–Mead algorithm is effective for incorporating constraint functions, which is essential for ensuring that the parameters remain within physically meaningful bounds
during optimization.

While the Nelder–Mead algorithm is known for its robustness, it is also well documented that its performance can be hindered by convergence to local minima, especially in high-dimensional or non-convex optimization landscapes. To assess the influence of local minima on the solution, two distinct initial parameter sets, denoted by $\boldsymbol{\theta}^1$ and $\boldsymbol{\theta}^2$, are employed in the optimization procedure, as summarized in Table~\ref{tab:Table 1}. These parameter sets differ in both the hardening behavior and the Hill48 yield parameters, with $\boldsymbol{\theta}^1$ initialized to represent the isotropic limit of the Hill48 yield surface. Lastly, a termination criterion was set for when the relative change in
parameter values between successive iterations reached a threshold of $10^{-8}$.

\subsection{Test specimen geometry and FE simulation}
\label{subsec:Test Geometry & FE Simulation}
\begin{figure}[H]
    \centering
    \includegraphics[width=0.9\columnwidth]{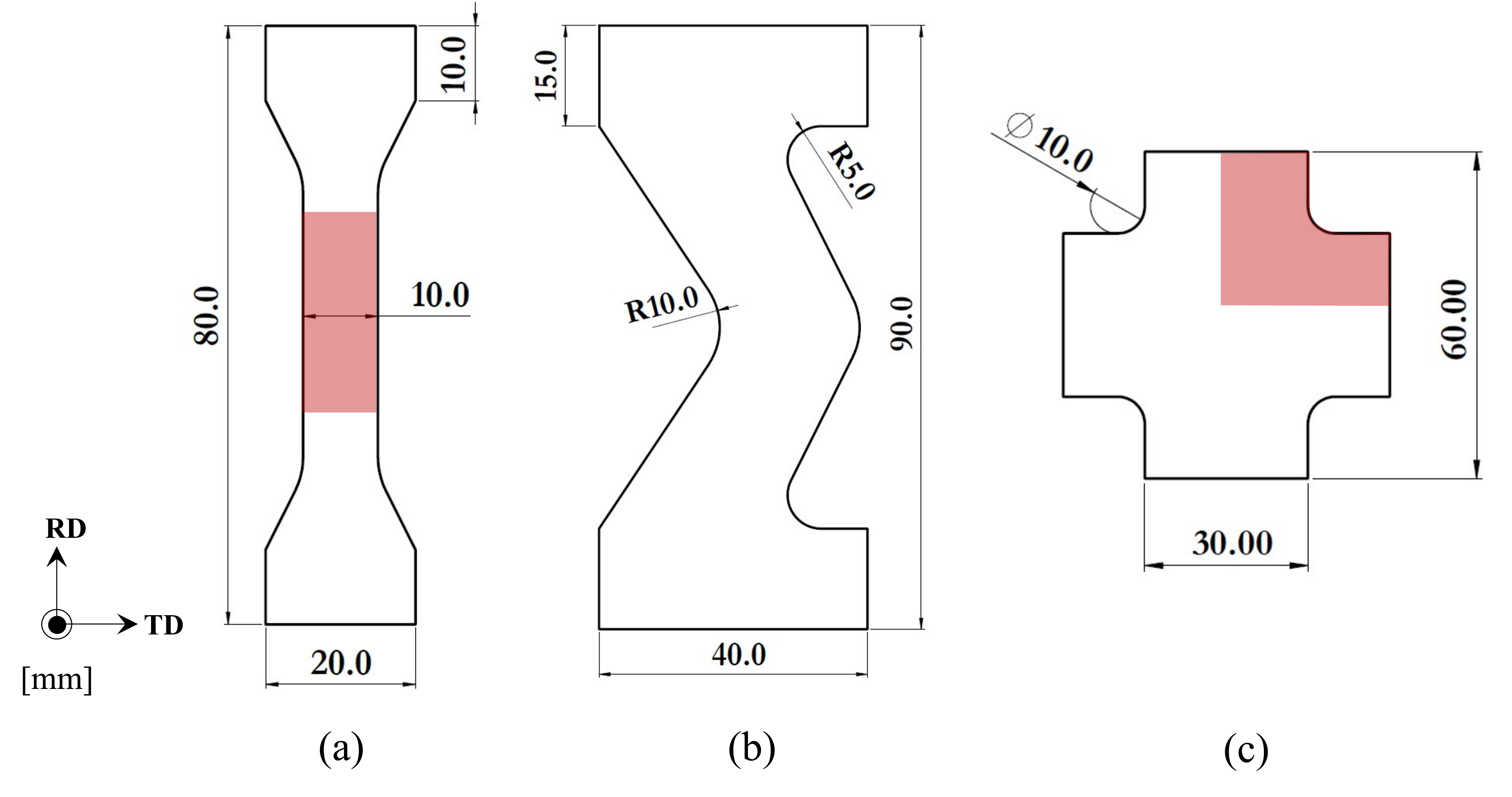} 
    \vspace{-1em}
    \caption{Test specimen geometries used in inverse learning: (a) uniaxial tension specimen, (b) $\Sigma$-shaped specimen \citep{kim2014determination}, and (c) cruciform specimen.
}
    \label{fig: Figure 2}
\end{figure}

To investigate the influence of test specimen information content on accuracy in learning the material law, selected test specimen geometries are considered. The specimens are chosen based on the variations in their geometric complexity, presenting test cases for specimens with different stress state information content. The first two specimens are designed for uniaxial testing conditions, where one end is fixed, and a uniform displacement is applied at the opposite end in the RD. These specimens, subjected to uniaxial tensile boundary conditions are the ubiquitous ``\textit{dog-bone}" uniaxial tension specimen and the $\Sigma$-shaped specimen intuitively designed by \citep{kim2014determination}. We then consider a central region of the biaxial cruciform specimen, subjected to a biaxial tension loading state with the sample stretched in the RD and TD.
Figure \ref{fig: Figure 2} illustrates the specimen geometries 
 highlighting their corresponding regions of interest (ROI) and their dimensions in millimeters. The ROI for the dog-bone specimen consists of the area within its gage length experiencing a uniform uniaxial stress state. The entire area of the $\Sigma$-shaped specimen constitutes its ROI, while one-quarter of the cruciform specimen is considered as its ROI due to its geometric symmetry.

FE simulations were carried out using the ABAQUS/EXPLICIT \textsuperscript{\textregistered} solver, employing 8-noded brick elements with reduced integration and hourglass control to ensure numerical stability. A fine mesh was adopted that strikes a balance between computational efficiency and the smoothness of the resulting strain fields. All simulations were designed to meet the specific boundary condition requirements of the specimens under study. For the uniaxial tension and $\Sigma$-shaped specimens, one end was fixed while a displacement of 2mm was applied along the RD. For the cruciform specimen, a displacement of 2mm was applied simultaneously along both the RD and the TD. The magnitude of the displacement was kept consistent between all specimens to maintain comparability.

\subsection{Specimen information content}
\label{subsection: Specimen Information Content}
Here, we analyze the stress state information content of the considered specimen geometries through the stress state entropy computed using Equation \ref{eq: Stress State Entropy}, according to the steps delineated in the work by Ihuaenyi et al. \citep{ihuaenyi2024seeking}. A natural concern arises regarding the use of model-based stress computations to quantify the information content of a specimen, especially when the model parameters are not known a priori. To address this, the stress field is computed using a set of parameters, denoted by $\boldsymbol{\theta}^1$, which serve as the initial point for the learning procedure. As listed in Table~\ref{tab:Table 1}, the initial yield surface parameters correspond to an isotropic von Mises yield surface. The resulting predicted stress field is then used to evaluate the information content of the specimen. While the exact stress values depend on the chosen constitutive model, the qualitative features of the information content, such as the presence of normal and shear stress components across multiple directions, are primarily determined by the specimen geometry and boundary conditions. A comparative analysis supporting this observation, using the cruciform specimen as a case study, is provided in ~\ref{Appendix: A}.

\begin{figure}[H]
    \centering

    \includegraphics[width=0.8\textwidth]{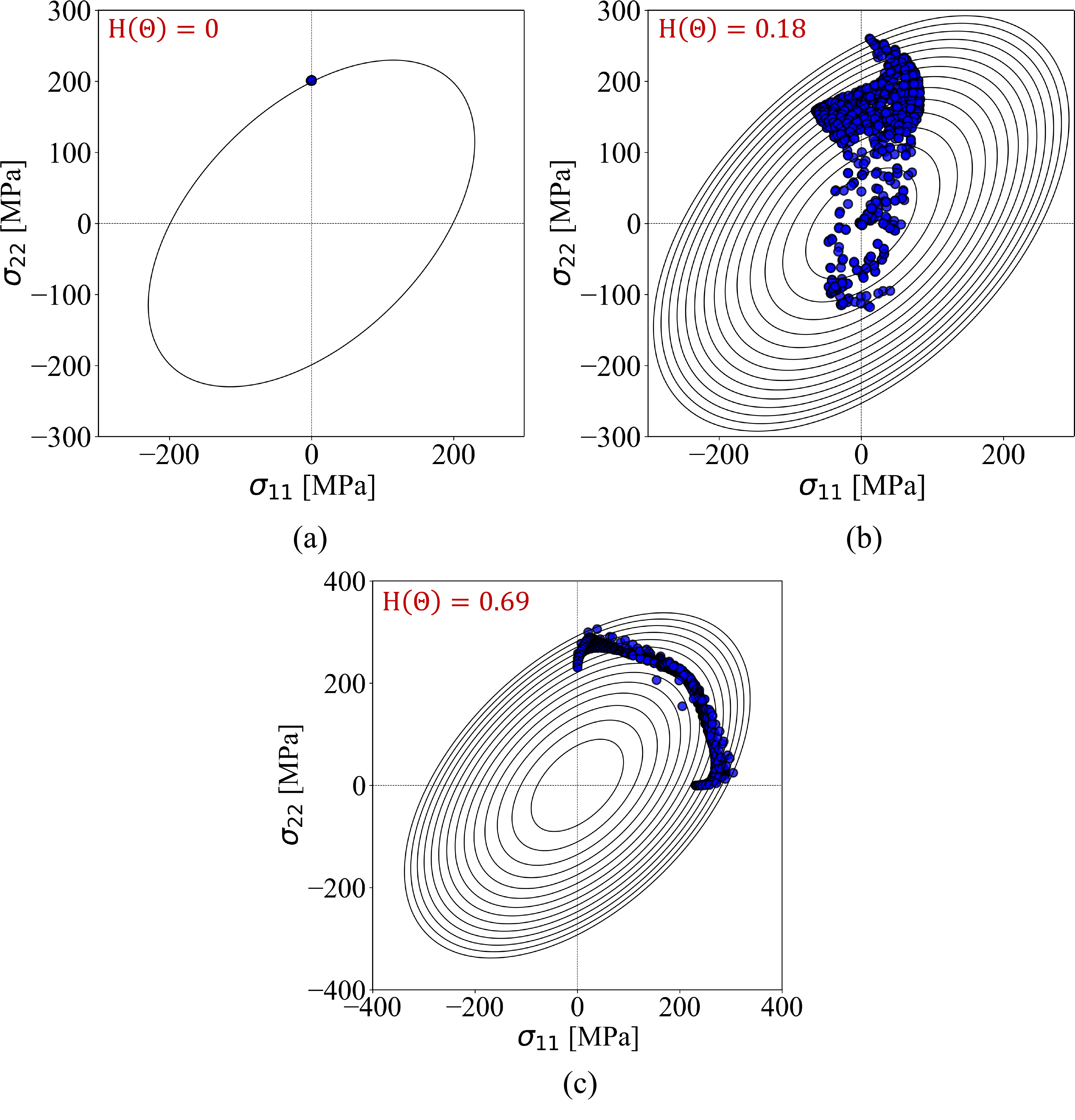} 
    \vspace{-1em}
    \caption{Stress distributions on the $\sigma_{11}$ and $\sigma_{22}$ yield loci, annotated with stress state entropy values for the (a) uniaxial tension, (b) $\Sigma$-shaped, and (c) cruciform specimens.
}
    \label{fig: Figure 3}
\end{figure}

To qualitatively assess the stress state richness of each test specimen, Figure \ref{fig: Figure 3} presents a visual representation of the stress distribution of the specimen geometries, plotted on the $\sigma_{11}$ and $\sigma_{22}$ yield loci. 

\begin{figure}[H]
    \centering
\centering
    \includegraphics[width=0.8\textwidth]{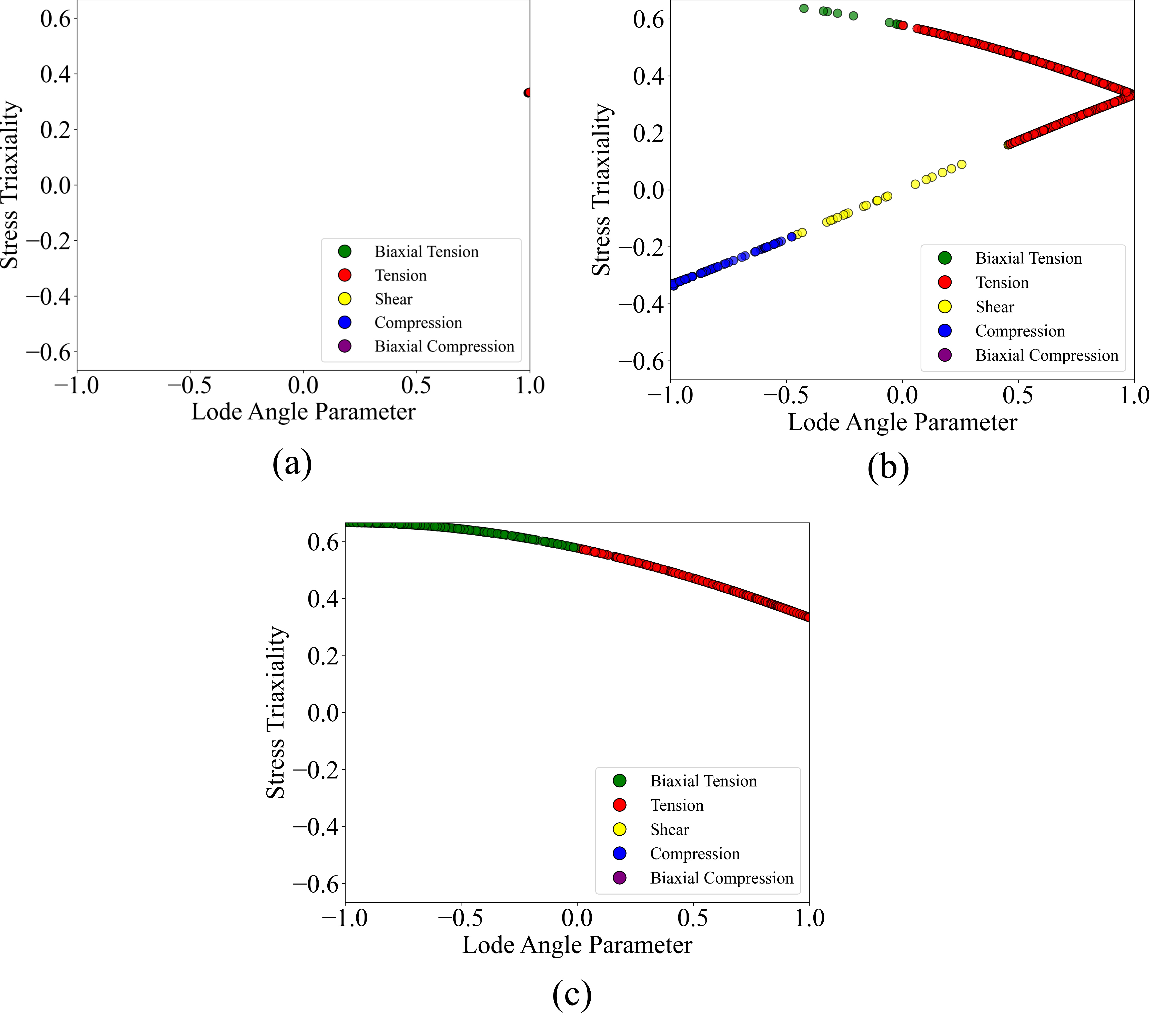} 
        \vspace{-1em}
    \caption{Stress state distributions on the 2D Lode angle parameter--stress triaxiality coordinate system for the (a) uniaxial tension, (b) $\Sigma$-shaped, and (c) cruciform specimens.
}
    \label{fig: Figure 4}
\end{figure}

To offer further qualitative insights into the stress state richness, Figure \ref{fig: Figure 4} illustrates the locations of the generated stress states of each specimen on the 2D Lode angle parameter--stress triaxiality coordinate system. This depiction highlights the diversity, or lack thereof, in the stress states generated by each of the considered test specimens.

From Figures \ref{fig: Figure 3}(a) and \ref{fig: Figure 4}(a), the stress state generated by the uniaxial tension specimen is distinctly uniform, concentrated at a single point on the yield surface. This corresponds to a purely uniaxial tensile response along the RD, with a stress state entropy of $\text{H}(\Theta) = 0$. This entropy value indicates the specimen provides no information about the broader spectrum of stress states required to learn the material law accurately. 

In contrast, the $\Sigma$-shaped specimen demonstrates a significantly more diverse stress state distribution, as evidenced by its complex profile on the yield surface (Fig. \ref{fig: Figure 3}(b)). Mapping this distribution in the 2D Lode angle parameter--stress triaxiality space (Fig. \ref{fig: Figure 4}(b)) reveals a combination of tension, compression, and shear states. The enhanced information content is quantified by a stress state entropy of $\text{H}(\Theta) = 0.18$, highlighting the specimen’s capability to capture a wider range of stress states relevant to the stress space under consideration. Nevertheless, the tensile stress state along the RD remains the predominant feature, limiting the overall stress state entropy value.

Lastly, the cruciform specimen generates stress states characterized by uniaxial tension in its arms and biaxial tension near its center. This specimen geometry effectively captures the uniaxial tensile response along both the RD and the TD. As a result, the stress state entropy for the cruciform specimen is markedly higher, with $\text{H}(\Theta) = 0.69$. Which is the maximum stress state entropy for a stress state space with cardinality $n=2$. This high entropy value reflects the specimen's capability to represent tensile stress states in the RD and TD with equal distributions. However, despite its comprehensive coverage of tensile states, the cruciform specimen provides no information on shear stress state, limiting its utility in fully exploring the stress space under consideration.

\subsection{Parameter identification}
\label{subsec:identification}
The three test specimens are evaluated for their effectiveness in accurately learning the anisotropic inelastic material law. To learn the material law, six parameters, ($A$, $\sigma_0$, $n$, $F$, $G$, and $N$), are to be identified. An attempt is made to simultaneously identify all parameters from the strain fields generated by each geometry. The resulting identified parameters, along with the absolute error for each learnt parameter, are presented in Table \ref{tab:Table 1}, highlighting the distinct contributions of each specimen's information content to the accuracy of parameter identification.

\begin{table}[H]
    \centering
    \caption{Identified parameter sets using the uniaxial tension, $\Sigma$-shaped, and cruciform specimens}.
    \vspace{0.2cm} 
    \label{tab:Table 1}
    {\footnotesize
    \begin{tabular}{llllllll} 
        \hline
          &&  $A$ [MPa] & $\sigma_0$ [MPa]& $n$ & $F$ & $G$ & $N$ \\ 
        \hline
         Ground Truth&& 471.92& 123.4& 0.29 & 0.278 & 0.373 & 2.340 \\
Initial 1 & $\boldsymbol{\theta}^1$ & 600 & 90 & 0.4 & 0.5 & 0.5 & 1.5\\
 Initial 2& $\boldsymbol{\theta}^2$ & 300& 150& 0.1& 0.13& 0.13&3.5\\ 
         Uniaxial Tension& $\boldsymbol{\theta}^1$& 481.22& 111.67& 0.32& 0.210& 0.413& 1.947\\
  &Abs. Error (\%)& 1.97& 9.51& 10.34& 24.46& 10.72&16.79\\
  &$\boldsymbol{\theta}^2$& 466.83& 111.66& 0.32& 0.212& 0.421&2.056\\ 
         &Abs. Error (\%)&  1.08&  9.51&  10.34& 23.74& 12.87& 12.14\\ 
         $\Sigma$-Shaped &$\boldsymbol{\theta}^1$& 471.92& 111.80& 0.33& 0.304& 0.367& 2.350\\ 
         &Abs. Error (\%)
&  0&  9.4&  13.79& 9.35& 1.61& 0.43\\
 & $\boldsymbol{\theta}^2$
& 476.15& 111.54& 0.34& 0.303& 0.369&2.370\\
 & Abs. Error (\%)& 0.90& 9.61& 17.24& 8.99& 1.07&1.28\\ 
         Cruciform &$\boldsymbol{\theta}^1$& 464.07& 113.61& 0.30& 0.275& 0.375& 2.120\\
 & Abs. Error (\%)
& 1.66& 7.93& 3.45& 1.08& 0.54&9.40\\
 & $\boldsymbol{\theta}^2$
& 460.4& 109.00& 0.30& 0.274& 0.375&2.100\\ 
         &Abs. Error (\%)&  2.44&  11.67&  3.45& 1.44& 0.54& 10.26\\ 
        \hline
    \end{tabular}
    }
\end{table}

 Convergence to the global minimum is achieved after approximately 100 iterations for all considered specimen as shown in Figure \ref{fig: Figure B1}. Furthermore, Figures ~\ref{fig: Figure B2}, ~\ref{fig: Figure B3}, and ~\ref{fig: Figure B4} present a visual comparison between the ground truth and identified strain fields for all three specimens. The figures also highlight error plots ($\hat{\boldsymbol{\varepsilon}}-\boldsymbol{\varepsilon(\boldsymbol{\theta})}$), indicating minimal errors in the strain field reconstructed using the identified parameters.  The agreement between the ground truth and reconstructed strain fields shows the effectiveness of the optimization procedure in minimizing the objective function. Also, as shown in Table~\ref{tab:Table 1} the optimization scheme showed minimal sensitivity to the initialization. However, the accuracy of the identified material law parameters varies across specimens. This discrepancy arises because no single specimen provides sufficient information to accurately identify all material parameters.

\begin{figure}[H]
    \centering
    \includegraphics[width=0.85\textwidth]{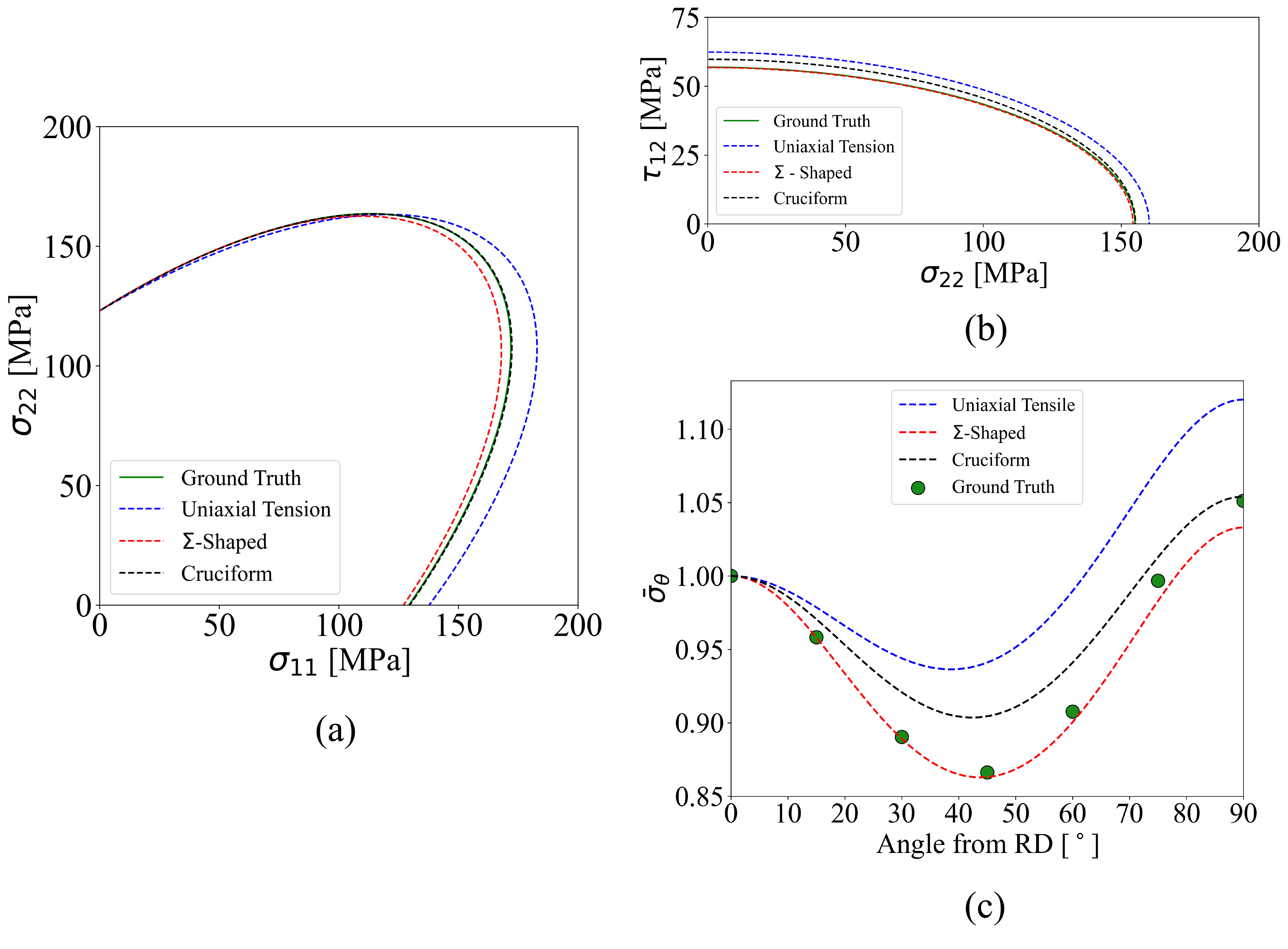} 
        \vspace{-1em}
    \caption{Comparison between the ground truth and identified anisotropic yield response in key stress planes: (a) $\sigma_{11}$--$\sigma_{22}$ plane, (b) $\sigma_{22}$--$\tau_{12}$ plane, and (c) normalized yield stress as a function of the orientation angle relative to the rolling direction (RD).}
    \label{fig:Figure 5}
\end{figure}

To further evaluate how closely the identified parameters reflect the material law, we analyze the predicted yield surface and yield stress anisotropy in Figure~\ref{fig:Figure 5}. Figures~\ref{fig:Figure 5}(a) and \ref{fig:Figure 5}(b) depict the first quadrant of the symmetric yield surface in the $\sigma_{11}$--$\sigma_{22}$ and $\sigma_{22}$--$\tau_{12}$ planes, respectively, comparing the known yield surface with the predicted yield surfaces. Additionally, Figure~\ref{fig:Figure 5}(c) illustrates the evolution of the normalized yield stress as a function of the angle of orientation from the RD. The yield stresses obtained using the ground truth parameters are compared against those predicted using the identified parameters. A quantitative comparison of the predicted yield stresses with respect to angles of orientation to the RD, ranging from 15$^\circ$ to 90$^\circ$ is presented in Table \ref{tab:Table D1}.

The results reveal that the limited stress state information of the uniaxial tension specimen, essentially confined to simple tension along the RD, restricts its effectiveness in accurately learning the material law. As this specimen does not generate a sufficiently informative data set to fully constrain the optimization, parameter adjustments compensate to approximate the ground truth data as closely as possible. Consequently, all material parameters, except for the hardening parameter ($A$), are identified with significant errors. This limitation is evident in the discrepancies observed between the predicted and ground truth yield surfaces (Figures~\ref{fig:Figure 5}a and \ref{fig:Figure 5}(b)). As a result, the identified parameters fail to accurately learn the hardening behavior and yield stress anisotropy. While the predicted yield surface along the RD and the normalized yield stress at the RD (0$^\circ$) align well with the ground truth value (Figure~\ref{fig:Figure 5}(c)), the variation in yield stress across different angles of orientation is poorly represented (Table \ref{tab:Table D1}). These findings are consistent with those of \citep{Pottier2011}, which emphasizes the necessity of incorporating data from uniaxial tensile test specimens oriented along the RD, TD, and at a 45$^\circ$ angle to adequately capture such a material law. In essence, enhancing the information content of the data through multiple tests.

As the information content of the test specimen increases, the material parameters become progressively decoupled, revealing their sensitivity to the complexity of strain fields associated with specific stress states. The $\Sigma$-shaped specimen provides stress state information relevant to both uniaxial tension in the RD and shear. This additional information imposes constraints on the optimization process, enabling accurate identification of the parameters \(G\) and \(N\), which govern the inelastic uniaxial response in the RD and the shear response, respectively. This accuracy is reflected in the agreement between the predicted and ground truth yield surfaces in the $\sigma_{22}$--$\tau_{12}$ plane, as well as the close alignment of normalized yield stresses across orientations from 0$^\circ$ to 60$^\circ$. However, the limited stress state information for uniaxial tension in the TD led to inaccuracies in identifying the parameter \(F\). This error resulted in poor agreement with the yield surface in the $\sigma_{11}$--$\sigma_{22}$ plane and in the normalized yield stress predictions as the orientation angle approached 90$^\circ$ (Figure~\ref{fig:Figure 5}(c) and  Table \ref{tab:Table D1}). Additionally, the data from this specimen was insufficient to fully decouple all parameters, leaving some of the hardening parameters ($\sigma_0$ and $n$) inaccurately identified. To accurately learn the material law using the $\Sigma$-shaped specimen, a combination of multiple tests is also required. As demonstrated by \citep{kim2014determination}, combining two tests with $\Sigma$-shaped specimens oriented in the RD and TD is necessary to overcome its limited tensile stress information in the TD for accurately learning the material law.

Lastly, the cruciform specimen generates uniaxial stress states in both the RD and TD, enabling the accurate identification of the parameters \(F\) and \(G\), which govern the material's inelastic tensile response in the RD and TD, respectively. The accuracy of these predictions is evident in the close alignment between the predicted and ground truth yield surfaces in the $\sigma_{11}$--$\sigma_{22}$ plane, as well as in the normalized yield stress predictions at the 0$^\circ$ and 90$^\circ$ orientations. However, the absence of shear stress state information in this specimen limits its accuracy in identifying the parameter \(N\). This limitation is reflected in the poor agreement between the predicted and ground truth yield surfaces in the $\sigma_{22}$--$\tau_{12}$ plane, as well as in the inaccurate prediction of normalized yield stresses at orientations between 0$^\circ$ and 90$^\circ$, where the shear response plays a significant role. Also, since the cruciform specimen does not generate all required stress states to fully decouple the material parameters, the hardening parameters ($\sigma_0$ and $n$) were identified with errors. This outcome is consistent with the findings of \citep{Martins2019}, which suggests that a modified cruciform specimen capable of generating shear stress response is necessary for accurately learning the material law. 

These results highlight the crucial role of stress state information in accurately learning material laws. Specimens capable of generating multiple stress states, such as the $\Sigma$-shaped and cruciform specimens, provide richer datasets that enable the accurate identification of certain material law parameters. The diversity of stress state information in the data effectively constrains the optimization process, reducing parameter coupling and improving prediction accuracy. However, none of the test specimens examined provided information on the stress states ($\Theta = \{ \sigma_{\text{UT}}^{\text{RD}}, \sigma_{\text{UT}}^{\text{TD}}, \sigma_{\text{S}} \}$) required accurately learn the material law in a single test. This raises an essential question: \textit{``How can we enhance the information content of our test data to accurately learn the constitutive material law?"}. Two potential strategies emerge. The first approach involves integrating data from multiple tests conducted under varied loading states to ensure a diverse range of stress conditions. Alternatively, a single test could be designed to achieve an optimal stress state entropy, maximizing the information richness for parameter identification. This study focuses on implementing the second strategy.

\section{Informative specimen design}
\label{sec:Design}
The preliminary attempts to learn the material law using a single test raised a crucial question: \textit{``How can we enhance the information content of our test data to accurately learn the constitutive material law?"}. To address this challenge, we implement a methodology proposed in our previous work \citep{ihuaenyi2024seeking} that incorporates stress state entropy as an objective function within a Bayesian optimization (BO) framework for designing test specimens. This approach facilitates the inverse design of specimen geometries by optimizing the stress state entropy. Thus, we can enrich the data available for accurately learning the material law. The biaxial cruciform specimen (Fig. \ref{fig: Figure 2}(c)) serves as the base geometry for the inverse design, as it generates two of the three stress states required and has the highest stress state entropy among the current candidates ($\text{H}(\Theta) = 0.69$).

The optimization process utilizes the Tree-Structured Parzen Estimator (TPE) algorithm \citep{Bergstra2011}, to optimally introduce and modify geometric features in the base geometry to achieve the stress state entropy requirement. TPE is chosen for its ability to perform an informed and adaptive search for optimal parameters, leveraging insights from previously evaluated parameter configurations to select new promising configurations. The parameter space $\mathbf{P} = \{x_i, y_i, r_i\}$ is explored, where $x_i$ and $y_i$ define the positions of geometric inclusions, and $r_i$ represents the radius of any notches or holes introduced. In this study, the number of holes ($i$) was randomly generated while respecting the specimen's geometric constraints, with an upper limit of six holes imposed. To increase the information content of the designed specimen, the design process is formulated as:
\begin{equation}
\textbf{P}^* = \arg\max_{\textbf{P}} \text{H}(\Theta).
\end{equation}
Additionally, the BO framework employs Sequential Model-Based Global Optimization (SMBO) as an iterative method \citep{hutter2011smbo,shahriari2015bayesian}. SMBO initializes by building a probability model for the objective function and identifies locally optimal parameters through an acquisition function. This function guides the search by striking a balance between the \textit{exploration} of the search space and the \textit{exploitation} of promising regions in the search space. The expected improvement (EI) is used as the acquisition function, defined as:
\begin{equation}
\text{EI}_{(\text{H}^*)} (\mathbf{P}) = \int_{-\infty}^{\text{H}^*} (\text{H}^* - \text{H})p(\text{H}\mid\mathbf{P}) \, \text{d}\text{H},
\end{equation}
where \( \text{H}^* \) is a threshold entropy. Given the parameter space \( \mathbf{P} \), \( \text{EI} \) is the expectation that \( \text{H} \) will exceed \( \text{H}^* \). Furthermore, TPE uses Bayes’ theorem to decompose \( p(\text{H}\mid \mathbf{P}) \) into \( p(\mathbf{P}\mid \text{H}) \) and \( p(\mathbf{P}) \) in the form:
\begin{equation}
p(\text{H}\mid\mathbf{P}) = \frac{p(\mathbf{P}\mid \text{H})p(\text{H})}{p(\mathbf{P})},
\end{equation}
\begin{equation}
p(\mathbf{P}\mid\text{H}) = 
\begin{cases}
    l(\mathbf{P}), & \text{H} < \text{H}^* \\
    g(\mathbf{P}), & \text{H} \geq \text{H}^*
\end{cases},
\end{equation}
where \( l(\mathbf{P}) \) is a Gaussian Mixture Model (GMM) fitted by the TPE to the set of parameter values associated with the best objective values, and \( g(\mathbf{P}) \) is another GMM constructed simultaneously using the remaining parameter values.

The algorithm was integrated into a comprehensive framework that couples the optimization scheme with ABAQUS\textsuperscript{\textregistered} FE software. The FE software was responsible for generating the specimen model given the parameters from the BO algorithm, applying the mesh and the necessary boundary conditions, running the simulation, and outputting the stress field. The stress field is then discretized to evaluate the stress state entropy values at each iteration point. The optimization process demonstrated high efficiency, consistently generating a posterior parameter distribution for specimen geometries with optimal stress state entropy (Eq. \ref{eq:entropy_criterion}) within 200 iterations

It is worth noting that there exists a trade-off between the number of geometric features introduced (e.g., holes) and the resulting stress state entropy in the specimen. As demonstrated through detailed case studies \citep{ihuaenyi2024seeking}, an optimal configuration of features can maximize the stress state entropy. Beyond this point, adding more features does not lead to further improvements and may even reduce the stress state entropy due to the reduced number of material points available to carry mechanical information.

\begin{figure}[H]
    \centering
    \includegraphics[width=0.8\textwidth]{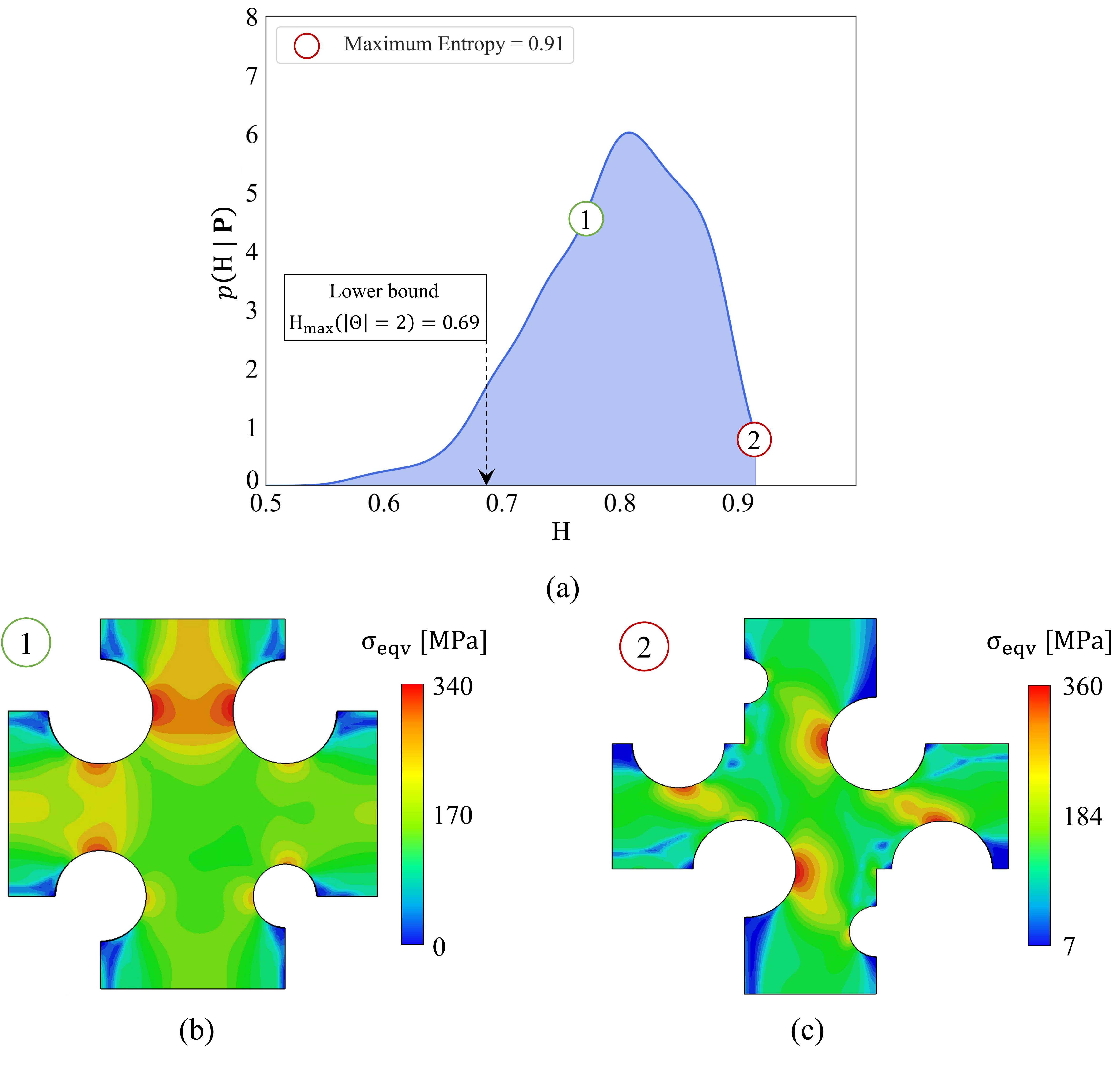} 
        \vspace{-1em}
    \caption{Informative specimen design: (a) Posterior density of the inverse design space, highlighting two sampled points (1 and 2) with stress state entropy values of 0.78 and 0.91, respectively. The corresponding specimen geometries and their equivalent stress fields: (b) Cruciform 1 and (c) Cruciform 2.}

    \label{fig:Figure 6}
\end{figure}

Figure \ref{fig:Figure 6}(a) illustrates the posterior distribution of the stress state entropy given the geometric parameter vector, where each point in the distribution represents a uniquely designed geometry defined by a specific parameter set and its associated stress state entropy. The majority of the designed geometries exhibited entropy values of 0.81, with the maximum stress state entropy for a designed geometry reaching 0.91. Additionally, Figure \ref{fig:Figure 6}(b) and Figure \ref{fig:Figure 6}(c) highlight the two geometries sampled at points 1 and 2 (beyond the lower bound of the maximum entropy hypothesis) with respective stress state entropy values of 0.78 and 0.91. These specimens are selected to assess their stress state information content, viability for learning the material law, and to verify the optimal entropy hypothesis (Eq. ~\ref{eq:entropy_criterion}). Notably, despite the difference in their entropy values, both specimens fall within the hypothesized optimal entropy range. Detailed geometric dimensions for these optimal specimens are illustrated in  Figure \ref{fig: Figure C1}.

\begin{figure}[H]
    \centering
    \includegraphics[width=0.9\textwidth]{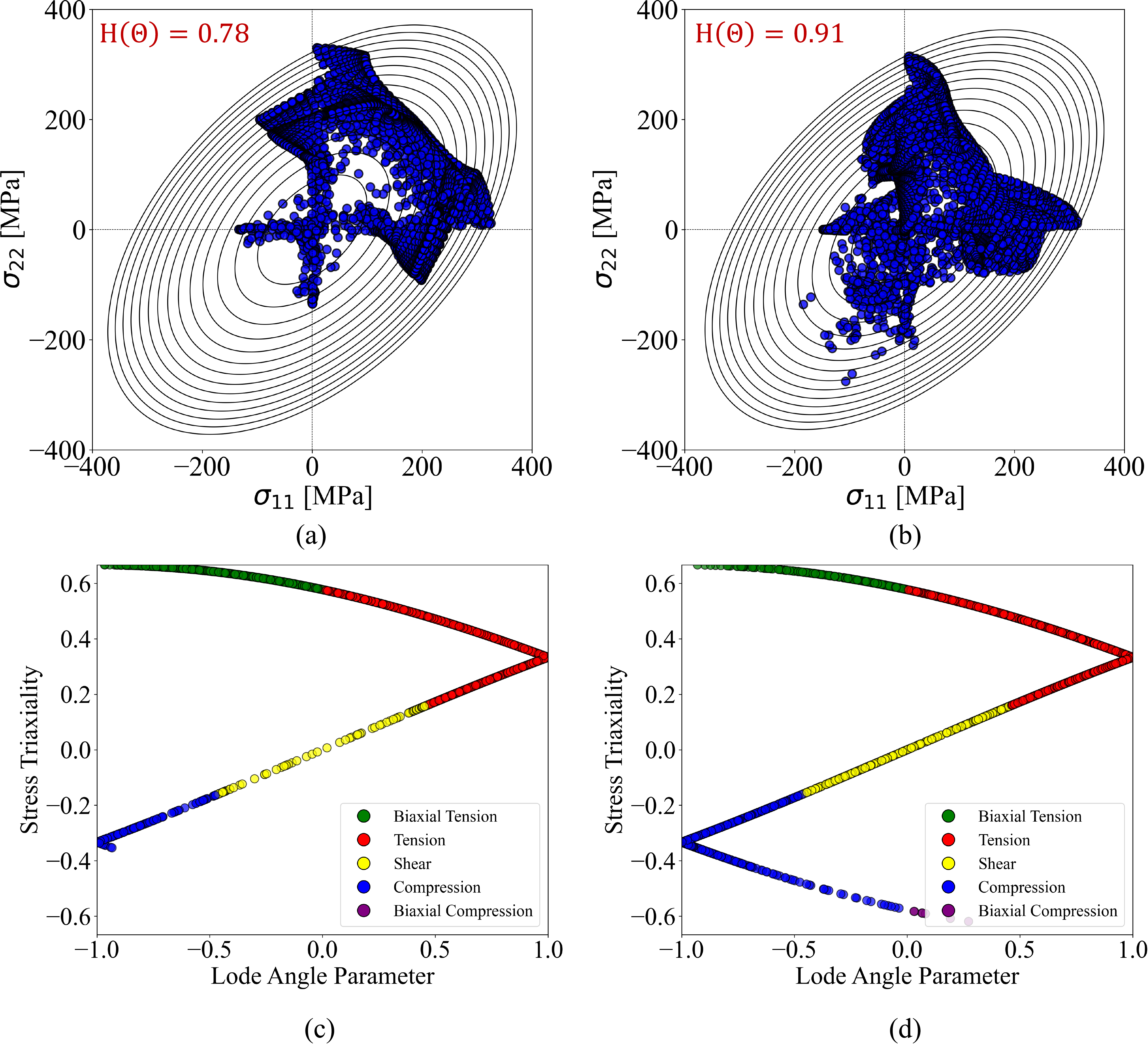} 
    \vspace{-1em}
    \caption{Stress distributions on the $\sigma_{11}$ and $\sigma_{22}$ yield loci, annotated with stress state entropy values for (a) Cruciform 1 and (b) Cruciform 2. Additionally, stress state distributions on the 2D Lode angle parameter–stress triaxiality coordinate system for (c) Cruciform 1 and (d) Cruciform 2.
}
    \label{fig: Figure 7}
\end{figure}

Figure \ref{fig: Figure 7} gives a qualitative illustration of the stress state information of the optimized specimens, which exhibit increased stress state entropy compared to the uniaxial tension, $\Sigma$-shaped, and cruciform specimens. In Figures \ref{fig: Figure 7}(a) and \ref{fig: Figure 7}(b), the stress distributions along the $\sigma_{11}$--$\sigma_{22}$ yield loci for the optimized specimens reveal a broader coverage of the yield surface. Furthermore, their enhanced stress state entropy is evident in the 2D Lode angle parameter – stress triaxiality space, as shown in Figures \ref{fig: Figure 7}(c) and \ref{fig: Figure 7}(d).

From Figure \ref{fig: Figure 7}, it is evident that the optimized specimens generate a diverse range of stress states, including uniaxial tension along the RD, biaxial tension compensating for the TD, and shear. Hence, the optimized specimens provide sufficient stress state information for accurately learning the material law. Although not considered in our stress state space, the test specimens also show compressive stress state information. The difference in the stress state entropy between the optimized specimens is also qualitatively evident in Figure \ref{fig: Figure 7}, where cruciform 2 with a larger stress state entropy value exhibits a broader distribution of stress states, both across the yield surface and within the 2D Lode angle parameter--stress triaxiality space.

\section{Performance of informative specimen in inverse learning}
\label{sec:Performance of Informative Specimen}
Here, we evaluate the optimized cruciform specimens for their effectiveness in accurately learning the material law. Additionally, to assess the robustness of these specimens under realistic testing conditions, Gaussian white noise, ${\mathcal{N}}(\mu = 0, \sigma^2)$, is added to the experimental strain fields at two distinct levels. The first noise level, with a standard deviation of $\sigma=10^{-3}$, represents the typical uncertainty observed in carefully conducted large strain measurements \citep{Sutton2009}. The second noise level, characterized by a standard deviation of $\sigma = 5 \times 10^{-3}$, represents an elevated noise scenario, enabling a rigorous evaluation of specimen performance under challenging experimental conditions. Table \ref{tab:Table 2} provides a summary of the material parameters identified across these cases.

\begin{table}[H]
    \centering
    \caption{Identified parameter sets for the optimized specimen geometries.}
    \vspace{0.2cm} 
    \label{tab:Table 2}
    {\footnotesize
    \begin{tabular}{cccccccc} 
    \hline
          &${\mathcal{N}}(0, \sigma^2)$&  $A$ [MPa] & $\sigma_0$ & $n$ & $F$ & $G$ & $N$ \\ 
    \hline    
          Ground Truth&-& 471.92& 123.4& 0.29 & 0.278 & 0.373 & 2.340 \\
 Initial & $\boldsymbol{\theta}$& 600 & 90 & 0.4 & 0.5 & 0.5 &1.5\\ 
          Cruciform 1&Noise-free& 465.05& 121.07& 0.29& 0.276& 0.374& 2.350\\ 
          &Abs. Error (\%)&  1.46&  1.89&  0.00& 0.72& 0.27& 0.43\\ 
          &${\mathcal{N}}(0, (10^{-3})^2)$& 468.75& 121.77& 0.29& 0.275& 0.377& 2.350\\ 
          &Abs. Error (\%)&  0.67&  1.32&  0.00& 1.08& 1.07& 0.43\\ 
          &${\mathcal{N}}(0, (5\times 10^{-3})^2)$& 491.96& 118.65& 0.28& 0.275& 0.365& 2.383\\ 
          &Abs. Error (\%)&  4.25&  3.85&  3.45& 1.08& 2.14& 1.84\\ 
        
 Cruciform 2& Noise-free & 462.75& 121.00& 0.29& 0.278& 0.373&2.339\\ 
 & Abs. Error (\%)& 1.94& 1.94& 0.00& 0.00& 0.00&0.04\\ 
 & ${\mathcal{N}}(0, (10^{-3})^2)$& 466.24& 120.65& 0.29& 0.281& 0.380&2.360\\ 
 & Abs. Error (\%)& 1.20& 2.23& 0.00& 1.08& 1.88&0.85\\ 
 & ${\mathcal{N}}(0, (5\times 10^{-3})^2)$& 502.71& 125.52& 0.26& 0.269& 0.361&2.447\\ 
 & Abs. Error (\%)& 6.52& 1.72& 10.34& 3.24& 3.22&4.57\\ 
    \hline
    \end{tabular}%
    }
    
\end{table}

The results demonstrate that the optimized test specimens generate sufficient stress state information for accurate parameter identification. The model parameters are all identified with minimal errors, as summarized in Table~\ref{tab:Table 2}. Notably, the performance of test specimens with stress state entropy within the optimal range verifies the optimal entropy hypothesis. As long as the stress state entropy of the test specimen falls within the bounds defined by Equation~\ref{eq:entropy_criterion}, accurate parameter identification is achievable.

\begin{figure}[H]
    \centering
    \includegraphics[width=0.9\textwidth]{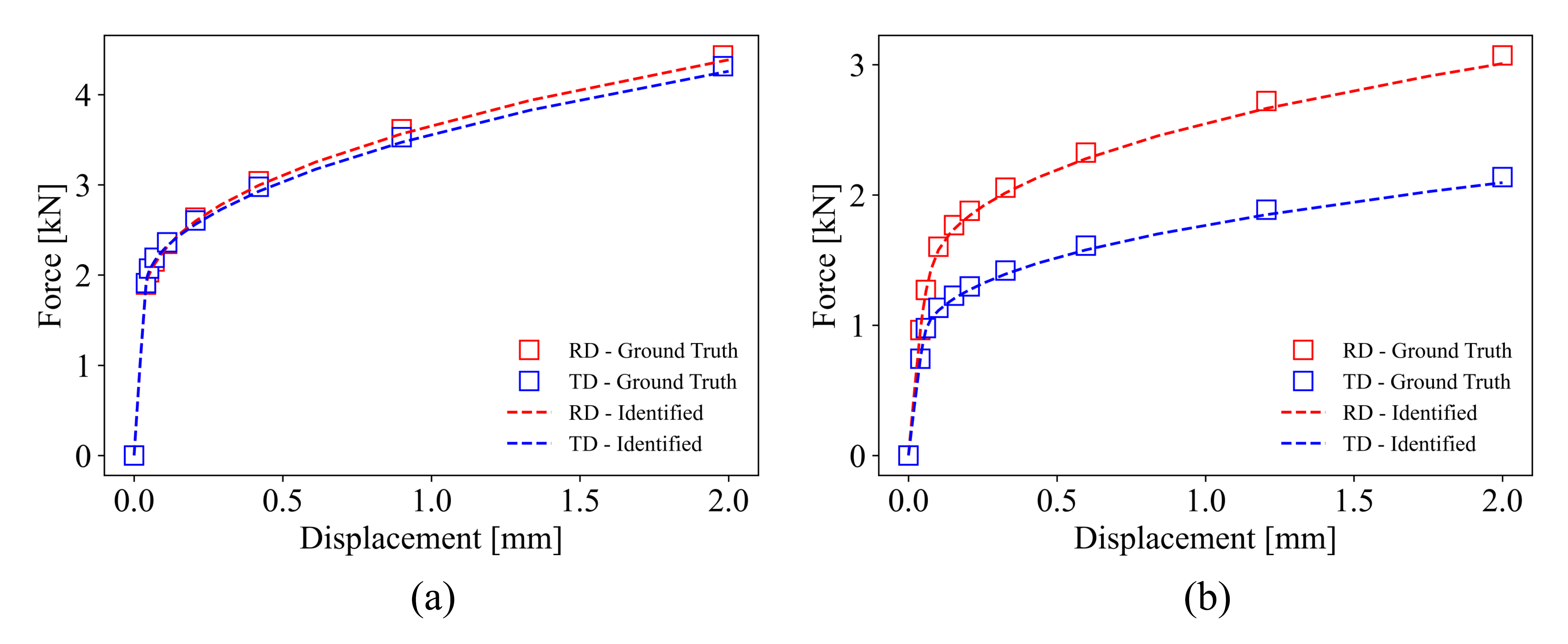} 
    \vspace{-1em}
    \caption{Comparison between the ground truth and identified force–displacement responses for the optimally designed specimens, evaluated in both the rolling direction (RD) and transverse direction (TD): (a) Cruciform 1, and (b) Cruciform 2.}
    \label{fig: Figure 8}
\end{figure}

\begin{figure}[H]
    \centering
    \includegraphics[width=0.85\textwidth]{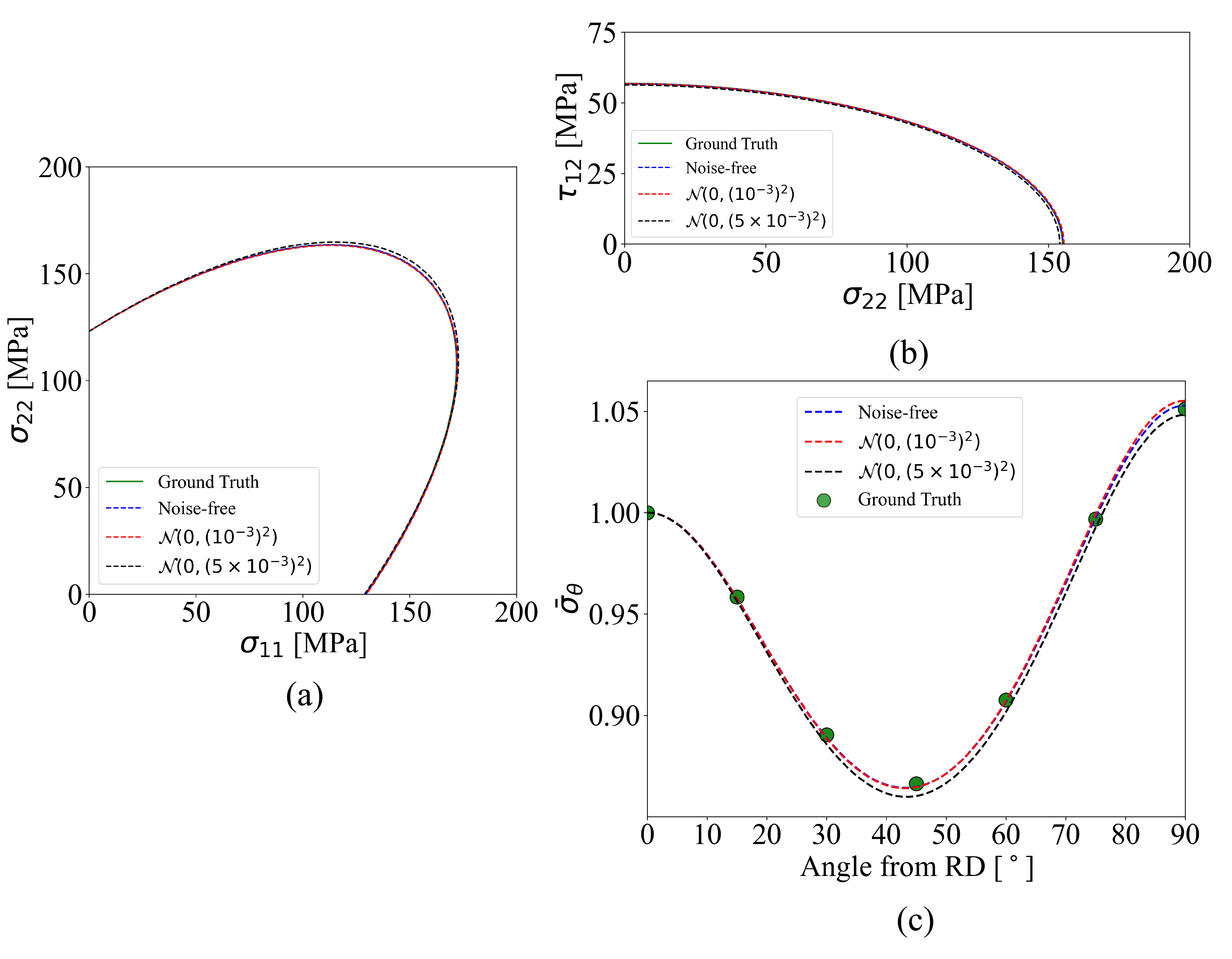} 
        \vspace{-1em}
    \caption{Comparison between the ground truth and identified anisotropic yield response for the optimally designed Cruciform 1 specimen across key stress planes: (a) yield contours in the $\sigma_{11}$--$\sigma_{22}$ plane, (b) yield contours in the $\sigma_{22}$--$\tau_{12}$ plane, and (c) normalized yield stress as a function of the orientation angle relative to the rolling direction (RD).
}
    \label{fig: Figure 9}
\end{figure}

\begin{figure}[H]
    \centering
    \includegraphics[width=0.85\textwidth]{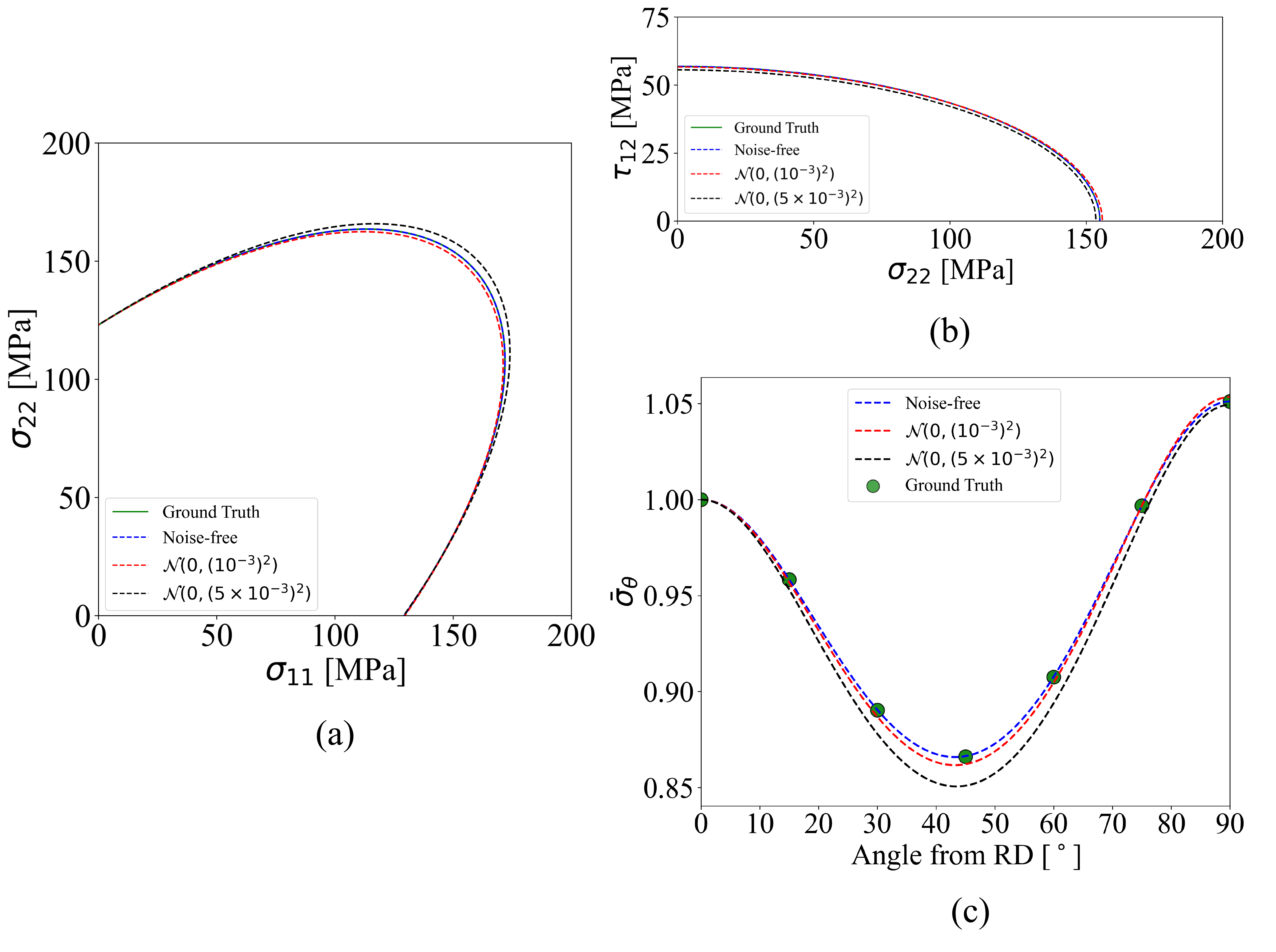} 
        \vspace{-1em}
    \caption{Comparison between the ground truth and identified anisotropic yield behavior for the optimally designed Cruciform 2 specimen across key stress planes: (a) yield contours in the $\sigma_{11}$--$\sigma_{22}$ plane, (b) yield contours in the $\sigma_{22}$--$\tau_{12}$ plane, and (c) normalized uniaxial yield stress as a function of orientation angle relative to the rolling direction (RD).
}
    \label{fig: Figure 10}
\end{figure}

To visualize the accuracy of the identified parameters and their robustness to experimental noise, Figure~\ref{fig: Figure 8} compares ground truth and predicted force--displacement response in the RD and TD, reconstructed from the identified parameters for the noise-free case. Additionally, Figures~\ref{fig: Figure 9}(a) and~\ref{fig: Figure 9}(b), along with Figures~\ref{fig: Figure 10}(a) and~\ref{fig: Figure 10}(b), provide a detailed comparison between the identified and ground truth yield surfaces. Lastly, the evolution of the normalized yield stress as a function of the orientation angle relative to the RD is presented in Figures~\ref{fig: Figure 9}c and ~\ref{fig: Figure 10}(c). A quantitative comparison between known and predicted yield stresses with respect to orientation angle is also presented in Table \ref{tab:Table D1} for the noise-free case. These results demonstrate that the identified parameters accurately capture the evolution of yield stress and the shape of the yield surface.

The results also reveal that parameter identification using the optimally designed specimens is largely robust to the considered experimental noise levels. However, an interesting trend emerges among the test specimens. Cruciform 2 with a larger stress state entropy value demonstrates greater sensitivity to noise. This observation is expected, as specimens with more informative data, while providing richer point-to-point information content, are inherently more vulnerable to experimental noise. This delicate trade-off between maximizing the information content and maintaining robustness to experimental uncertainties highlights a crucial consideration in test specimen design. It emphasizes the need to balance data richness with robustness to noise in the context of inverse learning and uncertainty quantification.

\section{Uncertainty quantification}
\label{subsec:uncertainty}
Uncertainty Quantification (UQ) is essential for ensuring both accuracy and reliability in inverse learning, particularly when employing geometrically complex and highly informative test specimens. Although these specimens generate informative data, due to their geometric complexities, they often introduce measurement uncertainties. This is especially true when using full-field techniques such as DIC, and quantifying such uncertainties is generally ignored in the deterministic parameter identification framework. Here, we leverage Bayesian UQ to rigorously quantify the aleatoric uncertainty associated with the full-field data. We also assume that the model fully captures the material behavior, thus excluding epistemic uncertainty. By establishing credible intervals for the identified parameters, this approach assesses the effect of experimental noise on parameter estimation. The Bayesian framework enables us to infer credible bounds in the parameter space, providing a robust framework for inverse learning and evaluating the robustness of test data in the presence of uncertainties.

In the Bayesian inverse learning framework, the model parameters are treated as random variables, denoted by \( \boldsymbol{\theta} \). These parameters are assumed to follow independent prior distributions, each characterized by a probability density function, \( p(\boldsymbol{\theta}) \). The primary objective of this approach is to identify the parameters governing the material behavior while rigorously accounting for uncertainties inherent in the experimental data. This goal is achieved by updating the prior knowledge of the parameters through the incorporation of observed data and can be formally expressed as:
\begin{equation}
\mathbf{y} = \mathcal{G}({\boldsymbol\theta}) + {\mathcal{N}},
\end{equation}
where \(\mathbf{y}\) represents the full-field experimental data, \(\mathcal{G}(\boldsymbol{\theta})\) is the forward operator or model, and \({\mathcal{N}}\) represents the uncertainty in the data. Applying Bayes' theorem to update the parameter distribution \(\boldsymbol{\theta}\) is fundamental to the framework, and it is expressed as:
\begin{equation}
p(\boldsymbol{\theta} \mid \mathbf{y}) = \frac{p(\mathbf{y} \mid \boldsymbol{\theta}) \cdot p(\boldsymbol{\theta})}{p(\mathbf{y})}.
\label{eq:Bayes}
\end{equation}
The probability density function for the prior distribution \( p(\boldsymbol{\theta}) \) of the model parameters can either be informative or uninformative, depending on the level of confidence in the available prior knowledge regarding the parameter distribution. The prior distribution encodes the assumptions or knowledge about the parameters before any data is observed, thus influencing the posterior parameter estimates. In cases where limited or no prior knowledge is available, an uninformative prior is typically used. A uniform distribution is often adopted as the uninformative prior, as it represents a state of complete uncertainty about the parameters. In this study, we select a uniform distribution as the prior, expressed as:
\begin{equation}
p(\boldsymbol{\theta}) = \prod_{i=1}^n \frac{1}{b_i - a_i},
\end{equation}

where \(n\) denotes the total number of parameters, and \(a_i\) and \(b_i\) represent the lower and upper bounds of the \(i\)-th parameter, respectively. Furthermore, derived from Equation~\ref{eq:Bayes}, the term \( p(\mathbf{y} \mid \boldsymbol{\theta}) \) corresponds to the likelihood, which quantifies the probability of observing the data \(\mathbf{y}\) given the model predictions parameterized by \( \boldsymbol{\theta} \). The likelihood is modeled as a Gaussian process, which provides a probabilistic measure of model--data agreement. For computational efficiency, and exploiting the monotonicity of the logarithmic function, the likelihood is expressed in logarithmic form as:
\begin{equation}
\ln p(\mathbf{y} \mid \boldsymbol{\theta}) = -\frac{n}{2} \ln(2\pi) - \frac{1}{2} \ln |\boldsymbol{\Sigma}| - \frac{1}{2} \sum_{i=1}^{n} \left( \mathbf{y}_i - \mathcal{G}_i(\boldsymbol{\theta}) \right)^\mathrm{T} \boldsymbol{\Sigma}^{-1} \left( \mathbf{y}_i - \mathcal{G}_i(\boldsymbol{\theta}) \right),
\end{equation}
where \( \boldsymbol{\Sigma} \) is the diagonal covariance matrix that captures the aleatoric uncertainty in the data, \( \mathcal{G}_i(\boldsymbol{\theta}) \) represents the model prediction corresponding to the \( i \)-th experimental observation \( \mathbf{y}_i \) for the parameter set \(\boldsymbol{\theta}\), \( n \) is the total number of experimental data points, and the superscript \( \mathrm{T} \) indicates the transpose operation.

The final term in Equation~\ref{eq:Bayes}, \( p(\mathbf{y}) \) is the model evidence, also known as the marginal likelihood, which serves as a normalizing factor to ensure that the posterior distribution is a valid probability distribution. The evidence is expressed as:
\begin{equation}
p(\mathbf{y}) = \int p(\mathbf{y} \mid \boldsymbol{\theta}) \, p(\boldsymbol{\theta}) \, \text{d}\boldsymbol{\theta}.
\end{equation}

When the posterior distribution \(p(\boldsymbol{\theta} \mid \mathbf{y})\) is high-dimensional and/or exhibits significant complexity, the marginal likelihood \(p(\mathbf{y})\) becomes intractable and lacks a closed-form solution. Hence, approximate approaches are required to determine the posterior distribution. Typically, sampling techniques are employed to evaluate the posterior distribution indirectly, circumventing the direct computation of the marginal likelihood. Specifically, Markov Chain Monte Carlo (MCMC) methods \citep{robert1999} are utilized to construct a Markov chain whose stationary distribution corresponds to the target posterior distribution. In this study, the random walk MCMC method is employed, wherein samples are iteratively drawn from a Gaussian proposal distribution. The acceptance of each proposed sample is determined using the Metropolis-Hastings (MH) criterion, \(\alpha\), defined as:
\begin{equation}
\alpha(\boldsymbol{\theta}^{(i-1)}, \boldsymbol{\theta}^{(*)}) = \min \left\{ 1, \frac{p(\boldsymbol{\theta}^{(*)} \mid \mathbf{y})}{p(\boldsymbol{\theta}^{(i-1)} \mid \mathbf{y})} \cdot \frac{q(\boldsymbol{\theta}^{(*)}, \boldsymbol{\theta}^{(i-1)})}{q(\boldsymbol{\theta}^{(i-1)}, \boldsymbol{\theta}^{(*)})} \right\},
\end{equation}
where \( \boldsymbol{\theta}^{(i-1)} \) represents the current state of the chain, \( \boldsymbol{\theta}^{(*)} \) is the proposed state, and \( q \) is the proposal distribution. In this work, the proposal distribution is Gaussian, facilitating efficient exploration of the parameter space.

\begin{figure}[H]
    \centering
    \includegraphics[width=0.85\textwidth]{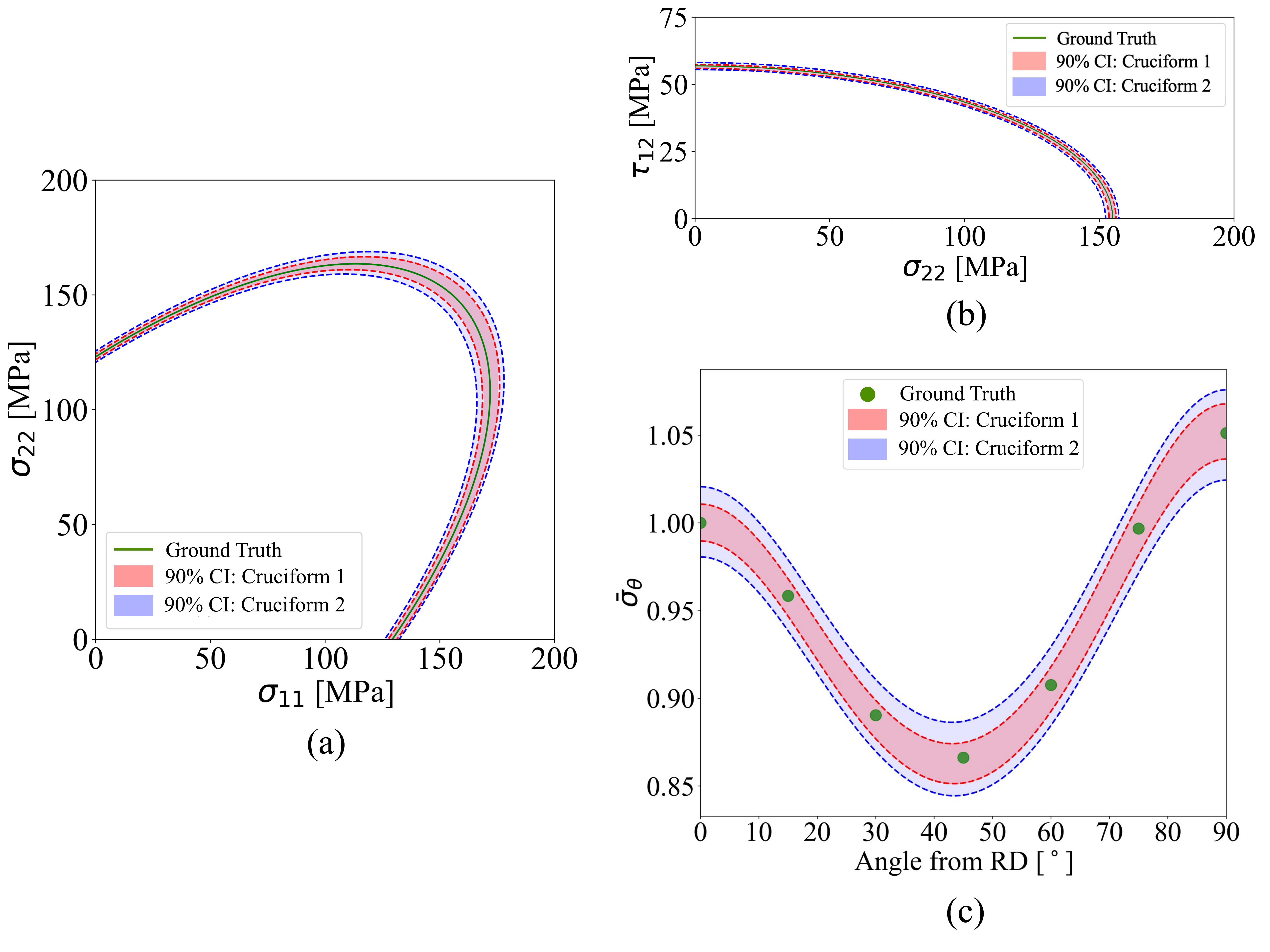} 
    \vspace{-1em}
    \caption{Yield surface reconstruction from noisy test data for Cruciform 1 and Cruciform 2, incorporating uncertainty quantification via 90\% credible interval bounds across key stress planes: (a) $\sigma_{11}$--$\sigma_{22}$ plane, (b) $\sigma_{11}$--$\tau_{12}$ plane, and (c) normalized yield stress as a function of the orientation angle relative to the rolling direction (RD).
}
    \label{fig: Figure 11}
\end{figure}

This framework is applied to perform UQ in learning the material law using the optimally designed specimens, incorporating a noise level of \({\mathcal{N}}(0, (10^{-3})^2)\) in the experimental data. The results, shown in Figure \ref{fig: Figure 11} illustrate the 90\% credible interval uncertainty bounds, showing the impact of experimental noise on inverse learning using the optimized test specimens.

The results presented in Figure~\ref{fig: Figure 11} demonstrate that the optimized specimens exhibit robustness to the prescribed noise level, as evidenced by the distribution of the confidence bounds around the ground truth. Notably, Cruciform 2, characterized by the highest stress state entropy (\(\text{H}(\Theta) = 0.91\)), exhibits a broader CI, reflecting greater uncertainties in learning the material law under noisy conditions. This observation is consistent with findings from the deterministic framework, which showed that specimens with higher stress state entropy, are less robust to experimental noise. These results highlight the trade-off between maximizing the information content of the test data and ensuring robustness to experimental uncertainties.

\section{Discussions}
\subsection{Efficiency vs. robustness in learning constitutive models}
\label{subsec: Efficiency vs robustness}

In contrast to the conventional human-centered learning paradigm, which heavily relies on a multi-experiment approach, the proposed informatics-driven framework offers significant advantages in learning efficiency. First, it eliminates the necessity for maintaining a pure stress state in the specimen, thereby eliminating the need for costly specimen optimization, control, and refinement. While an informative specimen is ideal, the framework does not impose strict requirements on specimen geometry, affording greater flexibility in the selection of testing equipment and fixtures. Second, the framework significantly reduces the number of required tests. The examples presented in this study demonstrate that a 6-parameter, 2D material model can be learned accurately using data from a single test. Third, the learning process has the potential to be fully automated, minimizing the need for user expertise and intervention. Collectively, these advantages directly address the limitations discussed in Section \ref{sec:intro}.

Despite its advantages, it is unlikely that the proposed informatics-led framework will completely replace the conventional human-centered learning paradigm. The primary reason is the robustness of the conventional approach, which allows for identifying parameters associated with a single stress state from a single test. Although specimens in this paradigm carry limited information, resulting in a stress state entropy close to zero, all material points in the specimen convey the same information. This makes measurements simpler and more cost-effective because the process is more resilient to missing data or significant noise in localized regions, as measurements can still be reliable.

\begin{figure}[H]
    \centering
    \includegraphics[width=0.85\textwidth]{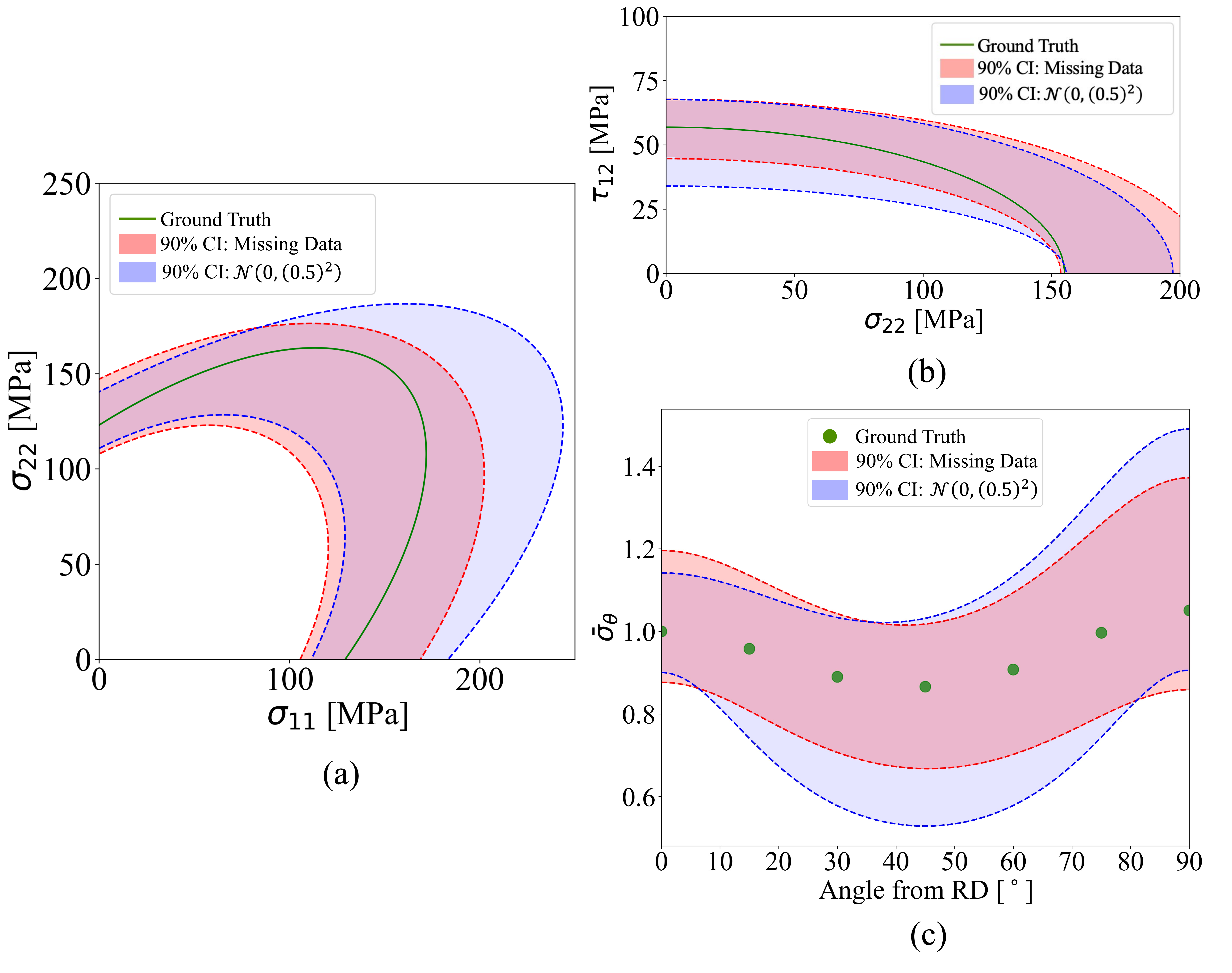} 
    \vspace{-1em}
    \caption{Yield surface reconstruction from Cruciform 2 test data, illustrating the effect of missing data and severe localized noise on uncertainty quantification. Shaded regions represent 90\% credible intervals across key stress planes: (a) $\sigma_{11}$--$\sigma_{22}$ plane, (b) $\sigma_{22}$--$\tau_{12}$ plane, and (c) normalized yield stress as a function of orientation angle relative to the rolling direction (RD).
}
    \label{fig: Figure 12}
\end{figure}

In contrast, the proposed informatics-driven framework relies on a single, highly informative test to calibrate multiple material parameters, which necessitates full-field measurements. This requirement introduces greater sensitivity to uncertainty in both experimental and modeling contexts. Firstly, in practical applications, the set of accessible stress states is inherently constrained by equilibrium conditions, material stability, and applied boundary conditions. These constraints can lead to non-uniform sampling in stress space, with some stress states being overrepresented. Secondly, uncertainties arising from manufacturing imperfections and the susceptibility of informative specimens to missing or noisy data—particularly common in digital image correlation (DIC) measurements of strain and displacement fields \citep{Rossi2015, Rossi2012}, further complicate the learning process. To mitigate these challenges, it is essential to pair the use of informative test specimens with appropriate uncertainty quantification strategies when learning constitutive models. Figure \ref{fig: Figure 12} illustrates the impact of some of these issues on learning accuracy using the optimized Cruciform 2. In Case 1, one-third of the data points are randomly missing, while in Case 2, a localized region of the specimen experiences an amplified characteristic noise level of ${\mathcal{N}}(0, ({0.5})^2)$. As shown, the resulting uncertainty in the learning process is significantly amplified, underscoring the sensitivity of the framework to data quality and completeness.

In summary, the process of learning constitutive models from experimental data inherently balances efficiency and robustness. State-of-the-art approaches often integrate elements of both paradigms to optimize performance. As experimental and computational techniques continue to evolve, achieving greater robustness and reliability, this trade-off is expected to shift progressively toward increased efficiency. Such advancements hold the potential to enable a more autonomous learning process.

\subsection{The least informative specimen}
\label{subsec:Least Informative Specimen}
In the informatics-led inverse learning framework, it is evident that specimens providing more information can facilitate the learning of a greater number of parameters. Therefore, finding the \textit{``most informative"} specimen becomes a priority to enhance learning efficiency. However, in many practical scenarios, the \textit{``least informative"} specimen is preferred, especially when robustness is a key consideration in traditional human-centered learning approaches. A notable example is the experimental determination of fracture strain, where maintaining a constant stress state throughout the test is essential. 

The experimental uncertainties associated with determining strain to fracture under a pure stress state have driven the development of strategies aimed at designing specimens that maintain such conditions within a defined gage section of a specimen approaching failure. Traditionally, these designs rely on trial-and-error approaches \citep{Peirs2012} or parametric studies of specimen geometries in attempts to ensure a constant stress state up to failure \citep{ROTH2016, Roth2018}. In this study, we demonstrate that stress state entropy offers a robust alternative, serving as an objective function for designing test specimens with uniform information content across the gage area. 

\begin{figure}[H]
    \centering
    \includegraphics[width=0.8\textwidth]{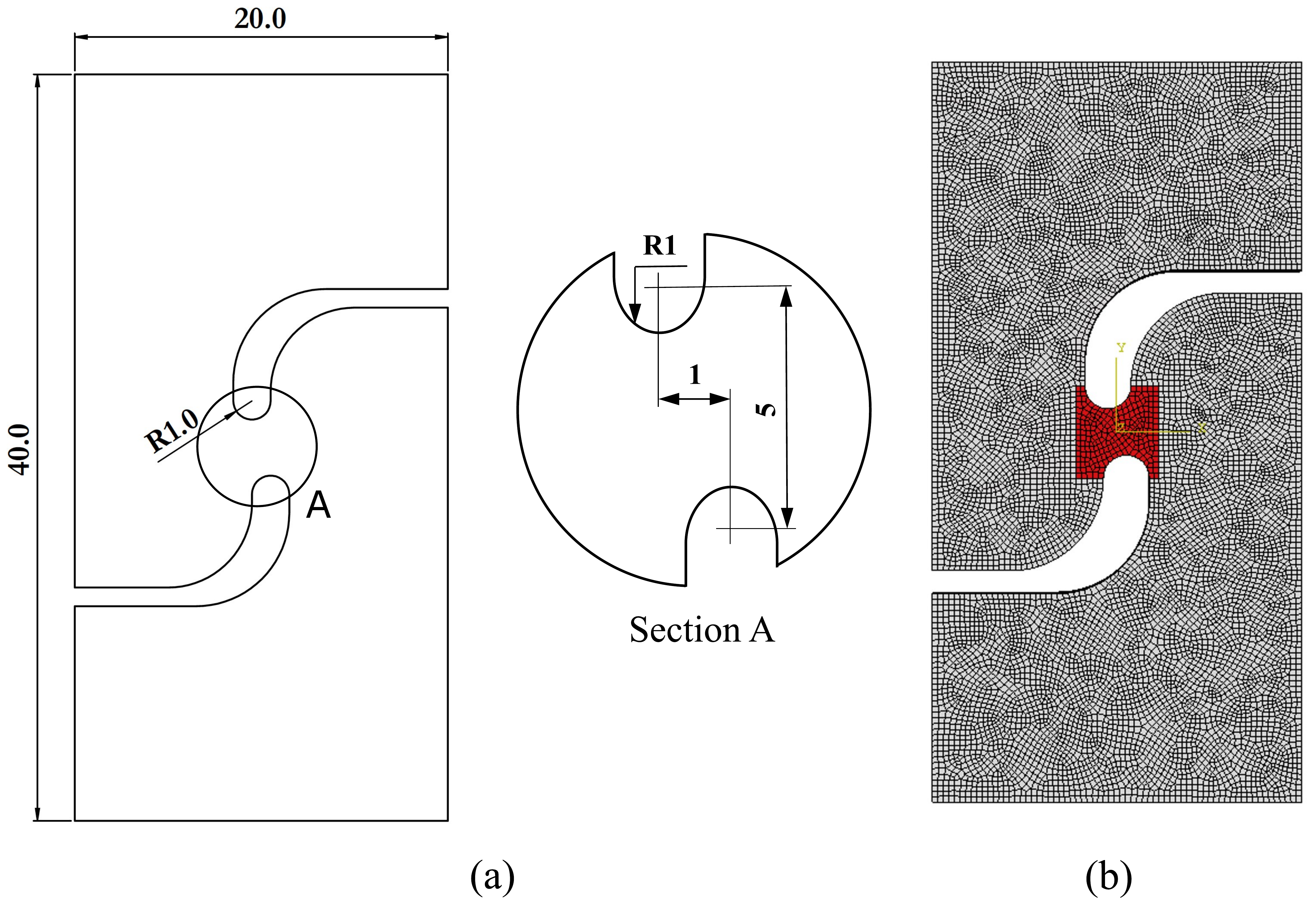} 
            \vspace{-1em}
    \caption{Shear specimen: (a) geometry of the base specimen adapted from \citep{Peirs2012}, and (b) finite element model showing the mesh configuration and highlighting the critical gage region used for analysis.
}
    \label{fig: Figure 13}
\end{figure}

As a case study, we focus on the optimal design of the Peirs' \citep{Peirs2012} shear test specimen (Fig. \ref{fig: Figure 13}) to maintain a shear stress state under large deformations in its gage section. Despite extensive numerical and experimental efforts \citep{Roth2018, Peirs2012}, achieving a consistent shear stress state in the specimen's critical gage region throughout its loading history has proven challenging. To ensure a constant pure stress state while accounting for the inelastic material behavior, both the stress triaxiality ($\eta$) and the Lode angle parameter ($\bar\theta$) at the shear line or critical gage regions must maintain a value of 0 throughout the deformation process. The design problem is mathematically formulated as a multi-objective minimization problem in the form:
\begin{equation}
\mathbf{P}^* = \arg\min_{\mathbf{P}} \left( \bar{\text{H}}(\Theta), \bar{\eta} \right).
\end{equation}
Here, $\mathbf{P}^*$ represents the vector space of geometric parameters that uniquely define the design of the specimen. The parameter space $\mathbf{P} = \{x_i, y_i, R\}$ is explored, where $x_i$ and $y_i$ define the center location of notches, and $R$ represents the notch radius, set to be the same for both notches. Also, $\bar{\text{H}}(\Theta)$ and $\bar{\eta}$ denote the average stress state entropy and stress triaxiality within the gage section of the specimen, respectively. The multi-objective optimization follows the BO framework detailed in Section~\ref{sec:Design}. The objective functions are minimized sequentially. First, the stress state entropy is minimized to ensure that the average information content in the gage section is $\bar{\text{H}}(\Theta) = 0$ bit. Subsequently, to achieve a shear stress state, the posterior distribution of the parameter space yielding this entropy is further refined to minimize the stress triaxiality. This approach defines a posterior distribution over the sampled design parameters that characterize the specimen geometry, ensuring a gage area with minimal information content and a stress state.

\begin{figure}[H]
    \centering
    \includegraphics[width=1\textwidth]{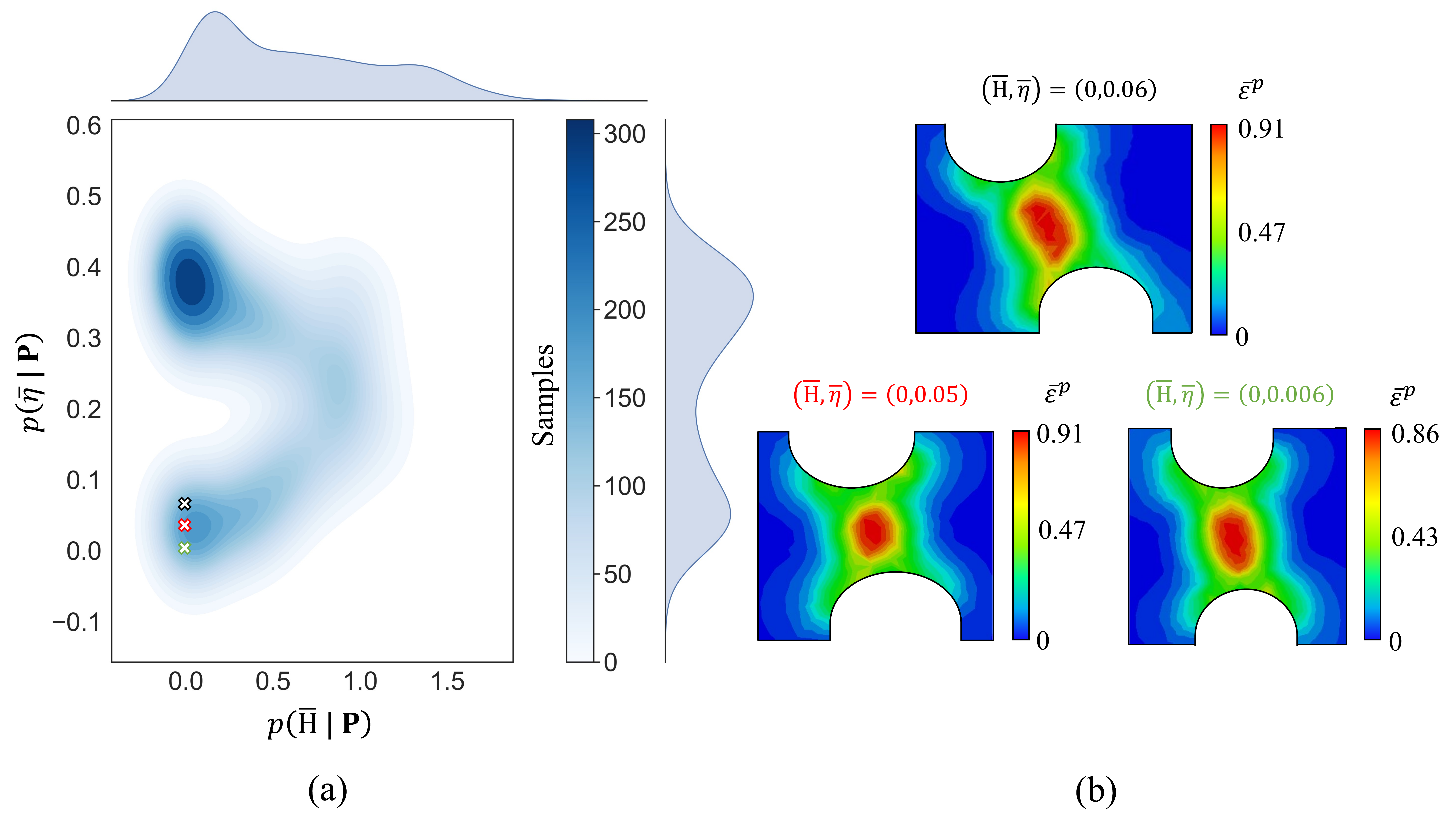} 
    \vspace{-3em}
     \caption{Specimen optimization results: (a) joint posterior distribution of stress state entropy and mean stress triaxiality, and (b) representative specimen geometries and corresponding equivalent plastic strain fields sampled from favorable regions of the $(\bar{\text{H}}(\Theta), \bar{\eta})$ design space.
}
    \label{fig: Figure 14}
\end{figure}

Following the specimen optimization framework, posterior distributions of the stress state entropy and the stress triaxiality are obtained. Figure \ref{fig: Figure 14}(a) presents the joint posterior distribution of these quantities on a contour plot, highlighting three distinct points that correspond to unique locations within the $(\bar{\text{H}}(\Theta),\bar{\eta})$ coordinate system. The geometries and equivalent plastic strain fields within the critical gage region of the specimens, designed based on the parameters associated with these locations, are shown in Figure \ref{fig: Figure 14}(b).

\begin{figure}[H]
    \centering
    \includegraphics[width=1\textwidth]{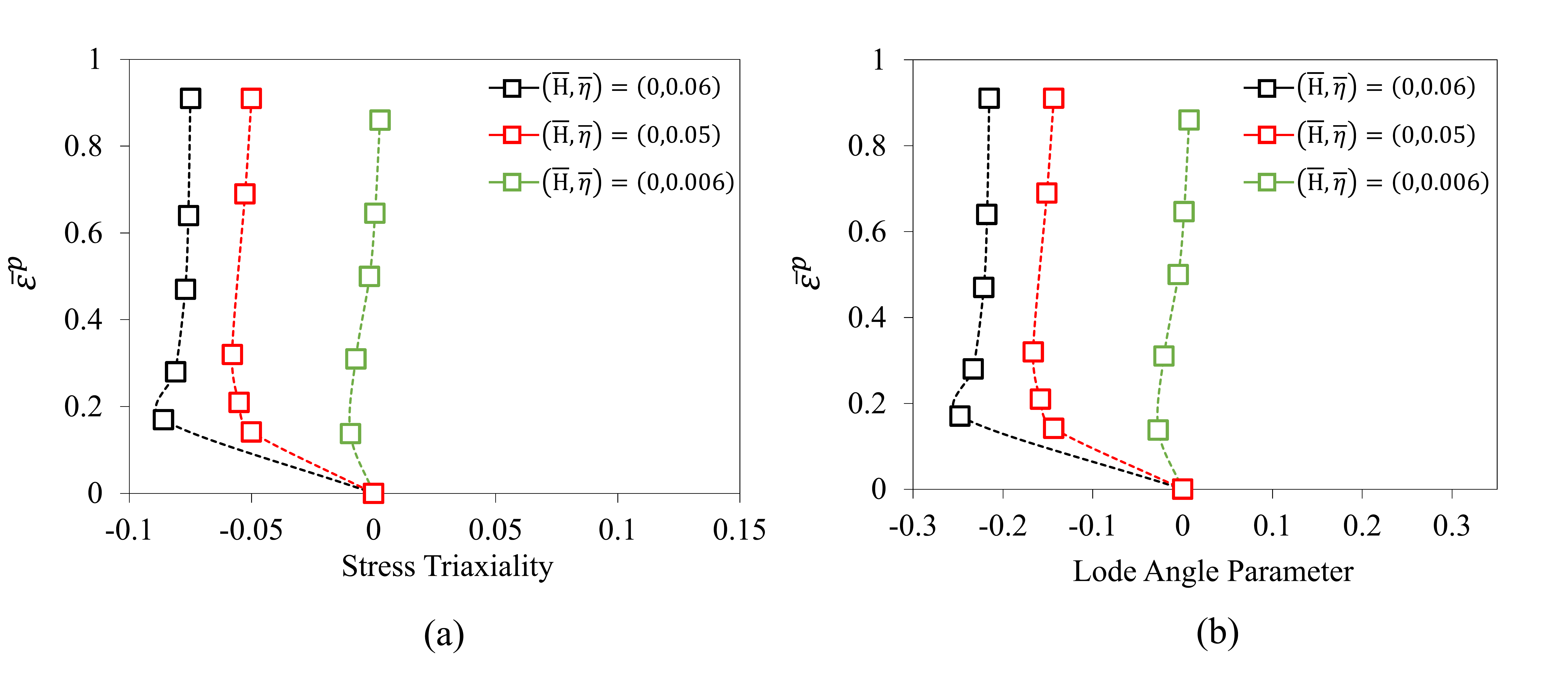} 
    \vspace{-3em}
    \caption{Loading paths for the critical element within the gauge section of the sampled geometries: (a) evolution of stress triaxiality with respect to equivalent plastic strain, and (b) evolution of the Lode angle parameter with respect to equivalent plastic strain.
}
    \label{fig: Figure 15}
\end{figure}

Furthermore, an analysis of the loading paths for the critical element within the gage section of the sampled geometries (Fig. \ref{fig: Figure 14}(b)) reveals different behaviors. As shown in Figure \ref{fig: Figure 15}, the geometries sampled at points $(0, 0.06)$ and $(0, 0.05)$ deviate from a pure shear state during their loading history. In contrast, the specimen corresponding to the location with minimal stress state entropy and stress triaxiality $(0, 0.006)$, maintains a pure shear stress state throughout its entire loading history. These results underscore the potential of the mechanics informatics paradigm in addressing challenges associated with determining strain to fracture under a pure stress state. By leveraging information-driven design, specimen geometries can be optimally tailored to ensure the required \textit{pure stress state} is maintained throughout the entire loading history.

\subsection{Transferability of mechanical tests for learning material laws}

Transfer learning, a powerful machine learning technique, involves re-purposing a model trained for one task as the foundation for solving a different but related task \citep{Torrey2010}. This strategy eliminates the need for extensive training from scratch, which typically demands large datasets and significant computational resources. Instead, it capitalizes on the knowledge embedded in a pre-trained model to improve learning efficiency and performance on the new task. Inspired by this paradigm, we pose a compelling question: \textit{``if two distinct mechanical tests provide stress state information of comparable quantity and quality within the mechanics informatics framework, can they be utilized to infer the same material law?"}. In other words, is the inverse learning process transferable across different mechanical tests?

In this study, we investigate the transferability of two tests for learning the considered material law. The first is the equibiaxial tension test and the second is the mini-punch test. Equibiaxial tension tests are a staple of traditional human-centered learning frameworks and typically rely on standard cruciform specimen geometries. However, performing equibiaxial tension tests presents significant challenges. The design of the loading fixture requires precise displacement control, while the fabrication of cruciform specimens demands meticulous considerations to ensure equibiaxial deformation at the center without premature failure in the arms. Due to these complexities, equibiaxial tension tests are often replaced by bulge or punch tests, where a thin sheet specimen is either deformed using hydraulic pressure~\citep{plancher2020tracking} or by a spherical rigid punch~\citep{zhu2014influence}. 

\begin{figure}[H]
    \centering
    \includegraphics[width=0.9\textwidth]{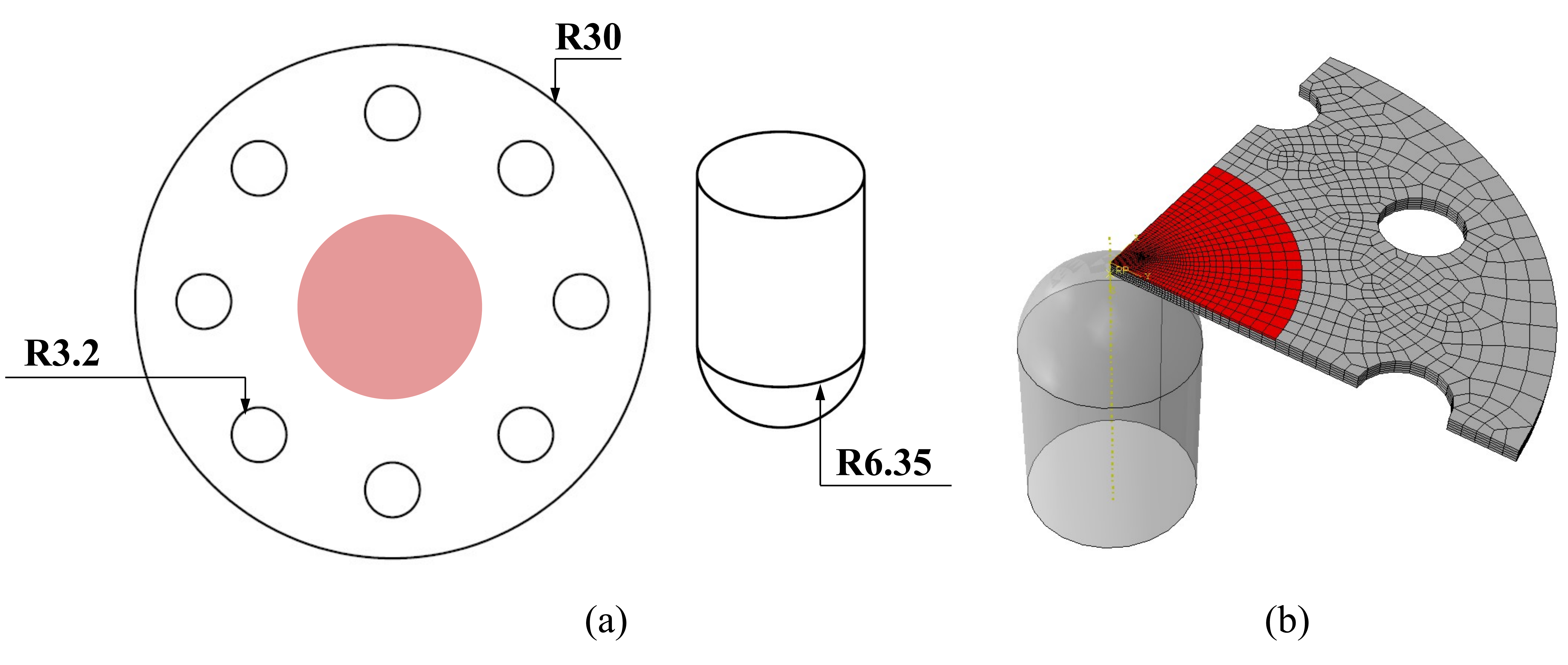} 
    \vspace{-1em}
    \caption{Mini-punch test: (a) specimen geometry with the region of interest highlighted, and (b) finite element model of one-quarter of the specimen showing mesh refinement in the region of interest. All dimensions are in millimeters (mm).}
    \label{fig: Figure 16}
\end{figure}

Figure \ref{fig: Figure 16} illustrates the geometries of the punch and specimen, highlighting the ROI and the FE model used for the mini-punch simulation. Notably, the central region in a mini-punch test achieves an ideal equibiaxial stress state, although deviations occur in the surrounding areas due to geometric constraints and frictional effects between the specimen and the punch~\citep{Zhang2017}. Simulations involving the cruciform specimen were conducted as detailed in Subsection \ref{subsec:Test Geometry & FE Simulation}. For the mini-punch test, the model consisted of a sheet metal specimen constrained by a pair of annular flanges with an inner diameter of 25 mm. Nodes located within a radial distance of 12.5 mm to 30 mm from the center were fully constrained. The punch was modeled as an analytical rigid shell with a radius of 6.35 mm and assigned a prescribed velocity of 0.033 mm/s in the z-direction. To minimize the influence of friction on the resulting strain field, a static coefficient of friction of 0.001 was employed. Symmetry boundary conditions were applied to the specimen geometry, enabling the simulations to be performed on a quarter of the specimen. This symmetry-based approach significantly reduced computational costs.

\begin{figure}[H]
    \centering
    \includegraphics[width=0.8\textwidth]{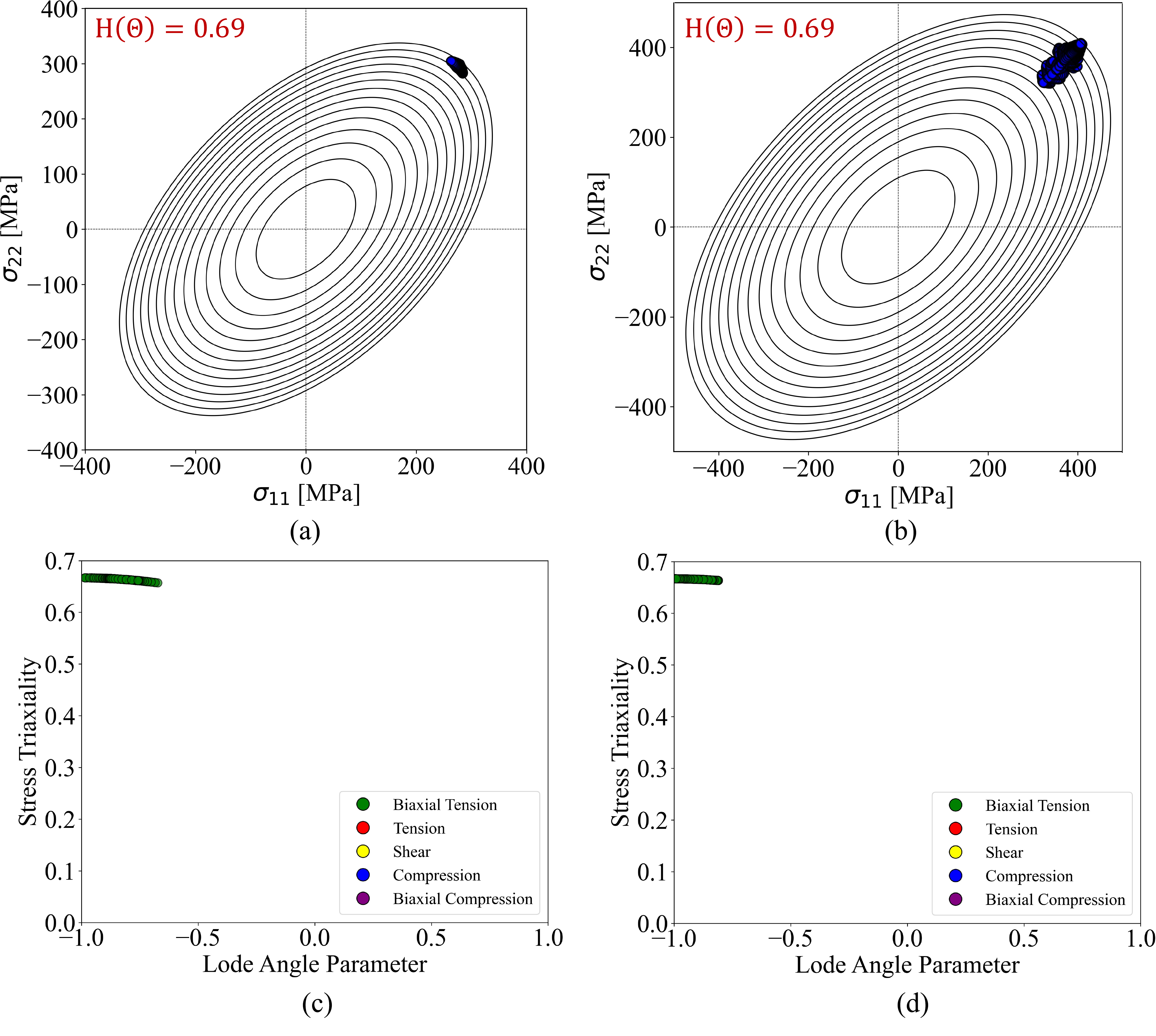} 
    \vspace{-1em}
    \caption{Stress distributions on the $\sigma_{11}$–$\sigma_{22}$ yield loci, annotated with stress state entropy values for the (a) cruciform and (b) mini-punch specimens. Stress state distributions in the 2D Lode angle parameter–stress triaxiality coordinate space for the (c) cruciform and (d) mini-punch specimens.
}
    \label{fig: Figure 17}
\end{figure}

From the results, Figure \ref{fig: Figure 17} compares the stress distributions in the central regions of the two tests, on the $\sigma_{11}-\sigma_{22}$ yield surface, as well as the distribution of the stress states on the Lode angle parameter--stress triaxiality coordinate system. The figure illustrates that both tests are qualitatively identical.

Furthermore, given the material law and stress state space considered in this study ($\Theta = \{ \sigma_{\text{UT}}^{\text{RD}}, \sigma_{\text{UT}}^{\text{TD}}, \sigma_{\text{S}} \}$), an equibiaxial tensile stress state provides equivalent stress state information in tension along both the RD and TD. This equivalence implies that the stress state entropy of the ROI for both specimen geometries is identical, with a value of $\text{H}(\Theta) = 0.69$. This indicates that, in theory, the two tests are equivalent. 

\begin{figure}[H]
    \centering
    \includegraphics[width=0.85\textwidth]{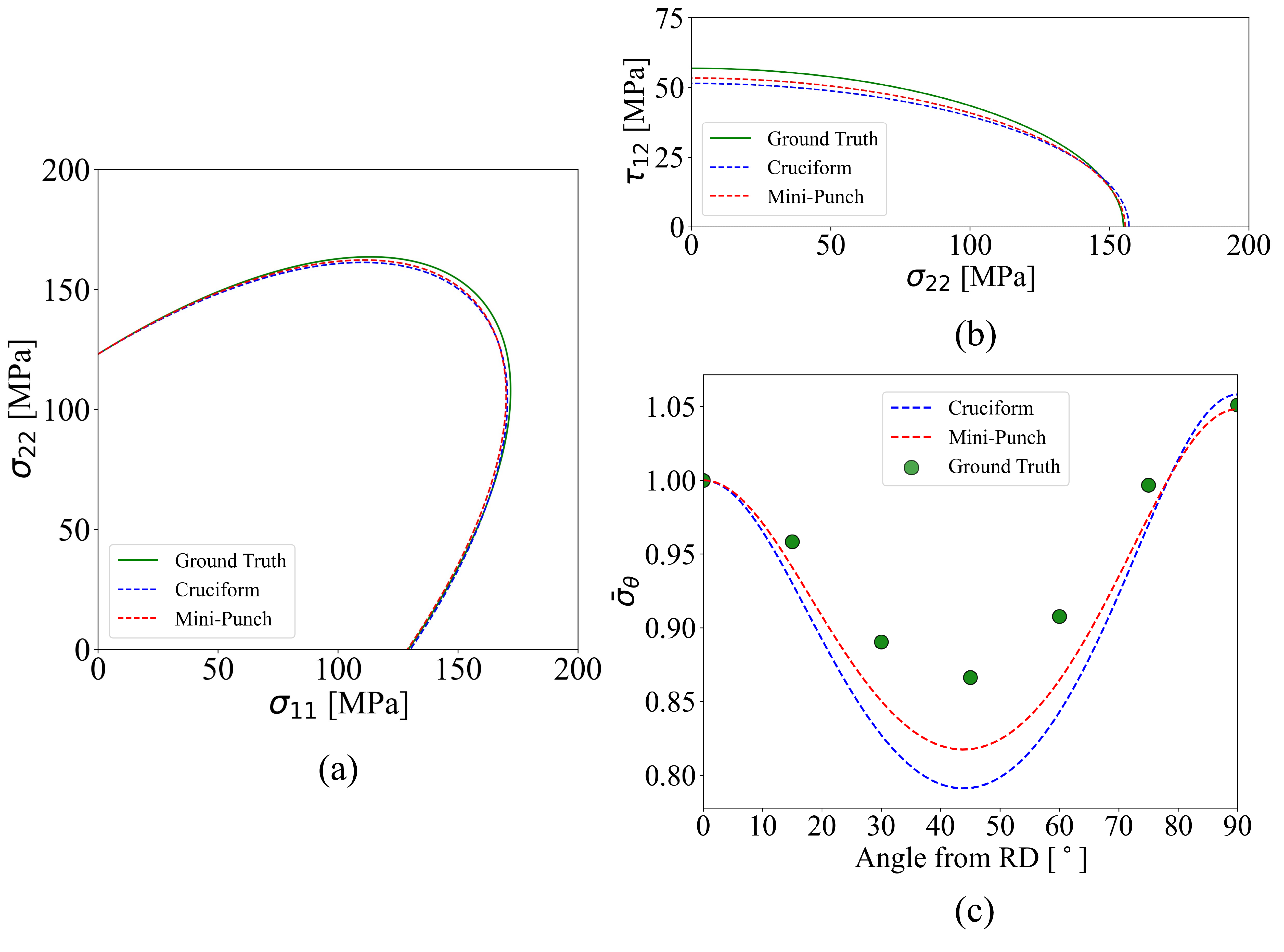} 
    \vspace{-1em}
    \caption{Comparison between the ground truth and identified anisotropic yield responses in key stress planes for the cruciform and mini-punch tests: (a) yield contours in the $\sigma_{11}$--$\sigma_{22}$ plane, (b) yield contours in the $\sigma_{22}$--$\tau_{12}$ plane, and (c) normalized uniaxial yield stress as a function of orientation angle relative to the rolling direction (RD).}
    \label{fig: Figure 18}
\end{figure}

Using the test data generated from both specimens in inverse learning, and following the procedures outlined in Section \ref{Section:Inverse learning}, Table \ref{tab:Table 3} summarizes the identified material law parameters obtained from the two specimens. Figure \ref{fig: Figure 18} compares the yield surfaces reconstructed from the ground truth with those predicted using the identified parameters. The results indicate that both specimens accurately identified parameters sensitive to the uniaxial tensile response in the RD and TD, which are well-represented by the specimens. This is evident from the good agreement in the yield surface in the $\sigma_{11}$--$\sigma_{22}$ plane (Fig. \ref{fig: Figure 18}(a)) and the accurate prediction of the normalized yield stress at angles $0^\circ$ and $90^\circ$ from the RD (Fig. \ref{fig: Figure 18}(c)). However, the results also highlight the inaccuracy in learning the parameter associated with shear response, reflected in the significant error in identifying the parameter $N$ in both cases from Table \ref{tab:Table 3} and as shown in the inaccurate prediction of the yield stresses within the intermediate angles of orientation governed by shear response (Figs. \ref{fig: Figure 18}(b) and \ref{fig: Figure 18}(c)).

\begin{table}[H]
    \centering
    \caption{Identified parameter sets for cruciform and mini-Punch test specimens with equivalent stress state information.}
    \vspace{0.2cm} 
    \label{tab:Table 3}
    {\footnotesize
    \begin{tabular}{lllllll} 
        \hline
         &  $A$ [MPa] & $\sigma_0$ & $n$ & $F$ & $G$ & $N$ \\ 
        \hline
         Ground truth & 471.92 & 123.4 & 0.29 & 0.278 & 0.373 & 2.340 \\
 Initial $\boldsymbol{\theta}$& 600 & 90 & 0.4 & 0.5 & 0.5 &1.5\\ 
         Cruciform &471.11  &148.60& 0.25 & 0.282 & 0.389 & 2.859  \\ 
         Abs. Error (\%)& 0.04 & 20.42& 13.79 &1.44 & 4.29& 22.18\\ 
         Mini-Punch & 473.31 & 121.63& 0.285 & 0.288 & 0.378 & 2.659\\ 
         Abs. Error (\%)& 0.43 &1.43& 1.72 &3.59 &1.35 & 13.63\\ 
        \hline
    \end{tabular}
    }
\end{table}

This example demonstrates the concept of \textit{transfer learning} within the context of mechanics informatics. It offers practical relevance in experimental studies, where a challenging test can be substituted with a more feasible alternative that retains equivalent information content. In inverse learning, unknown model parameters can be accurately identified, nonetheless, ensuring reliable outcomes despite the test constraints.

\subsection{Extension to different constitutive models}
\label{subsec:extension to different constitutive models}

In this section, we explore the extensibility of the mechanics informatics framework to different classes of constitutive behavior. Specifically, we apply the framework to two representative material models under plane stress. First, we consider anisotropic linear elasticity, which is representative of polymeric and composite materials in the small deformation regime. Second, we address anisotropic plasticity using Barlat’s (YLD2000-2d) yield criterion, which maintains convexity at moderate anisotropy and overcomes known limitations of the Hill48 model in modeling sheet metals.

In both cases, we employ the same optimized specimen geometry (Cruciform 2), previously shown to satisfy the optimal criterion (\ref{eq:entropy_criterion}). This is because the stress state information required to accurately learn both material laws requires test data containing information from uniaxial tension in both the RD and TD, as well as shear ($\Theta = \{ \sigma_{\text{UT}}^{\text{RD}}, \sigma_{\text{UT}}^{\text{TD}}, \sigma_{\text{S}} \}$). For the linear elastic case, the stress state entropy is evaluated based on the ratio of the principal stresses, as described in Subsection~\ref{Subsection:stres sstate}. The stress fields used to compute the specimen’s information content were generated under the assumption of linear elastic behavior, using a prescribed set of material parameters as an initial guess for learning the true constitutive response. This approach resulted in an entropy value of $\text{H}(\Theta) = 0.80$. Furthermore, within an elasto-plastic framework, the same specimen yields an entropy of $\text{H}(\Theta) = 0.91$, as reported in Section~\ref{sec:Design}. In both instances, the information content lies within the optimal entropy range defined for $n = 3$ (Eq.~\ref{eq:entropy_criterion}), supporting the use of this geometry across the considered constitutive formulations. The learning procedure follows the methodology described in Section~\ref{Section:Inverse learning}. For the anisotropic plasticity model, Barlat’s YLD2000-2D yield criterion was implemented through a user material (UMAT) subroutine in \textsc{ABAQUS/Standard}\textsuperscript{\textregistered}.

\subsubsection{Anisotropic linear elasticity}

In this example, we apply the mechanics informatics framework to learn the anisotropic linear elastic material law with the plane stress assumption. The constitutive response of such materials is expressed by Eq. \ref{eq:2-7}. The corresponding stiffness tensor, with the material axes aligned to the coordinate system, is given by:

\begin{equation}
\mathbb{C} =
\begin{bmatrix}
C_{11} & C_{12} & 0 \\
C_{12} & C_{22} & 0 \\
0 & 0 & C_{66}
\end{bmatrix}
\end{equation}

Here, the stiffness components are defined in terms of engineering constants: 
\begin{equation}
    C_{11} = \frac{E_1}{1 - \nu_{12} \nu_{21}}, \quad C_{22} = \frac{E_2}{1 - \nu_{12} \nu_{21}}, \quad C_{12} = \frac{\nu_{12} E_2}{1 - \nu_{12} \nu_{21}}, \quad C_{66} = G_{12}.
\end{equation}

In these expressions, \( E_1 \) and \( E_2 \) are the elastic moduli in the RD and TD directions, \( \nu_{12} \) and \( \nu_{21} \) are the major and minor Poisson's ratios, and \( G_{12} \) represents the in-plane shear modulus. The relationship \( -\frac{\nu_{12}}{E_1} = -\frac{\nu_{21}}{E_2} \) ensures symmetry in the stiffness matrix.

\begin{figure}[H]
    \centering
    \includegraphics[width=0.85\textwidth]{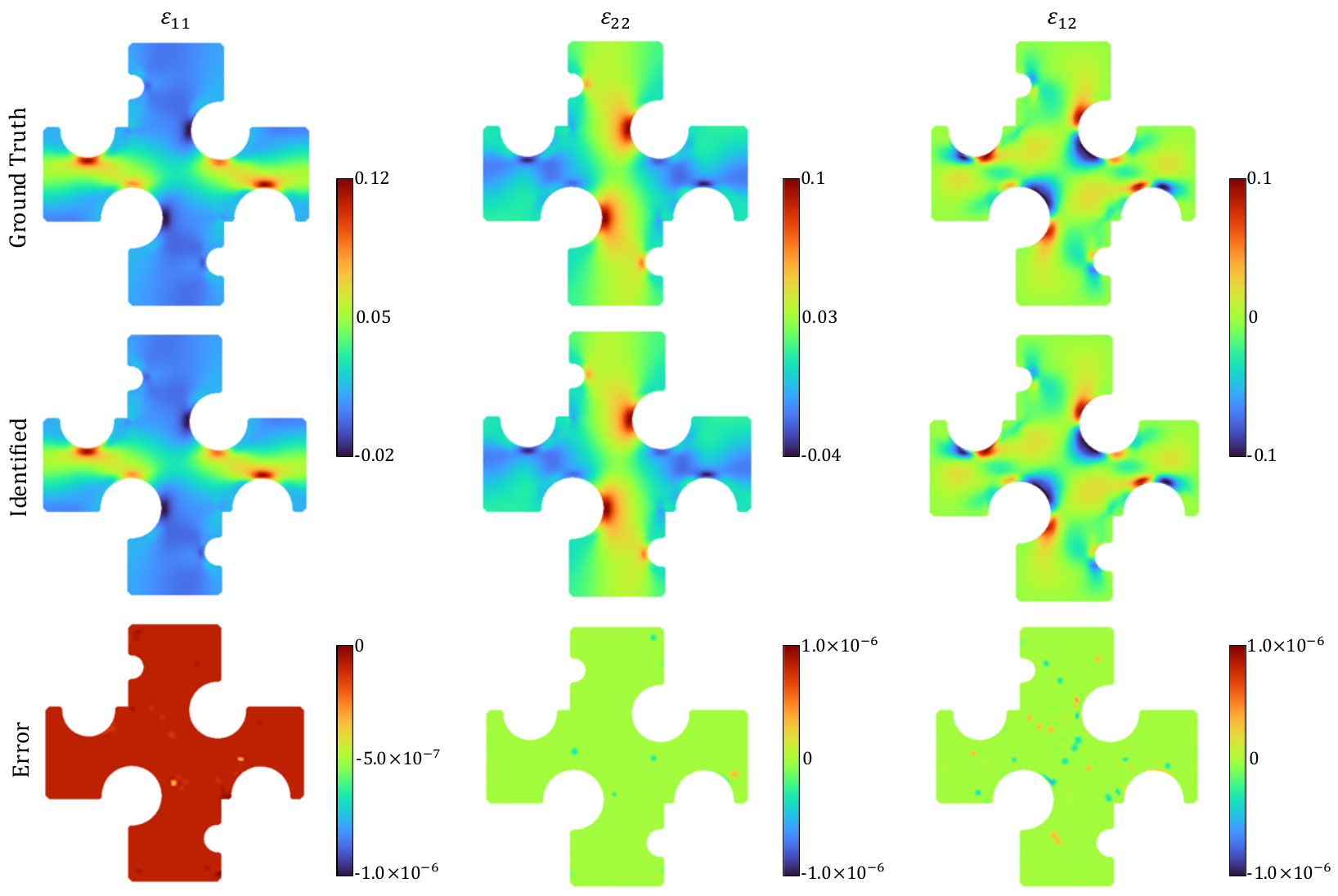} 
    \vspace{-1em}
    \caption{Comparison of strain fields for the optimized Cruciform 2 specimen. The figures show the ground truth strain fields, the reconstructed fields based on the identified material parameters, and the corresponding error plots.}
    \label{fig: Figure 19}
\end{figure}

Thus, to learn the anisotropic material law, the parameters \( E_1 \), \( E_2 \), \( \nu_{12} \), and \( G_{12} \) must be determined. Furthermore, the target material parameters, which represent the experimentally determined in-plane anisotropic elastic response of injection-molded low-density polyethylene, are presented as the ground truth in Table 4 \citep{Kroon2018}. In the learning procedure, the initialization is a parameter space for an isotropic linear elastic case.
\begin{table}[H]
    \centering
    \caption{Identified parameter sets for anisotropic elastic law using the optimized cruciform 2 specimen.}
    \vspace{0.2cm} 
    \label{tab:Table 4}
    {\footnotesize
    \begin{tabular}{lllll} 
        \hline
          &  $E_1$ [MPa]& $E_2$ [MPa]& $\nu_{12}$& $G_{12}$ [MPa]\\ 
        \hline
         Ground Truth
& 210& 150& 0.49& 46\\
 $\boldsymbol{\theta}$& 100& 100& 0.33& 37.59\\
  Identified& 205.43& 146.74& 0.489& 45.00\\ 
         Abs. Error (\%)&  2.18&  2.17&  0.20& 2.17\\ 
        \hline
    \end{tabular}
    }
\end{table}
Table~\ref{tab:Table 4} compares the ground truth and identified material parameters, demonstrating that the constitutive law is learned with high accuracy. Figure~\ref{fig: Figure 19} further illustrates this agreement by comparing the ground truth strain field with the predicted strain field, along with the corresponding prediction error. The reconstructed strain fields exhibit excellent agreement with the reference data, confirming that the modest errors in identifying $E_1$, $E_2$, and $G_{12}$ (all below $3\%$) have negligible impact on the predictions of the the learned anisotropic elastic law.
 
\subsubsection{Anisotropic plasticity with Barlat (YLD2000-2d) yield criterion}

The accurate modeling of anisotropic plastic behavior in metallic sheets, particularly under plane stress conditions, is critical for forming simulations and mechanical analysis. Classical quadratic criteria, such as Hill's 1948 model~\cite{hill1948theory}, offer a simple formulation but often fail to capture the directional variation of yield stresses and r-values when anisotropy becomes moderate or severe. In particular, the quadratic form imposes an elliptical yield surface, which limits its flexibility in matching experimental observations for materials with complex texture or evolving anisotropy.

To address these shortcomings, Barlat et al.~\cite{barlat2003plane} proposed the YLD2000-2d criterion, a non-quadratic yield function specifically formulated for plane stress states. By using two distinct linear transformations of the stress tensor, YLD2000-2d introduces a higher degree of freedom to represent asymmetric and highly directional yield behavior.

The anisotropic plane stress yield function is expressed as:
\begin{equation}
2\sigma_a = \left| X'_1 - X'_2 \right|^a + \left| 2X''_2 + X''_1 \right|^a + \left| 2X''_1 + X''_2 \right|^a,
\label{eq:yld2000}
\end{equation}
where $\sigma_a$ is the equivalent stress and $a$ is a material-specific exponent mainly associated with the crystal structure, with $a=8$ for FCC materials and $a=6$ for BCC materials. Furthermore, $X'_i$, $X''_i$ are the principal values of two linear transformations on the stress deviator $\mathbf{X'} = \mathbf{L'} \boldsymbol{\sigma}$ and $\mathbf{X''} = \mathbf{L''} \boldsymbol{\sigma}$, respectively. The components of the linear transformation are expressed as:
\begin{equation}
\begin{bmatrix}
L^{'}_{11} \\
L^{'}_{12} \\
L^{'}_{21} \\
L^{'}_{22} \\
L^{'}_{66}
\end{bmatrix}
= 
\frac{2}{3}
\begin{bmatrix}
0 & 0 & 1/3 & 0 & 0 & 1/3 & 0 & 0 \\
0 & 1/3 & 0 & 0 & 2/3 & 0 & 0 & 0 \\
0 & 2/3 & 0 & 0 & 0 & 1/3 & 0 & 0 \\
0 & 0 & 0 & 2/3 & 0 & 0 & 1/3 & 0 \\
0 & 0 & 0 & 0 & 0 & 0 & 0 & 1
\end{bmatrix}
\begin{bmatrix}
\alpha_1 \\ \alpha_2 \\ \alpha_7
\end{bmatrix},
\label{eq:L'}
\end{equation}

\begin{equation}
\begin{bmatrix}
L''_{11} \\
L''_{12} \\
L''_{21} \\
L''_{22} \\
L''_{66}
\end{bmatrix}
= 
\frac{1}{9}
\begin{bmatrix}
22 & 8 & 2 & 0 & 0 \\
1 & 4 & 4 & 4 & 0 \\
4 & 4 & 4 & 1 & 0 \\
28 & 2 & 2 & 2 & 0 \\
0 & 0 & 0 & 0 & 9
\end{bmatrix}
\begin{bmatrix}
\alpha_3 \\ \alpha_4 \\ \alpha_5 \\ \alpha_6 \\ \alpha_8
\end{bmatrix}
\label{eq:L''}
\end{equation}

These transformations are governed by two linear mappings $\mathbf{L'}$ and $\mathbf{L''}$ (Eqs. \ref{eq:L'} and \ref{eq:L''}), each parameterized by sets of coefficients $\alpha_k$ ($k=1,\dots,8$), which are identified from experimental yield stresses and r-values in multiple directions. The identification of $\alpha_k$ relies on fitting to the simple tensile yield stresses and r-values in the RD, TD and $45^\circ$ orientation ($\sigma_0$, $\sigma_{45}$, $\sigma_{90}$, $r_0$, $r_{45}$, $r_{90}$) and the yield stress and r-value ($\sigma_b$ and$r_b$ ) under balanced biaxial tension. In this study, we enforce a normalization condition to ensure a yield stress of unity in the rolling direction:
\begin{equation}
\left| \frac{2\alpha_1 + \alpha_2}{3} \right|^a + \left| \frac{2(\alpha_3 - \alpha_4)}{3} \right|^a + \left| \frac{4\alpha_5 - \alpha_6}{3} \right|^a  = 2.
\label{eq:normalization}
\end{equation}
In this study, the known YLD2000-2d parameters are those of
 AA6016-T4 aluminium alloy, determined conventionally in the work by ~\cite{guner2012}. Since AA6016-T4 is has an FCC crystal structure, $a=8$ and from Eq.\ref{eq:normalization}, the parameters to be identified are the hardening parameters($A,\sigma_0$ and $n$) as well as 7 yield surface parameters, $\alpha_k$. The initialization point is set as a parameter space for an isotropic
von Mises yield surface. Table \ref{tab:Table 5} compares the ground truth
and identified parameters. 

\begin{table}[H]
    \centering
    \caption{Identified material parameters for the Swift hardening law and the YLD2000-2D yield criterion obtained using the optimized Cruciform 2 specimen.}
    \vspace{0.2cm} 
    \label{tab:Table 5}
    {\footnotesize
    \begin{tabular}{llllllllllll} 
        \hline
         &  $A$ [MPa] & $\sigma_0$ & $n$ & $\alpha_1$& $\alpha_2$& $\alpha_3$ & $\alpha_4$& $\alpha_5$&$\alpha_6$ & $\alpha_7$&$\alpha_8$\\ 
        \hline
         Ground truth & 471.92 & 123.4 & 0.29 & 0.979& 0.998& 0.885& 1.008& 1.001& 0.965& 0.953&1.242\\
 Initial $\boldsymbol{\theta}$& 600 & 90 & 0.4 & 1& 1&1& 1& 1& 1& 1&1\\ 
         Identified&460.00&119.14& 0.28 & 0.979& 1.006&  0.871& 0.995& 0.995& 0.947& 0.939&1.257\\ 
         Abs. Error (\%)& 2.53& 3.45& 3.45&0.00& 0.80& 1.58& 1.29& 0.60& 1.87& 1.47&1.21\\
         \hline
    \end{tabular}
    }
\end{table}

\begin{figure}[H]
    \centering
    \includegraphics[width=0.85\textwidth]{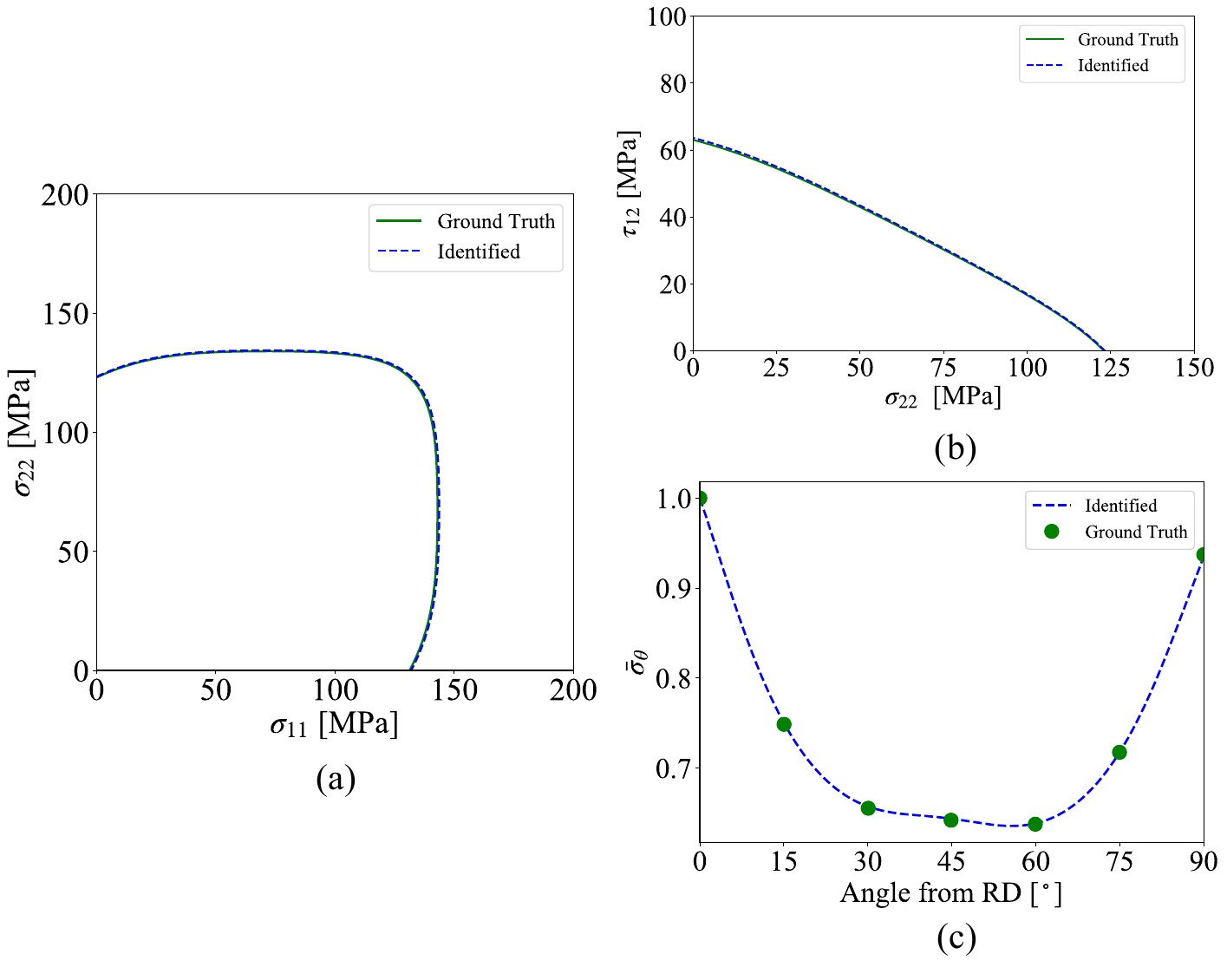} 
        \vspace{-1em}
    \caption{Comparison between the ground truth and identified anisotropic yield response in key stress planes for the optimally designed Cruciform 2 specimen: (a) yield contours in the $\sigma_{11}$--$\sigma_{22}$ plane, (b) yield contours in the $\sigma_{22}$--$\tau_{12}$ plane, and (c) normalized uniaxial yield stress as a function of loading orientation relative to the rolling direction (RD).}
    \label{fig: Figure 20}
\end{figure}

As summarized in Table~\ref{tab:Table 5}, the yield surface and hardening parameters are identified with high accuracy. The largest relative errors occur in the hardening parameters, specifically $\sigma_0$ and $n$, which exhibit a maximum deviation of $3.45\%$. In contrast, all yield surface parameter errors remain below $2\%$. The accuracy of the identified yield surface parameters is further corroborated by the close agreement between the ground truth and predicted yield loci in the $\sigma_{11}$--$\sigma_{22}$ and $\sigma_{22}$--$\tau_{12}$ planes (Figs.~\ref{fig: Figure 20}(a) and~\ref{fig: Figure 20}(b)). Moreover, the predicted normalized yield stress as a function of loading orientation relative to the rolling direction (RD), shown in Fig.~\ref{fig: Figure 20}(c), demonstrates excellent agreement with the ground truth data.

The presented results demonstrate the accuracy of learning both the anisotropic elastic material law and a more convex yield criterion within the YLD2000-2d framework using the informative optimized specimen. Furthermore, these results highlight the extensibility of the mechanics informatics framework to diverse material constitutive laws.

\section{Conclusions}  
This study introduces \textit{mechanics informatics} as a paradigm for efficiently and accurately learning constitutive models, addressing fundamental challenges in quantifying the information content of mechanical test data. By introducing \textit{stress state entropy} as a central metric, we have developed a framework for quantifying the information content of experimental data for efficiently and accurately learning constitutive models. We have explored the applicability of the uniaxial tension specimen, $\Sigma$-shaped specimen and cruciform specimen with different information content for accurately learning an anisotropic inelastic law in the Hill formulation. Notably, test specimens insufficient stress state information content partially constrained the optimization process, limiting accurate learning to only a few parameters, sensitive to the stress state information in the generated test data.  

The mechanics informatics paradigm was further extended to specimen design by incorporating stress state entropy within a Bayesian optimization scheme. In the first case, maximizing stress state entropy enabled the design of optimal cruciform specimens, which provided sufficient stress state information and ensured accurate parameter identification for the considered material law. The results also verified the optimal entropy criterion defined to bound the required stress state information for learning constitutive models between an upper and lower limit. Furthermore, our findings show that highly informative tests are susceptible to experimental uncertainties and such uncertainties can amplify errors in the learning process. Hence, we highlight the trade-offs between maximizing the information content of the test data and ensuring robustness to uncertainties in experimental data. This emphasizes the need for high-quality, complete full-field data when using highly informative data in the informatics-driven workflow.  In a contrasting specimen design case, the framework was applied to successfully minimize stress state entropy within the gage region of a Peirs shear specimen, achieving a shear stress state throughout the critical loading regime. The results show the flexibility of the mechanics informatics paradigm in enabling the targeted design of specimens for specific experimental goals, such as accurately determining strain-to-fracture under pure stress state conditions.  

Furthermore, we demonstrated the potential of leveraging transfer learning within the mechanics informatics framework, providing a pathway for overcoming practical limitations in experimental workflows. By efficiently learning material parameters using diverse test types that yield equivalent stress state information, we showcased the capability to adapt to constraints such as specimen design complexity or difficulties in precise displacement control. This adaptability reduces the dependency on exhaustive experimental testing while maintaining the fidelity required for accurate material parameter identification, thus streamlining computational and experimental workflows in constitutive model development.

Finally, employing an optimally designed informative specimen, we successfully applied the proposed framework to accurately identify both an anisotropic linear elastic constitutive law and a more convex anisotropic yield response within the YLD2000-2d framework. These results demonstrate the extensibility and broad applicability of the mechanics informatics approach to diverse constitutive models.

In summary, this work introduces the mechanics informatics framework, a paradigm for advancing accurate learning of constitutive models and experimental design. By developing a methodology to quantify the information content of test data, optimally design specimens in an information-driven manner, and address uncertainties, this paradigm offers a comprehensive toolkit for accurately learning constitutive models and enhancing the efficiency and reliability of experimental workflows.

\section* {CRediT authorship contribution statement}
Royal C. Ihuaenyi:  Writing – original draft, Writing – review \& editing, Visualization, Software,  Methodology, Investigation, Formal analysis, Data curation, Conceptualization. Wei Li: Writing – review \& editing, Software, Methodology. Martin Z. Bazant: Writing – review \& editing, Supervision, Methodology, Conceptualization. Juner Zhu: Writing – original draft Writing – review \& editing, Supervision, Resources, Project administration, Methodology, Investigation, Funding acquisition, Conceptualization.

\section* {Declaration of competing interest}
The authors declare that they have no known competing interests.

\section* {Acknowledgements}
The authors would like to thank Dr. Ruobing Bai for his constructive comments about the work. 

\section* {Funding sources}
This work is supported by the startup fund for J.Z. from Northeastern University. 

\appendix
\renewcommand{\thefigure}{\Alph{section}\arabic{figure}} 
\setcounter{figure}{0} %
\setcounter{table}{0} %

\section{Effect of the model parameter selection and boundary conditions on stress state entropy}
\label{Appendix: A}

In this section, we undertake a comparative investigation into the influence of initial parameter assumptions within the same constitutive framework and the imposed boundary conditions on the information content of test specimens. To this end, we examine the $\Sigma$-shaped specimen introduced in Subsection~\ref{subsec:Test Geometry & FE Simulation}, originally examined under a uniaxial loading state.

\begin{figure}[H]
    \centering
    \includegraphics[width=0.8\textwidth]{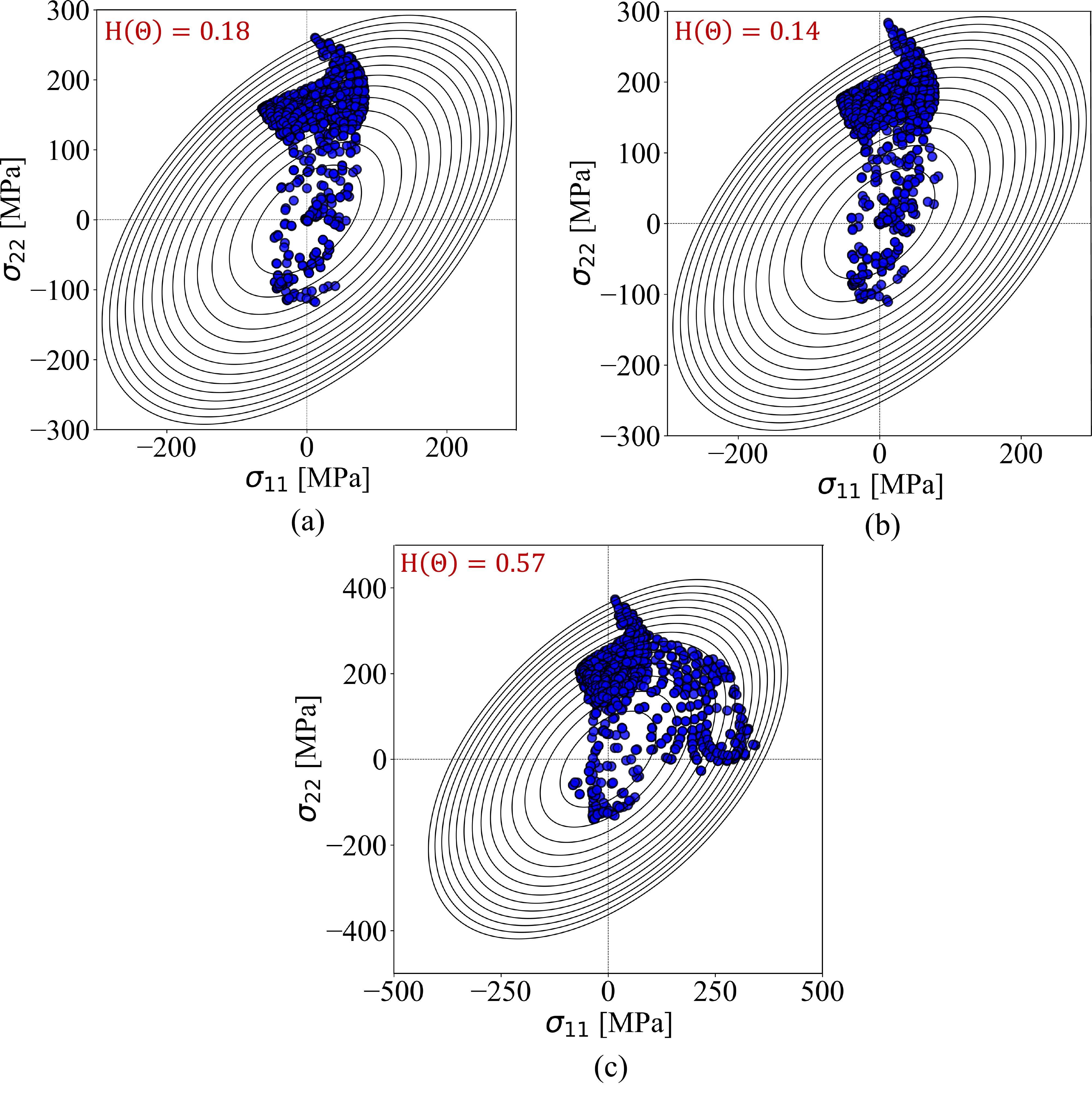}
     \vspace{-1em}
    \caption{Stress distributions projected onto the $\sigma_{11}$–$\sigma_{22}$ yield loci for the $\Sigma$-shaped specimen, annotated with corresponding stress state entropy values. Results are shown for: (a) the von Mises yield criterion under uniaxial loading, (b) the Hill48 anisotropic yield criterion under uniaxial loading, and (c) the von Mises yield criterion under biaxial loading.}
    \label{fig: Figure A1}
\end{figure}

To analyze the influence of the assumed initial parameter set, we evaluate the resulting stress distributions and associated stress state entropy under two distinct yield criteria, the isotropic von Mises and the anisotropic Hill48 formulation (Fig.~\ref{fig: Figure A1}(a) and Fig.~\ref{fig: Figure A1}(b)). The model parameters for the selected constitutive formulations are presented in Table \ref{tab:Table 1}. Despite the fundamental differences between the yield criteria, the resulting stress fields exhibit notable similarity. Quantitatively, the stress state entropy decreases marginally from $\text{H}(\Theta)=0.18$ for the von Mises criterion to $\text{H}(\Theta)=0.14$ for the Hill48 criterion, suggesting only a mild sensitivity of the stress state richness to the choice of the initial parameter set.

In contrast, the influence of the imposed boundary condition is more pronounced. When a biaxial loading condition is applied to the same $\Sigma$-shaped specimen with the parameter set representative of the von Mises yield criterion, a markedly broader distribution of stress states is observed on the yield surface (Fig.~\ref{fig: Figure A1}(c)). This broader distribution signifies an enhancement in the stress state richness, reflected in a substantial increase in the stress state entropy from $\text{H}(\Theta)=0.18$ to $\text{H}(\Theta)=0.57$.

As demonstrated in Subsection~\ref{subsection: Specimen Information Content}, the specimen geometry plays a significant role in its information content. Consequently, the present results highlight the significance of the imposed boundary condition and how negligible the assumed initial constitutive model parameters are to the stress state entropy. However, the closer we get to the true model parameters, the more accurate the quantified specimen information content becomes. Despite this, the findings suggest that, to analyze the information content of a specimen, a reasonable initial approximation of the constitutive response, combined with well-defined geometry and boundary conditions, is sufficient to yield meaningful insights. It is also noteworthy that while differences in stress state entropy due to changes in the initial model parameters may appear modest, comparable variations between different specimen geometries can reflect meaningful shifts in local deformation modes. Even a 5–10\% increase in stress state entropy may signal a transition from predominantly uniaxial to mixed-mode stress states, enhancing the richness of the mechanical response. We suggest using stress state entropy as a relative ranking metric for experimental design. Geometries with higher stress state entropy are more likely to generate diverse local stress conditions, improving parameter identifiability.

\section{Optimization descent and strain fields for the selected specimen geometries in inverse learning}
\label{Appendix: B}

\begin{figure}[H]
    \centering
    \includegraphics[width=0.8\textwidth]{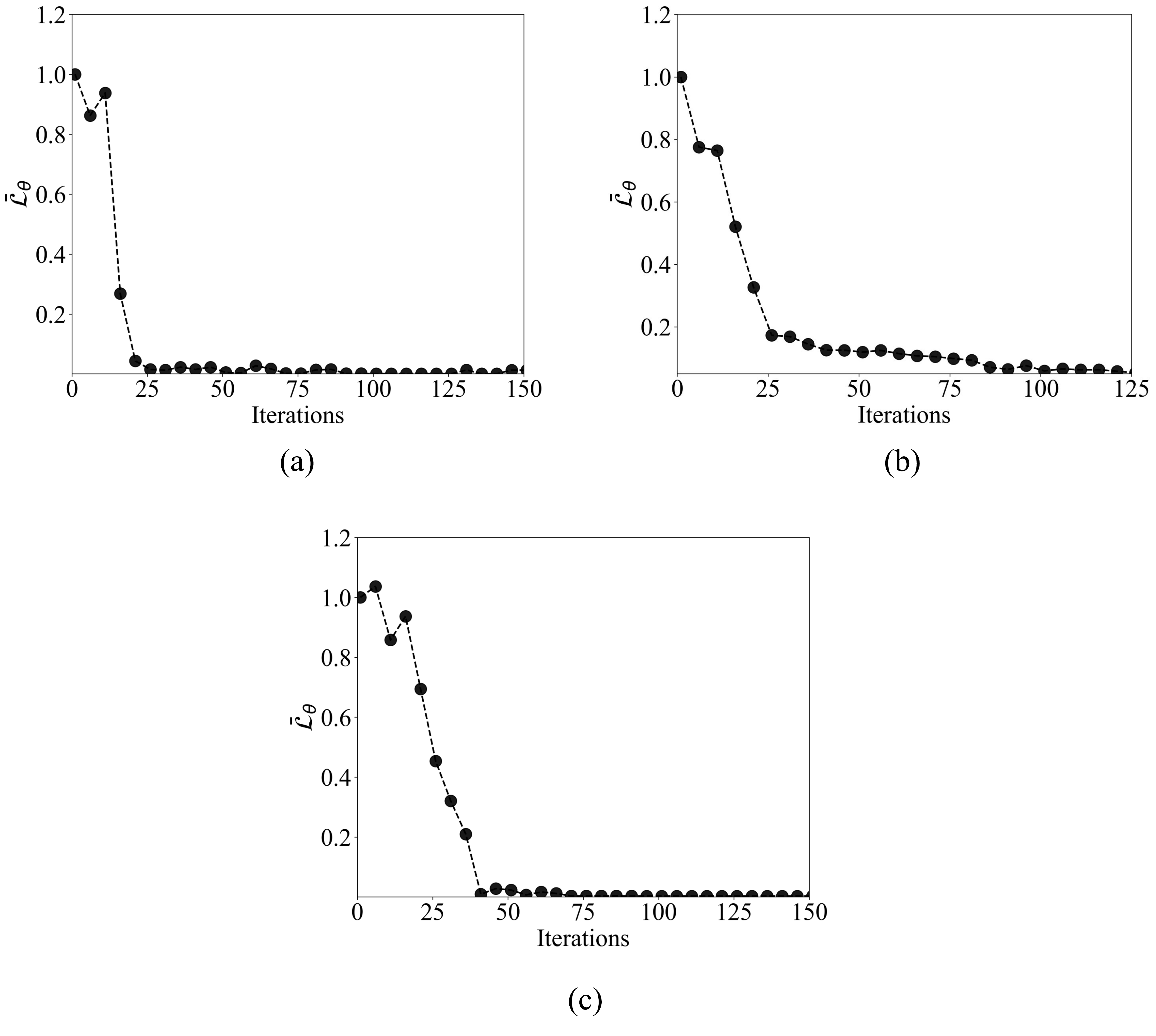} 
    \vspace{-1em}
    \caption{Evolution of the relative objective function with iterations for the (a) uniaxial tension, (b) $\Sigma$-shaped, and (c) cruciform specimens.}
    \label{fig: Figure B1}
\end{figure}

\begin{figure}[H]
    \centering
    \includegraphics[width=0.8\textwidth]{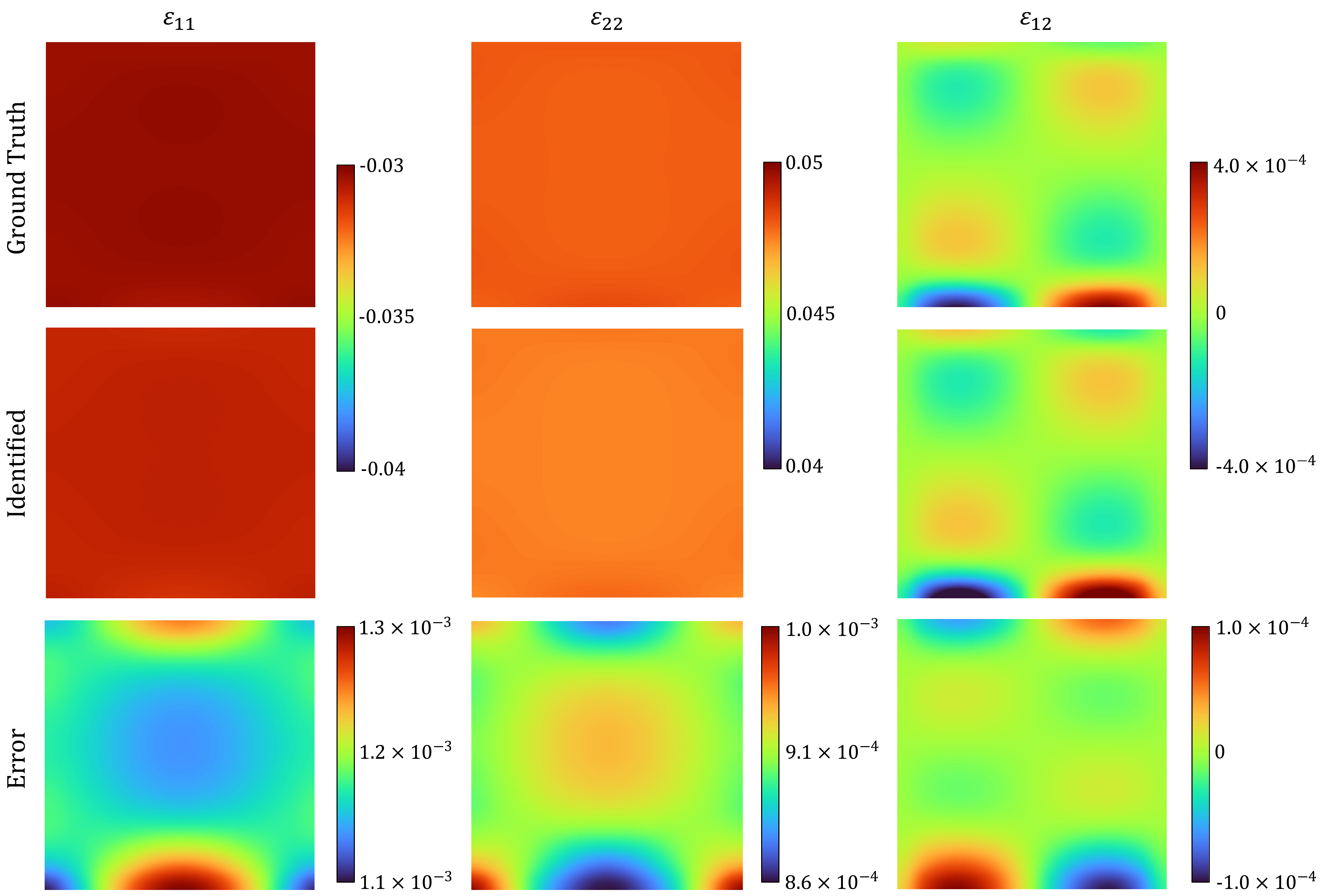} 
    \vspace{-1em}
    \caption{Comparison of strain fields within the region of interest (ROI) for the uniaxial tension specimen. The figures show the ground truth strain fields, the reconstructed fields based on the identified material parameters, and the corresponding error plots.}
    \label{fig: Figure B2}
\end{figure}

\begin{figure}[H]
    \centering
    \includegraphics[width=0.8\textwidth]{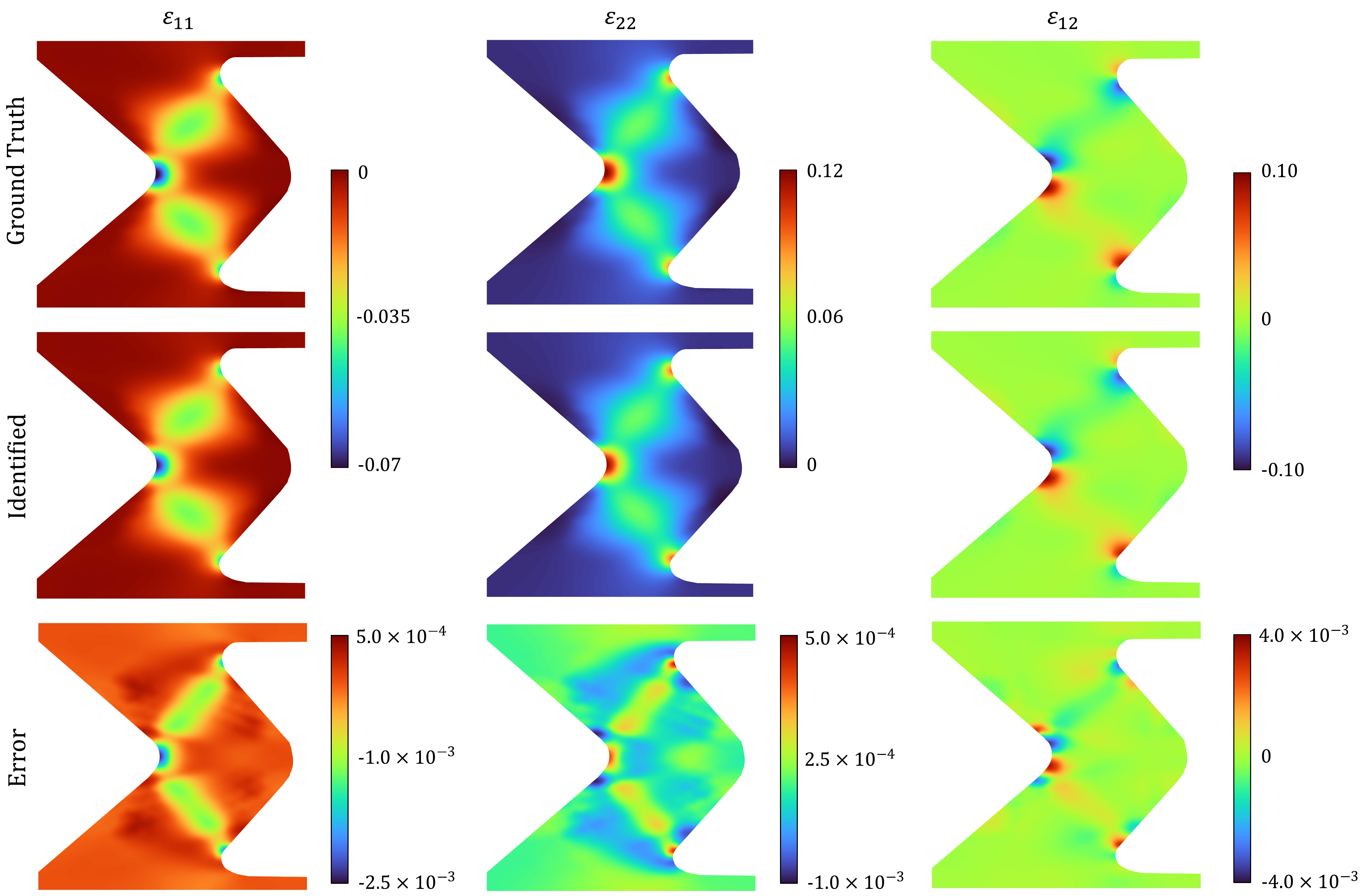} 
    \vspace{-1em}
    \caption{Comparison of strain fields within the region of interest (ROI) for the $\Sigma$-shaped specimen. The figures show the ground truth strain fields, the reconstructed fields based on the identified material parameters, and the corresponding error plots.}
    \label{fig: Figure B3}
\end{figure}

\begin{figure}[H]
    \centering
    \includegraphics[width=0.8\textwidth]{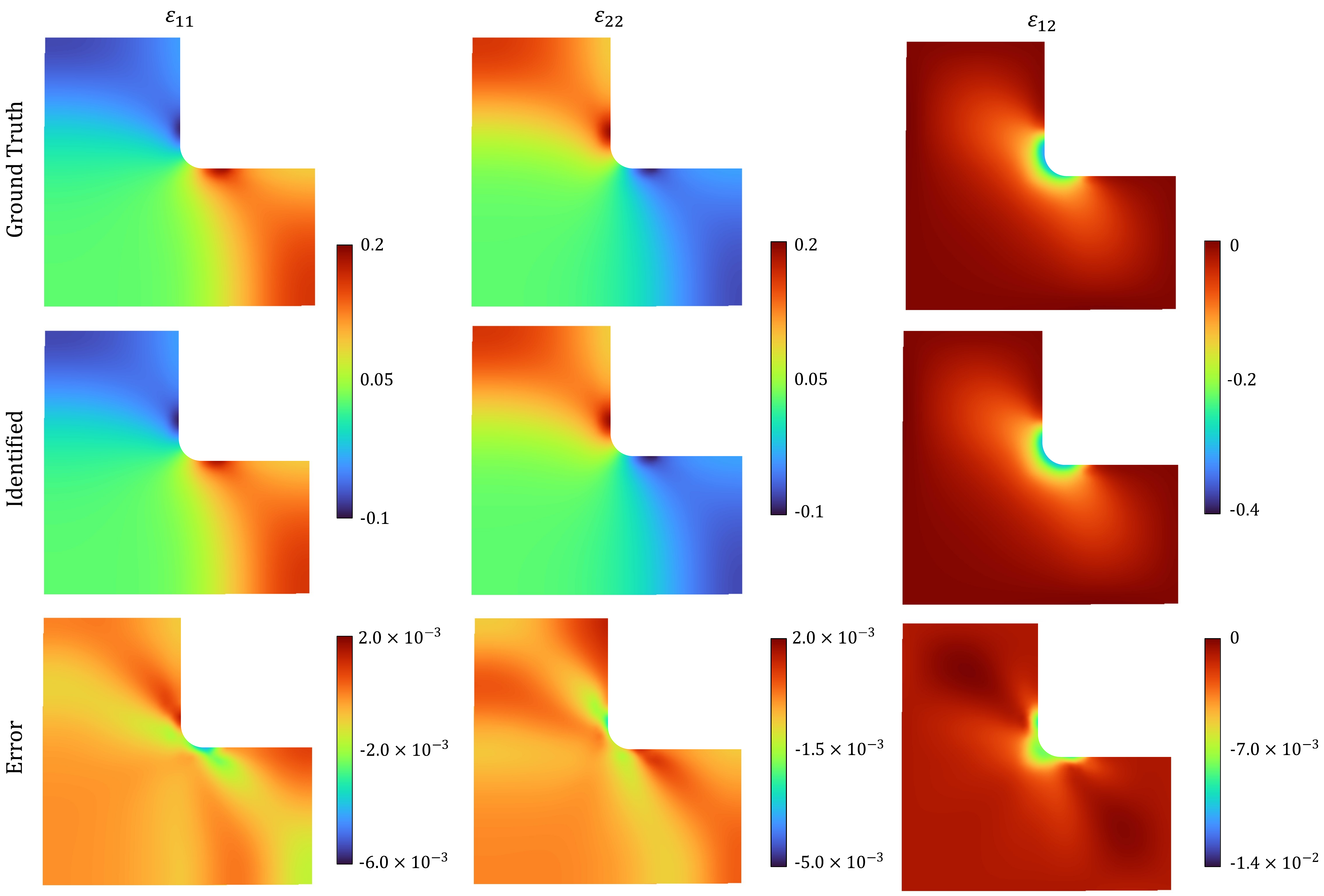} 
    \vspace{-1em}
    \caption{Comparison of strain fields within the region of interest (ROI) for the  cruciform specimen. The figures show the ground truth strain fields, the reconstructed fields based on the identified material parameters, and the corresponding error plots.}
    \label{fig: Figure B4}
\end{figure}

Figure \ref{fig: Figure B1} depicts the optimization descent for the considered test specimen, with all cases showing convergence within 100 iterations. The figure shows the evolution of the relative loss, ($\bar{\mathcal{L}}_{\boldsymbol{\theta}} = \frac{\mathcal{L}^i_{\boldsymbol{\theta}}}{\mathcal{L}^1_{\boldsymbol{\theta}}}$) with each successive iteration. Furthermore, Figures~\ref{fig: Figure B2}, \ref{fig: Figure B3}, and \ref{fig: Figure B4} illustrate the agreement between ground truth and identified strain fields for the uniaxial tension, $\Sigma$-shaped and cruciform specimens. The figures also highlight the error between the ground truth and identified strain fields. The convergence of the objective function and minimal errors in the strain field predictions highlight the robustness of the learning framework in accurately identifying parameter sets that minimize the objective function.

\section{Optimally designed cruciform specimen geometries}
\label{Appendix: C}
\setcounter{figure}{0}
Here, the geometries and dimensions of the optimally designed cruciform specimens are presented. These specimen geometries were designed to maximize the stress state information content in test data from a single experiment, enabling accurate and efficient learning of the considered constitutive model.

\begin{figure}[H]
    \centering
    \includegraphics[width=0.8\textwidth]{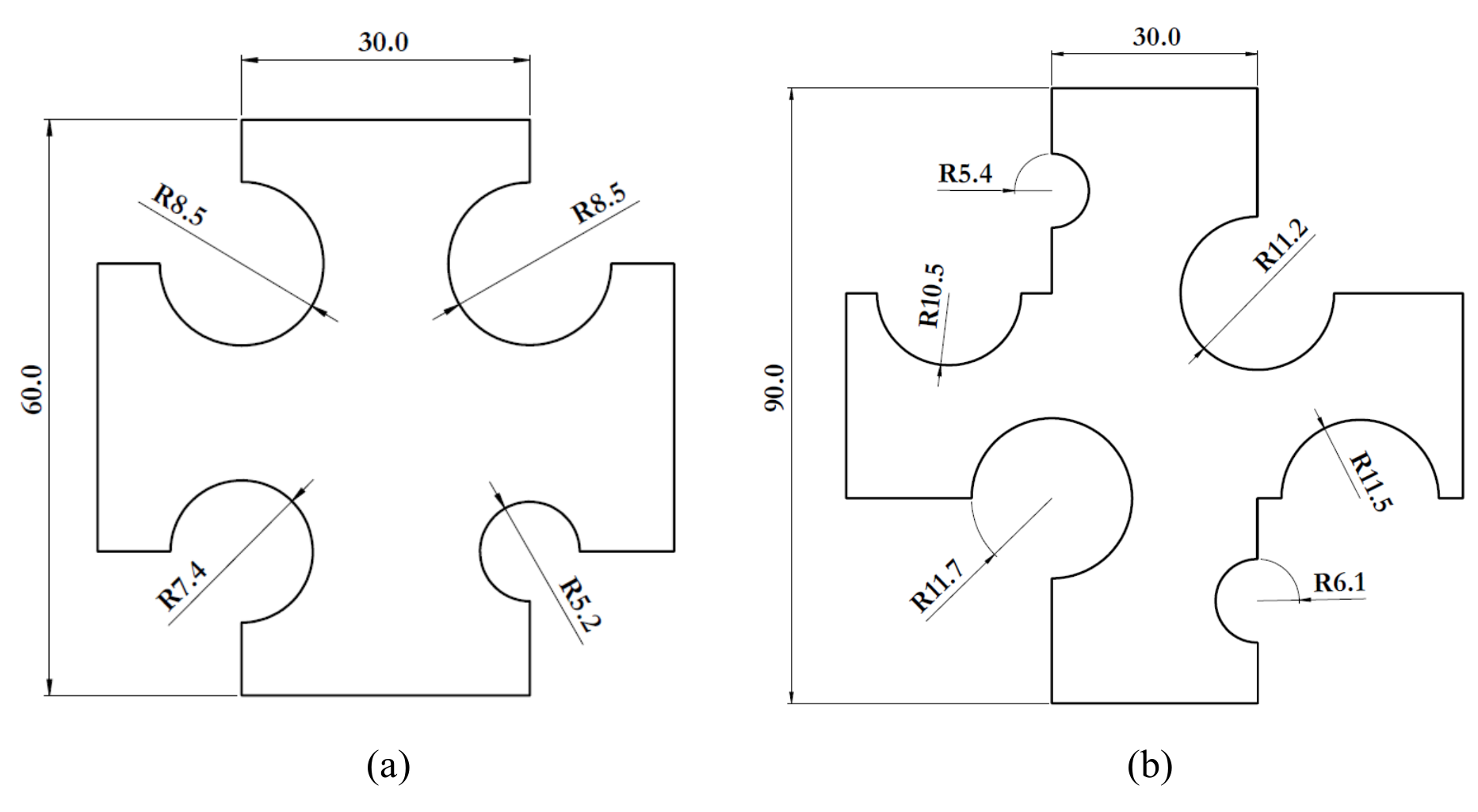} 
    \vspace{-1em}
    \caption{Geometric dimensions of the optimally designed cruciform specimens: (a) Cruciform 1 and (b) Cruciform 2. All dimensions are provided in millimeters (mm).}
    \label{fig: Figure C1}
\end{figure}

\section{Analytical expression of the Hill48 normalized yield stress as a function of loading orientation to the rolling direction.}
\label{Appendix: D}

The Hill48 yield function under the plane stress assumption is expressed as:
\begin{equation}
    2f(\boldsymbol{\sigma}) = F \sigma_{22}^2 + G \sigma_{11}^2 + H (\sigma_{11} - \sigma_{22})^2 + 2N \tau_{12}^2 = \sigma_Y^2,
    \label{eqn:hill48}
\end{equation}
where \(\sigma_0\) is the uniaxial yield stress along the RD.

For uniaxial tension at an angle \(\theta\) from the RD, we rotate the coordinate system. The stress components are related by:
\begin{equation}
    \sigma_{11} = \sigma_\theta \cos^2 \theta,
    \label{eqn:sigma11}
\end{equation}
\begin{equation}
    \sigma_{22} = \sigma_\theta \sin^2 \theta,
    \label{eqn:sigma22}
\end{equation}
\begin{equation}
    \tau_{12} = \sigma_\theta \sin \theta \cos \theta.
    \label{eqn:tau12}
\end{equation}
Substituting these expressions (Eq. \ref{eqn:sigma11} -Eq. \ref{eqn:tau12})  into the yield function gives:
\begin{equation}
    2f(\boldsymbol{\sigma}) = F (\sigma_\theta \sin^2 \theta)^2 + G (\sigma_\theta \cos^2 \theta)^2 + H (\sigma_\theta \cos^2 \theta - \sigma_\theta \sin^2 \theta)^2 + 2N (\sigma_\theta \sin \theta \cos \theta)^2 = \sigma_\theta^2,
    \label{eqn:plug-in}
\end{equation}
which simplifies to:
\begin{equation}
    2f(\boldsymbol{\sigma}) = \sigma_\theta^2 \left[ F \sin^4 \theta + G \cos^4 \theta + H (\cos^2 \theta - \sin^2 \theta)^2 + 2N \sin^2 \theta \cos^2 \theta \right].
    \label{eqn:simplified}
\end{equation}
Next, we use the identity:
\begin{equation}
    (\cos^2 \theta - \sin^2 \theta)^2 = \cos^4 \theta + \sin^4 \theta - 2 \sin^2 \theta \cos^2 \theta.
    \label{eqn:identity}
\end{equation}
Substituting Eq.\ref{eqn:identity} into Eq.\ref{eqn:simplified}, gives:
\begin{equation}
    2f(\boldsymbol{\sigma}) = \sigma_\theta^2 \left[ (F + H) \sin^4 \theta + (G + H) \cos^4 \theta + (2N - 2H) \sin^2 \theta \cos^2 \theta \right].
    \label{eqn:grouped}
\end{equation}
Thus, the formulation yield stress \(\sigma_\theta\) as a function of the orientation angle \(\theta\) is:
\begin{equation}
    \sigma_\theta = \sigma_0 \cdot \left[ (F + H) \sin^4 \theta + (G + H) \cos^4 \theta + (2N - 2H) \sin^2 \theta \cos^2 \theta \right]^{-1/2}.
    \label{eqn:final-yield-stress}
\end{equation}
Hence, the normalized yield stress as a function of the angle of orientation is given by:
\begin{equation}
    \bar\sigma_\theta =  \left[ (F + H) \sin^4 \theta + (G + H) \cos^4 \theta + (2N - 2H) \sin^2 \theta \cos^2 \theta \right]^{-1/2}.
    \label{eqn:normalized-yield-stress}
\end{equation}

Following the analytical formulation for the yield stress as a function of the angle of orientation, Table~\ref{tab:Table D1} provides a quantitative comparison between the ground truth and predicted yield stresses at various orientations with respect to the rolling direction. The results highlight the effectiveness of the optimized, information-rich test specimens (Cruciform 1 and 2) in accurately identifying the material law. In contrast, the table also underscores the limitations of the less informative test specimen, thereby supporting the qualitative trends observed in Figure~\ref{fig:Figure 5}.

\begin{table}[H]
    \centering
    \caption{Comparison of ground truth and predicted yield stresses at various orientations to the rolling direction}
    \label{tab:Table D1}
    \vspace{0.2cm}
    {\footnotesize
    \begin{tabular}{lllllll} 
        \hline
        Angle (°) & 15 & 30 & 45 & 60 & 75 & 90 \\ 
        \hline
        Ground Truth & 118.27 & 109.87 & 106.89 & 111.99 & 123.01 & 129.72 \\ 
        Uniaxial Tension & 120.77 & 116.48 & 116.12 & 122.14 & 132.43 & 138.22 \\ 
        Abs. Error (\%) & 2.11 & 6.02 & 8.64 & 9.06 & 7.66 & 6.55 \\ 
        $\Sigma$-Shaped & 118.23 & 109.72 & 106.49 & 111.11 & 121.31 & 127.48 \\ 
        Abs. Error (\%) & 0.03 & 0.14 & 0.37 & 0.79 & 1.38 & 1.73 \\ 
        Cruciform & 119.78 & 113.62 & 111.61 & 116.12 & 124.98 & 130.08 \\ 
        Abs. Error (\%) & 1.28 & 3.41 & 4.42 & 3.69 & 1.60 & 0.28 \\
        Cruciform 1 & 118.19 & 109.70 & 106.70 & 111.89 & 123.81 & 129.93 \\
        Abs. Error (\%) & 0.07 & 0.15 & 0.18 & 0.09 & 0.65 & 0.16 \\
        Cruciform 2 & 118.27 & 109.89 & 106.91 & 112.02 & 123.01 & 129.72 \\
        Abs. Error (\%) & 0.00 & 0.02 & 0.02 & 0.03 & 0.00 & 0.00 \\
        \hline
    \end{tabular}
    }
\end{table}

\section* {Data Availability}
Data will be made available on request.
\newpage
 \bibliographystyle{elsarticle-harv}
 \bibliography{cas-refs}

\begin{thebibliography}{100}
\expandafter\ifx\csname natexlab\endcsname\relax\def\natexlab#1{#1}\fi
\providecommand{\url}[1]{\texttt{#1}}
\providecommand{\href}[2]{#2}
\providecommand{\path}[1]{#1}
\providecommand{\DOIprefix}{doi:}
\providecommand{\ArXivprefix}{arXiv:}
\providecommand{\URLprefix}{URL: }
\providecommand{\Pubmedprefix}{pmid:}
\providecommand{\doi}[1]{\href{http://dx.doi.org/#1}{\path{#1}}}
\providecommand{\Pubmed}[1]{\href{pmid:#1}{\path{#1}}}
\providecommand{\bibinfo}[2]{#2}
\ifx\xfnm\relax \def\xfnm[#1]{\unskip,\space#1}\fi
\bibitem[{Alves et~al.(2011)Alves, Nielsen and Martins}]{alves2011revisiting}
\bibinfo{author}{Alves, L.M.}, \bibinfo{author}{Nielsen, C.V.}, \bibinfo{author}{Martins, P.A.}, \bibinfo{year}{2011}.
\newblock \bibinfo{title}{Revisiting the fundamentals and capabilities of the stack compression test}.
\newblock \bibinfo{journal}{Experimental Mechanics} \bibinfo{volume}{51}, \bibinfo{pages}{1565--1572}.
\newblock \DOIprefix\doi{https://doi.org/10.1007/s11340-011-9480-5}.
\bibitem[{Anghileri et~al.(2005)Anghileri, Chirwa, Lanzi and Mentuccia}]{anghileri2005inverse}
\bibinfo{author}{Anghileri, M.}, \bibinfo{author}{Chirwa, E.}, \bibinfo{author}{Lanzi, L.}, \bibinfo{author}{Mentuccia, F.}, \bibinfo{year}{2005}.
\newblock \bibinfo{title}{An inverse approach to identify the constitutive model parameters for crashworthiness modelling of composite structures}.
\newblock \bibinfo{journal}{Composite structures} \bibinfo{volume}{68}, \bibinfo{pages}{65--74}.
\newblock \DOIprefix\doi{https://doi.org/10.1016/j.compstruct.2004.03.001}.
\bibitem[{{ASTM International}(2000)}]{astm2000compression}
\bibinfo{author}{{ASTM International}}, \bibinfo{year}{2000}.
\newblock \bibinfo{title}{Standard test methods of compression testing of metallic materials at room temperature}.
\newblock \bibinfo{publisher}{ASTM International}.
\bibitem[{Attia et~al.(2020)Attia, Grover, Jin, Severson, Markov, Liao, Chen, Cheong, Perkins, Yang et~al.}]{attia2020Nature}
\bibinfo{author}{Attia, P.M.}, \bibinfo{author}{Grover, A.}, \bibinfo{author}{Jin, N.}, \bibinfo{author}{Severson, K.A.}, \bibinfo{author}{Markov, T.M.}, \bibinfo{author}{Liao, Y.H.}, \bibinfo{author}{Chen, M.H.}, \bibinfo{author}{Cheong, B.}, \bibinfo{author}{Perkins, N.}, \bibinfo{author}{Yang, Z.}, et~al., \bibinfo{year}{2020}.
\newblock \bibinfo{title}{Closed-loop optimization of fast-charging protocols for batteries with machine learning}.
\newblock \bibinfo{journal}{Nature} \bibinfo{volume}{578}, \bibinfo{pages}{397--402}.
\newblock \DOIprefix\doi{https://doi.org/10.1038/s41586-020-1994-5}.
\bibitem[{Avril et~al.(2008a)Avril, Bonnet, Bretelle, Gr{\'e}diac, Hild, Ienny, Latourte, Lemosse, Pagano, Pagnacco and Pierron}]{avril2008overview}
\bibinfo{author}{Avril, S.}, \bibinfo{author}{Bonnet, M.}, \bibinfo{author}{Bretelle, A.S.}, \bibinfo{author}{Gr{\'e}diac, M.}, \bibinfo{author}{Hild, F.}, \bibinfo{author}{Ienny, P.}, \bibinfo{author}{Latourte, F.}, \bibinfo{author}{Lemosse, D.}, \bibinfo{author}{Pagano, S.}, \bibinfo{author}{Pagnacco, E.}, \bibinfo{author}{Pierron, F.}, \bibinfo{year}{2008}a.
\newblock \bibinfo{title}{Overview of identification methods of mechanical parameters based on full-field measurements}.
\newblock \bibinfo{journal}{Experimental Mechanics} \bibinfo{volume}{48}, \bibinfo{pages}{381--402}.
\newblock \DOIprefix\doi{https://doi.org/10.1007/s11340-008-9148-y}.
\bibitem[{Avril et~al.(2008b)Avril, Pierron, Sutton and Yan}]{avril2008identification}
\bibinfo{author}{Avril, S.}, \bibinfo{author}{Pierron, F.}, \bibinfo{author}{Sutton, M.A.}, \bibinfo{author}{Yan, J.}, \bibinfo{year}{2008}b.
\newblock \bibinfo{title}{Identification of elasto-visco-plastic parameters and characterization of l{\"u}ders behavior using digital image correlation and the virtual fields method}.
\newblock \bibinfo{journal}{Mechanics of materials} \bibinfo{volume}{40}, \bibinfo{pages}{729--742}.
\newblock \DOIprefix\doi{https://doi.org/10.1016/j.mechmat.2008.03.007}.
\bibitem[{Aykol et~al.(2021)Aykol, Gopal, Anapolsky, Herring, van Vlijmen, Berliner, Bazant, Braatz, Chueh and Storey}]{Aykol2021}
\bibinfo{author}{Aykol, M.}, \bibinfo{author}{Gopal, C.B.}, \bibinfo{author}{Anapolsky, A.}, \bibinfo{author}{Herring, P.K.}, \bibinfo{author}{van Vlijmen, B.}, \bibinfo{author}{Berliner, M.D.}, \bibinfo{author}{Bazant, M.Z.}, \bibinfo{author}{Braatz, R.D.}, \bibinfo{author}{Chueh, W.C.}, \bibinfo{author}{Storey, B.D.}, \bibinfo{year}{2021}.
\newblock \bibinfo{title}{Perspective{\textemdash}combining physics and machine learning to predict battery lifetime}.
\newblock \bibinfo{journal}{Journal of The Electrochemical Society} \bibinfo{volume}{168}, \bibinfo{pages}{030525}.
\newblock \DOIprefix\doi{10.1149/1945-7111/abec55}.
\bibitem[{Bao and Wierzbicki(2004)}]{BAO2004}
\bibinfo{author}{Bao, Y.}, \bibinfo{author}{Wierzbicki, T.}, \bibinfo{year}{2004}.
\newblock \bibinfo{title}{On fracture locus in the equivalent strain and stress triaxiality space}.
\newblock \bibinfo{journal}{International Journal of Mechanical Sciences} \bibinfo{volume}{46}, \bibinfo{pages}{81--98}.
\newblock \DOIprefix\doi{10.1016/j.ijmecsci.2004.02.006}.
\bibitem[{Barlat et~al.(2003)Barlat, Brem, Yoon, Chung, Dick and Lege}]{barlat2003plane}
\bibinfo{author}{Barlat, F.}, \bibinfo{author}{Brem, J.C.}, \bibinfo{author}{Yoon, J.W.}, \bibinfo{author}{Chung, K.}, \bibinfo{author}{Dick, R.}, \bibinfo{author}{Lege, D.}, \bibinfo{year}{2003}.
\newblock \bibinfo{title}{Plane stress yield function for aluminum alloy sheets—part 1: theory}.
\newblock \bibinfo{journal}{International Journal of Plasticity} \bibinfo{volume}{19}, \bibinfo{pages}{1297--1319}.
\bibitem[{Barroqueiro et~al.(2020)Barroqueiro, Andrade-Campos, de~Oliveira and Valente}]{Barroqueiro2020Design}
\bibinfo{author}{Barroqueiro, B.}, \bibinfo{author}{Andrade-Campos, A.}, \bibinfo{author}{de~Oliveira, J.D.}, \bibinfo{author}{Valente, R.}, \bibinfo{year}{2020}.
\newblock \bibinfo{title}{Design of mechanical heterogeneous specimens using topology optimization}.
\newblock \bibinfo{journal}{International Journal of Mechanical Sciences} \bibinfo{volume}{181}, \bibinfo{pages}{105764}.
\newblock \DOIprefix\doi{https://doi.org/10.1016/j.ijmecsci.2020.105764}.
\bibitem[{Bergstra et~al.(2011)Bergstra, Bardenet, Bengio and K{\'e}gl}]{Bergstra2011}
\bibinfo{author}{Bergstra, J.}, \bibinfo{author}{Bardenet, R.}, \bibinfo{author}{Bengio, Y.}, \bibinfo{author}{K{\'e}gl, B.}, \bibinfo{year}{2011}.
\newblock \bibinfo{title}{Algorithms for hyper-parameter optimization}.
\newblock \bibinfo{journal}{Advances in Neural Information Processing Systems} \bibinfo{volume}{24}, \bibinfo{pages}{2546--2554}.
\bibitem[{Cameron and Tasan(2021)}]{cameron2021full}
\bibinfo{author}{Cameron, B.C.}, \bibinfo{author}{Tasan, C.C.}, \bibinfo{year}{2021}.
\newblock \bibinfo{title}{Full-field stress computation from measured deformation fields: A hyperbolic formulation}.
\newblock \bibinfo{journal}{Journal of the Mechanics and Physics of Solids} \bibinfo{volume}{147}, \bibinfo{pages}{104186}.
\newblock \DOIprefix\doi{https://doi.org/10.1016/j.jmps.2020.104186}.
\bibitem[{Cao et~al.(2024)Cao, Bambach, Merklein, Mozaffar and Xue}]{cao2024artificial}
\bibinfo{author}{Cao, J.}, \bibinfo{author}{Bambach, M.}, \bibinfo{author}{Merklein, M.}, \bibinfo{author}{Mozaffar, M.}, \bibinfo{author}{Xue, T.}, \bibinfo{year}{2024}.
\newblock \bibinfo{title}{Artificial intelligence in metal forming}.
\newblock \bibinfo{journal}{CIRP Annals} \DOIprefix\doi{https://doi.org/10.1016/j.cirp.2024.04.102}.
\bibitem[{Cao and Banu(2020)}]{Cao2020Opportunities}
\bibinfo{author}{Cao, J.}, \bibinfo{author}{Banu, M.}, \bibinfo{year}{2020}.
\newblock \bibinfo{title}{{Opportunities and Challenges in Metal Forming for Lightweighting: Review and Future Work}}.
\newblock \bibinfo{journal}{Journal of Manufacturing Science and Engineering} \bibinfo{volume}{142}, \bibinfo{pages}{110813}.
\newblock \DOIprefix\doi{https://doi.org/10.1115/1.4047732}.
\bibitem[{Carrara et~al.(2020)Carrara, {De Lorenzis}, Stainier and Ortiz}]{CARRARA2020}
\bibinfo{author}{Carrara, P.}, \bibinfo{author}{{De Lorenzis}, L.}, \bibinfo{author}{Stainier, L.}, \bibinfo{author}{Ortiz, M.}, \bibinfo{year}{2020}.
\newblock \bibinfo{title}{Data-driven fracture mechanics}.
\newblock \bibinfo{journal}{Computer Methods in Applied Mechanics and Engineering} \bibinfo{volume}{372}, \bibinfo{pages}{113390}.
\newblock \DOIprefix\doi{10.1016/j.cma.2020.113390}.
\bibitem[{Chang et~al.(2017)Chang, Li, Xu and Gao}]{chang2017lithiation}
\bibinfo{author}{Chang, C.}, \bibinfo{author}{Li, X.}, \bibinfo{author}{Xu, Z.}, \bibinfo{author}{Gao, H.}, \bibinfo{year}{2017}.
\newblock \bibinfo{title}{Lithiation-enhanced charge transfer and sliding strength at the silicon-graphene interface: A first-principles study}.
\newblock \bibinfo{journal}{acta mechanica solida sinica} \bibinfo{volume}{30}, \bibinfo{pages}{254--262}.
\newblock \DOIprefix\doi{https://doi.org/10.1016/j.camss.2017.03.011}.
\bibitem[{Deng et~al.(2022)Deng, Zhao, Jin, Hughes, Savitzky, Ophus, Fraggedakis, Borb{\'e}ly, Yu, Lomeli et~al.}]{deng2022correlative}
\bibinfo{author}{Deng, H.D.}, \bibinfo{author}{Zhao, H.}, \bibinfo{author}{Jin, N.}, \bibinfo{author}{Hughes, L.}, \bibinfo{author}{Savitzky, B.H.}, \bibinfo{author}{Ophus, C.}, \bibinfo{author}{Fraggedakis, D.}, \bibinfo{author}{Borb{\'e}ly, A.}, \bibinfo{author}{Yu, Y.S.}, \bibinfo{author}{Lomeli, E.G.}, et~al., \bibinfo{year}{2022}.
\newblock \bibinfo{title}{Correlative image learning of chemo-mechanics in phase-transforming solids}.
\newblock \bibinfo{journal}{Nature Materials} \bibinfo{volume}{21}, \bibinfo{pages}{547--554}.
\newblock \DOIprefix\doi{https://doi.org/10.1038/s41563-021-01191-0}.
\bibitem[{Deng et~al.(2015)Deng, Kuwabara and Korkolis}]{deng2015cruciform}
\bibinfo{author}{Deng, N.}, \bibinfo{author}{Kuwabara, T.}, \bibinfo{author}{Korkolis, Y.}, \bibinfo{year}{2015}.
\newblock \bibinfo{title}{Cruciform specimen design and verification for constitutive identification of anisotropic sheets}.
\newblock \bibinfo{journal}{Experimental Mechanics} \bibinfo{volume}{55}, \bibinfo{pages}{1005--1022}.
\newblock \DOIprefix\doi{https://doi.org/10.1007/s11340-015-9999-y}.
\bibitem[{Dreyer et~al.(2015)Dreyer, Janotti and Van~de Walle}]{dreyer2015brittle}
\bibinfo{author}{Dreyer, C.}, \bibinfo{author}{Janotti, A.}, \bibinfo{author}{Van~de Walle, C.}, \bibinfo{year}{2015}.
\newblock \bibinfo{title}{Brittle fracture toughnesses of gan and aln from first-principles surface-energy calculations}.
\newblock \bibinfo{journal}{Applied Physics Letters} \bibinfo{volume}{106}.
\newblock \DOIprefix\doi{https://doi.org/10.1063/1.4921855}.
\bibitem[{Dunand and Mohr(2010)}]{dunand2010hybrid}
\bibinfo{author}{Dunand, M.}, \bibinfo{author}{Mohr, D.}, \bibinfo{year}{2010}.
\newblock \bibinfo{title}{Hybrid experimental--numerical analysis of basic ductile fracture experiments for sheet metals}.
\newblock \bibinfo{journal}{International journal of solids and structures} \bibinfo{volume}{47}, \bibinfo{pages}{1130--1143}.
\newblock \DOIprefix\doi{https://doi.org/10.1016/j.ijsolstr.2009.12.011}.
\bibitem[{Dunand and Mohr(2011)}]{dunand2011optimized}
\bibinfo{author}{Dunand, M.}, \bibinfo{author}{Mohr, D.}, \bibinfo{year}{2011}.
\newblock \bibinfo{title}{Optimized butterfly specimen for the fracture testing of sheet materials under combined normal and shear loading}.
\newblock \bibinfo{journal}{Engineering Fracture Mechanics} \bibinfo{volume}{78}, \bibinfo{pages}{2919--2934}.
\newblock \DOIprefix\doi{https://doi.org/10.1016/j.engfracmech.2011.08.008}.
\bibitem[{Eggersmann et~al.(2019)Eggersmann, Kirchdoerfer, Reese, Stainier and Ortiz}]{EGGERSMANN2019}
\bibinfo{author}{Eggersmann, R.}, \bibinfo{author}{Kirchdoerfer, T.}, \bibinfo{author}{Reese, S.}, \bibinfo{author}{Stainier, L.}, \bibinfo{author}{Ortiz, M.}, \bibinfo{year}{2019}.
\newblock \bibinfo{title}{Model-free data-driven inelasticity}.
\newblock \bibinfo{journal}{Computer Methods in Applied Mechanics and Engineering} \bibinfo{volume}{350}, \bibinfo{pages}{81--99}.
\newblock \DOIprefix\doi{10.1016/j.cma.2019.02.016}.
\bibitem[{Eggersmann et~al.(2021)Eggersmann, Stainier, Ortiz and Reese}]{EGGERSMANN2021}
\bibinfo{author}{Eggersmann, R.}, \bibinfo{author}{Stainier, L.}, \bibinfo{author}{Ortiz, M.}, \bibinfo{author}{Reese, S.}, \bibinfo{year}{2021}.
\newblock \bibinfo{title}{Model-free data-driven computational mechanics enhanced by tensor voting}.
\newblock \bibinfo{journal}{Computer Methods in Applied Mechanics and Engineering} \bibinfo{volume}{373}, \bibinfo{pages}{113499}.
\newblock \DOIprefix\doi{10.1016/j.cma.2020.113499}.
\bibitem[{Flaschel et~al.(2021)Flaschel, Kumar and De~Lorenzis}]{flaschel2021unsupervised}
\bibinfo{author}{Flaschel, M.}, \bibinfo{author}{Kumar, S.}, \bibinfo{author}{De~Lorenzis, L.}, \bibinfo{year}{2021}.
\newblock \bibinfo{title}{Unsupervised discovery of interpretable hyperelastic constitutive laws}.
\newblock \bibinfo{journal}{Computer Methods in Applied Mechanics and Engineering} \bibinfo{volume}{381}, \bibinfo{pages}{113852}.
\newblock \DOIprefix\doi{https://doi.org/10.1016/j.cma.2021.113852}.
\bibitem[{Flaschel et~al.(2023)Flaschel, Kumar and De~Lorenzis}]{flaschel2023automated}
\bibinfo{author}{Flaschel, M.}, \bibinfo{author}{Kumar, S.}, \bibinfo{author}{De~Lorenzis, L.}, \bibinfo{year}{2023}.
\newblock \bibinfo{title}{Automated discovery of generalized standard material models with euclid}.
\newblock \bibinfo{journal}{Computer Methods in Applied Mechanics and Engineering} \bibinfo{volume}{405}, \bibinfo{pages}{115867}.
\newblock \DOIprefix\doi{https://doi.org/10.1016/j.cma.2022.115867}.
\bibitem[{Fu et~al.(2016)Fu, Barlat, Kim and Pierron}]{fu2016identification}
\bibinfo{author}{Fu, J.}, \bibinfo{author}{Barlat, F.}, \bibinfo{author}{Kim, J.H.}, \bibinfo{author}{Pierron, F.}, \bibinfo{year}{2016}.
\newblock \bibinfo{title}{Identification of nonlinear kinematic hardening constitutive model parameters using the virtual fields method for advanced high strength steels}.
\newblock \bibinfo{journal}{International Journal of Solids and Structures} \bibinfo{volume}{102}, \bibinfo{pages}{30--43}.
\newblock \DOIprefix\doi{https://doi.org/10.1016/j.ijsolstr.2016.10.020}.
\bibitem[{Fu et~al.(2020)Fu, Xie, Zhou and Qi}]{fu2020method}
\bibinfo{author}{Fu, J.}, \bibinfo{author}{Xie, W.}, \bibinfo{author}{Zhou, J.}, \bibinfo{author}{Qi, L.}, \bibinfo{year}{2020}.
\newblock \bibinfo{title}{A method for the simultaneous identification of anisotropic yield and hardening constitutive parameters for sheet metal forming}.
\newblock \bibinfo{journal}{International Journal of Mechanical Sciences} \bibinfo{volume}{181}, \bibinfo{pages}{105756}.
\newblock \DOIprefix\doi{https://doi.org/10.1016/j.ijmecsci.2020.105756}.
\bibitem[{Fuhg et~al.(2024)Fuhg, Anantha~Padmanabha, Bouklas, Bahmani, Sun, Vlassis, Flaschel, Carrara and De~Lorenzis}]{fuhg2024review}
\bibinfo{author}{Fuhg, J.N.}, \bibinfo{author}{Anantha~Padmanabha, G.}, \bibinfo{author}{Bouklas, N.}, \bibinfo{author}{Bahmani, B.}, \bibinfo{author}{Sun, W.}, \bibinfo{author}{Vlassis, N.N.}, \bibinfo{author}{Flaschel, M.}, \bibinfo{author}{Carrara, P.}, \bibinfo{author}{De~Lorenzis, L.}, \bibinfo{year}{2024}.
\newblock \bibinfo{title}{A review on data-driven constitutive laws for solids}.
\newblock \bibinfo{journal}{Archives of Computational Methods in Engineering} , \bibinfo{pages}{1--43}\DOIprefix\doi{https://doi.org/10.1007/s11831-024-10196-2}.
\bibitem[{Galilei(1638)}]{Galilei1638}
\bibinfo{author}{Galilei, G.}, \bibinfo{year}{1638}.
\newblock \bibinfo{title}{Discorsi e dimostrazioni matematiche intorno a due nuove scienze attinenti la meccanica e i movimenti locali}.
\newblock \bibinfo{publisher}{Leiden: appresso gli Elsevirii}.
\bibitem[{Gorji and Mohr(2017)}]{GORJI2017}
\bibinfo{author}{Gorji, M.B.}, \bibinfo{author}{Mohr, D.}, \bibinfo{year}{2017}.
\newblock \bibinfo{title}{Micro-tension and micro-shear experiments to characterize stress-state dependent ductile fracture}.
\newblock \bibinfo{journal}{Acta Materialia} \bibinfo{volume}{131}, \bibinfo{pages}{65--76}.
\newblock \DOIprefix\doi{10.1016/j.actamat.2017.03.034}.
\bibitem[{Guery et~al.(2016)Guery, Hild, Latourte and Roux}]{guery2016identification}
\bibinfo{author}{Guery, A.}, \bibinfo{author}{Hild, F.}, \bibinfo{author}{Latourte, F.}, \bibinfo{author}{Roux, S.}, \bibinfo{year}{2016}.
\newblock \bibinfo{title}{Identification of crystal plasticity parameters using dic measurements and weighted femu}.
\newblock \bibinfo{journal}{Mechanics of Materials} \bibinfo{volume}{100}, \bibinfo{pages}{55--71}.
\newblock \DOIprefix\doi{https://doi.org/10.1016/j.mechmat.2016.06.007}.
\bibitem[{Gurtin et~al.(2010)Gurtin, Fried and Anand}]{gurtin2010mechanics}
\bibinfo{author}{Gurtin, M.E.}, \bibinfo{author}{Fried, E.}, \bibinfo{author}{Anand, L.}, \bibinfo{year}{2010}.
\newblock \bibinfo{title}{The mechanics and thermodynamics of continua}.
\newblock \bibinfo{publisher}{Cambridge University Press}.
\bibitem[{Güner et~al.(2012)Güner, Soyarslan, Brosius and Tekkaya}]{guner2012}
\bibinfo{author}{Güner, A.}, \bibinfo{author}{Soyarslan, C.}, \bibinfo{author}{Brosius, A.}, \bibinfo{author}{Tekkaya, A.}, \bibinfo{year}{2012}.
\newblock \bibinfo{title}{Characterization of anisotropy of sheet metals employing inhomogeneous strain fields for yld2000-2d yield function}.
\newblock \bibinfo{journal}{International Journal of Solids and Structures} \bibinfo{volume}{49}, \bibinfo{pages}{3517--3527}.
\newblock \DOIprefix\doi{10.1016/j.ijsolstr.2012.07.013}.
\bibitem[{Haux(2010)}]{haux2010medical}
\bibinfo{author}{Haux, R.}, \bibinfo{year}{2010}.
\newblock \bibinfo{title}{Medical informatics: past, present, future}.
\newblock \bibinfo{journal}{International journal of medical informatics} \bibinfo{volume}{79}, \bibinfo{pages}{599--610}.
\newblock \DOIprefix\doi{https://doi.org/10.1016/j.ijmedinf.2010.06.003}.
\bibitem[{Hencky(1924)}]{hencky1924theorie}
\bibinfo{author}{Hencky, H.}, \bibinfo{year}{1924}.
\newblock \bibinfo{title}{Zur theorie plastischer deformationen und der hierdurch im material hervorgerufenen nachspannungen}.
\newblock \bibinfo{journal}{ZAMM-Journal of Applied Mathematics and Mechanics/Zeitschrift f{\"u}r Angewandte Mathematik und Mechanik} \bibinfo{volume}{4}, \bibinfo{pages}{323--334}.
\newblock \DOIprefix\doi{https://doi.org/10.1002/zamm.19240040405}.
\bibitem[{Hill(1948)}]{hill1948theory}
\bibinfo{author}{Hill, R.}, \bibinfo{year}{1948}.
\newblock \bibinfo{title}{A theory of the yielding and plastic flow of anisotropic metals}.
\newblock \bibinfo{journal}{Proceedings of the Royal Society of London. Series A. Mathematical and Physical Sciences} \bibinfo{volume}{193}, \bibinfo{pages}{281--297}.
\newblock \DOIprefix\doi{https://doi.org/10.1098/rspa.1948.0045}.
\bibitem[{Huang et~al.(2020)Huang, Xu, Farhat and Darve}]{HUANG2020JComP}
\bibinfo{author}{Huang, D.Z.}, \bibinfo{author}{Xu, K.}, \bibinfo{author}{Farhat, C.}, \bibinfo{author}{Darve, E.}, \bibinfo{year}{2020}.
\newblock \bibinfo{title}{Learning constitutive relations from indirect observations using deep neural networks}.
\newblock \bibinfo{journal}{Journal of Computational Physics} \bibinfo{volume}{416}, \bibinfo{pages}{109491}.
\newblock \DOIprefix\doi{10.1016/j.jcp.2020.109491}.
\bibitem[{Huber(1904)}]{huber1904}
\bibinfo{author}{Huber, M.T.}, \bibinfo{year}{1904}.
\newblock \bibinfo{title}{Właściwa praca odkształcenia jako miara wytezenia materiału. translated as ``specific work of strain as a measure of material effort"}.
\newblock \bibinfo{journal}{Archives of Mechanics} \bibinfo{volume}{56}, \bibinfo{pages}{173--190}.
\bibitem[{Hutter et~al.(2011)Hutter, Hoos and Leyton-Brown}]{hutter2011smbo}
\bibinfo{author}{Hutter, F.}, \bibinfo{author}{Hoos, H.H.}, \bibinfo{author}{Leyton-Brown, K.}, \bibinfo{year}{2011}.
\newblock \bibinfo{title}{Sequential model-based optimization for general algorithm configuration}, in: \bibinfo{editor}{Coello, C.A.C.} (Ed.), \bibinfo{booktitle}{Learning and Intelligent Optimization: 5th International Conference, LION 5, Rome, Italy, January 17--21, 2011, Selected Papers}. \bibinfo{publisher}{Springer}, \bibinfo{address}{Berlin, Heidelberg}. volume \bibinfo{volume}{6683} of \textit{\bibinfo{series}{Lecture Notes in Computer Science}}, pp. \bibinfo{pages}{507--523}.
\newblock \DOIprefix\doi{10.1007/978-3-642-25566-3_40}.
\bibitem[{Ihuaenyi et~al.(2023a)Ihuaenyi, Deng, Bae and Xiao}]{ihuaenyi2023orthotropic}
\bibinfo{author}{Ihuaenyi, R.}, \bibinfo{author}{Deng, J.}, \bibinfo{author}{Bae, C.}, \bibinfo{author}{Xiao, X.}, \bibinfo{year}{2023}a.
\newblock \bibinfo{title}{An orthotropic nonlinear thermoviscoelastic model for polymeric battery separators}.
\newblock \bibinfo{journal}{Journal of the Electrochemical Society} \bibinfo{volume}{170}, \bibinfo{pages}{010520}.
\newblock \DOIprefix\doi{https://doi.org/10.1149/1945-7111/acb178}.
\bibitem[{Ihuaenyi et~al.(2021)Ihuaenyi, Yan, Deng, Bae, Sakib and Xiao}]{ihuaenyi2021orthotropic}
\bibinfo{author}{Ihuaenyi, R.}, \bibinfo{author}{Yan, S.}, \bibinfo{author}{Deng, J.}, \bibinfo{author}{Bae, C.}, \bibinfo{author}{Sakib, I.}, \bibinfo{author}{Xiao, X.}, \bibinfo{year}{2021}.
\newblock \bibinfo{title}{Orthotropic thermo-viscoelastic model for polymeric battery separators with electrolyte effect}.
\newblock \bibinfo{journal}{Journal of The Electrochemical Society} \bibinfo{volume}{168}, \bibinfo{pages}{090536}.
\newblock \DOIprefix\doi{https://doi.org/10.1149/1945-7111/ac24b6}.
\bibitem[{Ihuaenyi et~al.(2023b)Ihuaenyi, Deng, Bae and Xiao}]{ihuaenyi2023coupled}
\bibinfo{author}{Ihuaenyi, R.C.}, \bibinfo{author}{Deng, J.}, \bibinfo{author}{Bae, C.}, \bibinfo{author}{Xiao, X.}, \bibinfo{year}{2023}b.
\newblock \bibinfo{title}{A coupled nonlinear viscoelastic--viscoplastic thermomechanical model for polymeric lithium-ion battery separators}.
\newblock \bibinfo{journal}{Batteries} \bibinfo{volume}{9}, \bibinfo{pages}{475}.
\newblock \DOIprefix\doi{https://doi.org/10.3390/batteries9090475}.
\bibitem[{Ihuaenyi et~al.(2024)Ihuaenyi, Luo, Li and Zhu}]{ihuaenyi2024seeking}
\bibinfo{author}{Ihuaenyi, R.C.}, \bibinfo{author}{Luo, J.}, \bibinfo{author}{Li, W.}, \bibinfo{author}{Zhu, J.}, \bibinfo{year}{2024}.
\newblock \bibinfo{title}{Seeking the most informative design of test specimens for learning constitutive models}.
\newblock \bibinfo{journal}{Extreme Mechanics Letters} \bibinfo{volume}{69}, \bibinfo{pages}{102169}.
\newblock \DOIprefix\doi{https://doi.org/10.1016/j.eml.2024.102169}.
\bibitem[{Iqbal et~al.(2023)Iqbal, Li, Sonta, Ihuaenyi and Xiao}]{iqbal2023probabilistic}
\bibinfo{author}{Iqbal, S.}, \bibinfo{author}{Li, B.}, \bibinfo{author}{Sonta, K.}, \bibinfo{author}{Ihuaenyi, R.}, \bibinfo{author}{Xiao, X.}, \bibinfo{year}{2023}.
\newblock \bibinfo{title}{Probabilistic finite element analysis of sheet molding compound composites with an extended strength distribution model}.
\newblock \bibinfo{journal}{Finite Elements in Analysis and Design} \bibinfo{volume}{214}, \bibinfo{pages}{p.103865}.
\bibitem[{Jones et~al.(2018)Jones, Carroll, Karlson, Kramer, Lehoucq, Reu and Turner}]{jones2018parameter}
\bibinfo{author}{Jones, E.}, \bibinfo{author}{Carroll, J.D.}, \bibinfo{author}{Karlson, K.N.}, \bibinfo{author}{Kramer, S.L.B.}, \bibinfo{author}{Lehoucq, R.B.}, \bibinfo{author}{Reu, P.L.}, \bibinfo{author}{Turner, D.Z.}, \bibinfo{year}{2018}.
\newblock \bibinfo{title}{Parameter covariance and non-uniqueness in material model calibration using the virtual fields method}.
\newblock \bibinfo{journal}{Computational Materials Science} \bibinfo{volume}{152}, \bibinfo{pages}{268--290}.
\newblock \DOIprefix\doi{https://doi.org/10.1016/j.commatsci.2018.05.037}.
\bibitem[{Kamaya and Kawakubo(2011)}]{KAMAYA2011243}
\bibinfo{author}{Kamaya, M.}, \bibinfo{author}{Kawakubo, M.}, \bibinfo{year}{2011}.
\newblock \bibinfo{title}{A procedure for determining the true stress–strain curve over a large range of strains using digital image correlation and finite element analysis}.
\newblock \bibinfo{journal}{Mechanics of Materials} \bibinfo{volume}{43}, \bibinfo{pages}{243--253}.
\newblock \DOIprefix\doi{10.1016/j.mechmat.2011.02.007}.
\bibitem[{Kim et~al.(2014)Kim, Barlat, Pierron and Lee}]{kim2014determination}
\bibinfo{author}{Kim, J.H.}, \bibinfo{author}{Barlat, F.}, \bibinfo{author}{Pierron, F.}, \bibinfo{author}{Lee, M.G.}, \bibinfo{year}{2014}.
\newblock \bibinfo{title}{Determination of anisotropic plastic constitutive parameters using the virtual fields method}.
\newblock \bibinfo{journal}{Experimental Mechanics} \bibinfo{volume}{54}, \bibinfo{pages}{1189--1204}.
\newblock \DOIprefix\doi{https://doi.org/10.1007/s11340-014-9879-x}.
\bibitem[{Kirchdoerfer and Ortiz(2016)}]{KIRCHDOERFER2016CMAME}
\bibinfo{author}{Kirchdoerfer, T.}, \bibinfo{author}{Ortiz, M.}, \bibinfo{year}{2016}.
\newblock \bibinfo{title}{Data-driven computational mechanics}.
\newblock \bibinfo{journal}{Computer Methods in Applied Mechanics and Engineering} \bibinfo{volume}{304}, \bibinfo{pages}{81--101}.
\newblock \DOIprefix\doi{10.1016/j.cma.2016.02.001}.
\bibitem[{Kirchdoerfer and Ortiz(2017)}]{KIRCHDOERFER2017CMAME}
\bibinfo{author}{Kirchdoerfer, T.}, \bibinfo{author}{Ortiz, M.}, \bibinfo{year}{2017}.
\newblock \bibinfo{title}{Data driven computing with noisy material data sets}.
\newblock \bibinfo{journal}{Computer Methods in Applied Mechanics and Engineering} \bibinfo{volume}{326}, \bibinfo{pages}{622--641}.
\newblock \DOIprefix\doi{10.1016/j.cma.2017.07.039}.
\bibitem[{Kirchdoerfer and Ortiz(2018)}]{KIRCHDOERFER2018NME}
\bibinfo{author}{Kirchdoerfer, T.}, \bibinfo{author}{Ortiz, M.}, \bibinfo{year}{2018}.
\newblock \bibinfo{title}{Data-driven computing in dynamics}.
\newblock \bibinfo{journal}{International Journal for Numerical Methods in Engineering} \bibinfo{volume}{113}, \bibinfo{pages}{1697--1710}.
\newblock \DOIprefix\doi{10.1002/nme.5716}.
\bibitem[{Kramer and Scherzinger(2014)}]{kramer2014implementation}
\bibinfo{author}{Kramer, S.L.}, \bibinfo{author}{Scherzinger, W.M.}, \bibinfo{year}{2014}.
\newblock \bibinfo{title}{Implementation and evaluation of the virtual fields method: determining constitutive model parameters from full-field deformation data}.
\newblock \bibinfo{type}{Technical Report} \bibinfo{number}{SAND2014-17871}. Sandia National Laboratories.
\newblock \DOIprefix\doi{https://doi.org/10.2172/1158669}.
\bibitem[{Kroon et~al.(2018)Kroon, Andreasson, Persson~Jutemar, Petersson, Persson, Dorn and Olsson}]{Kroon2018}
\bibinfo{author}{Kroon, M.}, \bibinfo{author}{Andreasson, E.}, \bibinfo{author}{Persson~Jutemar, E.}, \bibinfo{author}{Petersson, V.}, \bibinfo{author}{Persson, L.}, \bibinfo{author}{Dorn, M.}, \bibinfo{author}{Olsson, P.A.}, \bibinfo{year}{2018}.
\newblock \bibinfo{title}{Anisotropic elastic-viscoplastic properties at finite strains of injection-moulded low-density polyethylene}.
\newblock \bibinfo{journal}{Experimental Mechanics} \bibinfo{volume}{58}, \bibinfo{pages}{75--86}.
\newblock \DOIprefix\doi{10.1007/s11340-017-0317-4}.
\bibitem[{Kuwabara(2007)}]{KUWABARA2007}
\bibinfo{author}{Kuwabara, T.}, \bibinfo{year}{2007}.
\newblock \bibinfo{title}{Advances in experiments on metal sheets and tubes in support of constitutive modeling and forming simulations}.
\newblock \bibinfo{journal}{International Journal of Plasticity} \bibinfo{volume}{23}, \bibinfo{pages}{385--419}.
\newblock \DOIprefix\doi{10.1016/j.ijplas.2006.06.003}.
\bibitem[{Kuwabara et~al.(1998)Kuwabara, Ikeda and Kuroda}]{KUWABARA1998}
\bibinfo{author}{Kuwabara, T.}, \bibinfo{author}{Ikeda, S.}, \bibinfo{author}{Kuroda, K.}, \bibinfo{year}{1998}.
\newblock \bibinfo{title}{Measurement and analysis of differential work hardening in cold-rolled steel sheet under biaxial tension}.
\newblock \bibinfo{journal}{Journal of Materials Processing Technology} \bibinfo{volume}{80-81}, \bibinfo{pages}{517--523}.
\newblock \DOIprefix\doi{10.1016/S0924-0136(98)00155-1}.
\bibitem[{Lagarias et~al.(1998)Lagarias, Reeds, Wright and Wright}]{lagarias1998convergence}
\bibinfo{author}{Lagarias, J.}, \bibinfo{author}{Reeds, J.}, \bibinfo{author}{Wright, M.}, \bibinfo{author}{Wright, P.}, \bibinfo{year}{1998}.
\newblock \bibinfo{title}{Convergence properties of the nelder--mead simplex method in low dimensions}.
\newblock \bibinfo{journal}{SIAM Journal on Optimization} \bibinfo{volume}{9}, \bibinfo{pages}{112--147}.
\newblock \DOIprefix\doi{https://doi.org/10.1137/S1052623496303470}.
\bibitem[{Lattanzi et~al.(2020)Lattanzi, Barlat, Pierron, Marek and Rossi}]{LATTANZI2020}
\bibinfo{author}{Lattanzi, A.}, \bibinfo{author}{Barlat, F.}, \bibinfo{author}{Pierron, F.}, \bibinfo{author}{Marek, A.}, \bibinfo{author}{Rossi, M.}, \bibinfo{year}{2020}.
\newblock \bibinfo{title}{Inverse identification strategies for the characterization of transformation-based anisotropic plasticity models with the non-linear vfm}.
\newblock \bibinfo{journal}{International Journal of Mechanical Sciences} \bibinfo{volume}{173}, \bibinfo{pages}{105422}.
\newblock \DOIprefix\doi{10.1016/j.ijmecsci.2020.105422}.
\bibitem[{Lava et~al.(2020)Lava, Jones, Wittevrongel and Pierron}]{Lava2020}
\bibinfo{author}{Lava, P.}, \bibinfo{author}{Jones, E.}, \bibinfo{author}{Wittevrongel, L.}, \bibinfo{author}{Pierron, F.}, \bibinfo{year}{2020}.
\newblock \bibinfo{title}{Validation of finite‐element models using full‐field experimental data: Levelling finite‐element analysis data through a digital image correlation engine}.
\newblock \bibinfo{journal}{Strain} \bibinfo{volume}{56}, \bibinfo{pages}{e12350}.
\newblock \DOIprefix\doi{https://doi.org/10.1111/str.12350}.
\bibitem[{Li et~al.(2019)Li, Zhu, Xia, Gorji and Wierzbicki}]{LI2019Joule}
\bibinfo{author}{Li, W.}, \bibinfo{author}{Zhu, J.}, \bibinfo{author}{Xia, Y.}, \bibinfo{author}{Gorji, M.B.}, \bibinfo{author}{Wierzbicki, T.}, \bibinfo{year}{2019}.
\newblock \bibinfo{title}{Data-driven safety envelope of lithium-ion batteries for electric vehicles}.
\newblock \bibinfo{journal}{Joule} \bibinfo{volume}{3}, \bibinfo{pages}{2703--2715}.
\newblock \DOIprefix\doi{10.1016/j.joule.2019.07.026}.
\bibitem[{Marino et~al.(2023)Marino, Flaschel, Kumar and De~Lorenzis}]{marino2023automated}
\bibinfo{author}{Marino, E.}, \bibinfo{author}{Flaschel, M.}, \bibinfo{author}{Kumar, S.}, \bibinfo{author}{De~Lorenzis, L.}, \bibinfo{year}{2023}.
\newblock \bibinfo{title}{Automated identification of linear viscoelastic constitutive laws with euclid}.
\newblock \bibinfo{journal}{Mechanics of Materials} \bibinfo{volume}{181}, \bibinfo{pages}{104643}.
\newblock \DOIprefix\doi{https://doi.org/10.1016/j.mechmat.2023.104643}.
\bibitem[{Martins et~al.(2019)Martins, Andrade-Campos and Thuillier}]{Martins2019}
\bibinfo{author}{Martins, J.}, \bibinfo{author}{Andrade-Campos, A.}, \bibinfo{author}{Thuillier, S.}, \bibinfo{year}{2019}.
\newblock \bibinfo{title}{Calibration of anisotropic plasticity models using a biaxial test and the virtual fields method}.
\newblock \bibinfo{journal}{International Journal of Solids and Structures} \bibinfo{volume}{172}, \bibinfo{pages}{21--37}.
\newblock \DOIprefix\doi{https://doi.org/10.1016/j.ijsolstr.2019.05.019}.
\bibitem[{Hernandez-de Menendez et~al.(2020)Hernandez-de Menendez, Escobar~D{\'\i}az and Morales-Menendez}]{hernandez2020engineering}
\bibinfo{author}{Hernandez-de Menendez, M.}, \bibinfo{author}{Escobar~D{\'\i}az, C.A.}, \bibinfo{author}{Morales-Menendez, R.}, \bibinfo{year}{2020}.
\newblock \bibinfo{title}{Engineering education for smart 4.0 technology: a review}.
\newblock \bibinfo{journal}{International Journal on Interactive Design and Manufacturing} \bibinfo{volume}{14}, \bibinfo{pages}{789--803}.
\newblock \DOIprefix\doi{https://doi.org/10.1007/s12008-020-00672-x}.
\bibitem[{Meng et~al.(2016)Meng, Xia, Zhou and Lin}]{Meng2016}
\bibinfo{author}{Meng, Y.}, \bibinfo{author}{Xia, Y.}, \bibinfo{author}{Zhou, Q.}, \bibinfo{author}{Lin, S.}, \bibinfo{year}{2016}.
\newblock \bibinfo{title}{Identification of true stress-strain curve of thermoplastic polymers under biaxial tension}.
\newblock \bibinfo{journal}{SAE International Journal of Materials and Manufacturing} \bibinfo{volume}{9}, \bibinfo{pages}{768--775}.
\newblock \DOIprefix\doi{10.4271/2016-01-0514}.
\bibitem[{von Mises(1913)}]{mises1913mechanik}
\bibinfo{author}{von Mises, R.E.}, \bibinfo{year}{1913}.
\newblock \bibinfo{title}{Mechanik der festen k{\"o}rper im plastisch-deformablen zustand}.
\newblock \bibinfo{journal}{Nachrichten von der Gesellschaft der Wissenschaften zu G{\"o}ttingen, Mathematisch-Physikalische Klasse} \bibinfo{volume}{1913}, \bibinfo{pages}{582--592}.
\bibitem[{Morin et~al.(2017)Morin, Leblond, Mohr and Kondo}]{morin2017prediction}
\bibinfo{author}{Morin, L.}, \bibinfo{author}{Leblond, J.B.}, \bibinfo{author}{Mohr, D.}, \bibinfo{author}{Kondo, D.}, \bibinfo{year}{2017}.
\newblock \bibinfo{title}{Prediction of shear-dominated ductile fracture in a butterfly specimen using a model of plastic porous solids including void shape effects}.
\newblock \bibinfo{journal}{European Journal of Mechanics-A/Solids} \bibinfo{volume}{61}, \bibinfo{pages}{433--442}.
\newblock \DOIprefix\doi{https://doi.org/10.1016/j.euromechsol.2016.10.014}.
\bibitem[{Mulder et~al.(2015)Mulder, Vegter, Aretz, Keller and Van Den~Boogaard}]{mulder2015accurate}
\bibinfo{author}{Mulder, J.}, \bibinfo{author}{Vegter, H.}, \bibinfo{author}{Aretz, H.}, \bibinfo{author}{Keller, S.}, \bibinfo{author}{Van Den~Boogaard, A.}, \bibinfo{year}{2015}.
\newblock \bibinfo{title}{Accurate determination of flow curves using the bulge test with optical measuring systems}.
\newblock \bibinfo{journal}{Journal of Materials Processing Technology} \bibinfo{volume}{226}, \bibinfo{pages}{169--187}.
\newblock \DOIprefix\doi{https://doi.org/10.1016/j.jmatprotec.2015.06.034}.
\bibitem[{Ni et~al.(2021)Ni, Kopp, Kalfon-Cohen, Furtado, Lee, Arteiro, Borstnar, Mavrogordato, Helfen, Sinclair, Spearing, Camanho and Wardle}]{NI2021}
\bibinfo{author}{Ni, X.}, \bibinfo{author}{Kopp, R.}, \bibinfo{author}{Kalfon-Cohen, E.}, \bibinfo{author}{Furtado, C.}, \bibinfo{author}{Lee, J.}, \bibinfo{author}{Arteiro, A.}, \bibinfo{author}{Borstnar, G.}, \bibinfo{author}{Mavrogordato, M.N.}, \bibinfo{author}{Helfen, L.}, \bibinfo{author}{Sinclair, I.}, \bibinfo{author}{Spearing, S.M.}, \bibinfo{author}{Camanho, P.P.}, \bibinfo{author}{Wardle, B.L.}, \bibinfo{year}{2021}.
\newblock \bibinfo{title}{In situ synchrotron computed tomography study of nanoscale interlaminar reinforcement and thin-ply effects on damage progression in composite laminates}.
\newblock \bibinfo{journal}{Composites Part B: Engineering} \bibinfo{volume}{217}, \bibinfo{pages}{108623}.
\newblock \DOIprefix\doi{10.1016/j.compositesb.2021.108623}.
\bibitem[{Pan et~al.(2009)Pan, Qian, Xie and Asundi}]{pan2009review}
\bibinfo{author}{Pan, B.}, \bibinfo{author}{Qian, K.}, \bibinfo{author}{Xie, H.}, \bibinfo{author}{Asundi, A.}, \bibinfo{year}{2009}.
\newblock \bibinfo{title}{Two-dimensional digital image correlation for in-plane displacement and strain measurement: a review}.
\newblock \bibinfo{journal}{Measurement Science and Technology} \bibinfo{volume}{20}, \bibinfo{pages}{062001}.
\newblock \DOIprefix\doi{10.1088/0957-0233/20/6/062001}.
\bibitem[{Papasidero et~al.(2015)Papasidero, Doquet and Mohr}]{papasidero2015ductile}
\bibinfo{author}{Papasidero, J.}, \bibinfo{author}{Doquet, V.}, \bibinfo{author}{Mohr, D.}, \bibinfo{year}{2015}.
\newblock \bibinfo{title}{Ductile fracture of aluminum 2024-t351 under proportional and non-proportional multi-axial loading: Bao–wierzbicki results revisited}.
\newblock \bibinfo{journal}{International Journal of Solids and Structures} \bibinfo{volume}{69}, \bibinfo{pages}{459--474}.
\newblock \DOIprefix\doi{https://doi.org/10.1016/j.ijsolstr.2015.05.006}.
\bibitem[{Peirs et~al.(2012)Peirs, Verleysen and Degrieck}]{Peirs2012}
\bibinfo{author}{Peirs, J.}, \bibinfo{author}{Verleysen, P.}, \bibinfo{author}{Degrieck, J.}, \bibinfo{year}{2012}.
\newblock \bibinfo{title}{Novel technique for static and dynamic shear testing of ti6al4v sheet}.
\newblock \bibinfo{journal}{Experimental Mechanics} \bibinfo{volume}{52}, \bibinfo{pages}{729--741}.
\newblock \DOIprefix\doi{https://doi.org/10.1007/s11340-011-9541-9}.
\bibitem[{Plancher et~al.(2020)Plancher, Qu, Vonk, Gorji, Tancogne-Dejean and Tasan}]{plancher2020tracking}
\bibinfo{author}{Plancher, E.}, \bibinfo{author}{Qu, K.}, \bibinfo{author}{Vonk, N.H.}, \bibinfo{author}{Gorji, M.B.}, \bibinfo{author}{Tancogne-Dejean, T.}, \bibinfo{author}{Tasan, C.C.}, \bibinfo{year}{2020}.
\newblock \bibinfo{title}{Tracking microstructure evolution in complex biaxial strain paths: A bulge test methodology for the scanning electron microscope}.
\newblock \bibinfo{journal}{Experimental Mechanics} \bibinfo{volume}{60}, \bibinfo{pages}{35--50}.
\newblock \DOIprefix\doi{https://doi.org/10.1007/s11340-019-00538-8}.
\bibitem[{Pottier et~al.(2011)Pottier, Toussaint and Vacher}]{Pottier2011}
\bibinfo{author}{Pottier, T.}, \bibinfo{author}{Toussaint, F.}, \bibinfo{author}{Vacher, P.}, \bibinfo{year}{2011}.
\newblock \bibinfo{title}{Contribution of heterogeneous strain field measurements and boundary conditions modelling in inverse identification of material parameters}.
\newblock \bibinfo{journal}{European Journal of Mechanics-A/Solids} \bibinfo{volume}{30}, \bibinfo{pages}{373--382}.
\newblock \DOIprefix\doi{https://doi.org/10.1016/j.euromechsol.2010.10.001}.
\bibitem[{Qi et~al.(2014)Qi, Hector, James and Kim}]{qi2014lithium}
\bibinfo{author}{Qi, Y.}, \bibinfo{author}{Hector, L.G.}, \bibinfo{author}{James, C.}, \bibinfo{author}{Kim, K.J.}, \bibinfo{year}{2014}.
\newblock \bibinfo{title}{Lithium concentration dependent elastic properties of battery electrode materials from first principles calculations}.
\newblock \bibinfo{journal}{Journal of the Electrochemical Society} \bibinfo{volume}{161}, \bibinfo{pages}{F3010}.
\newblock \DOIprefix\doi{http://dx.doi.org/10.1149/2.0031411jes}.
\bibitem[{Ramprasad et~al.(2017)Ramprasad, Batra, Pilania, Mannodi-Kanakkithodi and Kim}]{ramprasad2017machine}
\bibinfo{author}{Ramprasad, R.}, \bibinfo{author}{Batra, R.}, \bibinfo{author}{Pilania, G.}, \bibinfo{author}{Mannodi-Kanakkithodi, A.}, \bibinfo{author}{Kim, C.}, \bibinfo{year}{2017}.
\newblock \bibinfo{title}{Machine learning in materials informatics: recent applications and prospects}.
\newblock \bibinfo{journal}{npj Computational Materials} \bibinfo{volume}{3}, \bibinfo{pages}{54}.
\newblock \DOIprefix\doi{https://doi.org/10.1038/s41524-017-0056-5}.
\bibitem[{Robert(1999)}]{robert1999}
\bibinfo{author}{Robert, C.P.}, \bibinfo{year}{1999}.
\newblock \bibinfo{title}{Monte Carlo Statistical Methods}.
\newblock \bibinfo{publisher}{Springer}.
\bibitem[{Rossi et~al.(2022)Rossi, Lattanzi, Morichelli, Martins, Thuillier, Andrade-Campos and Coppieters}]{rossi2022testing}
\bibinfo{author}{Rossi, M.}, \bibinfo{author}{Lattanzi, A.}, \bibinfo{author}{Morichelli, L.}, \bibinfo{author}{Martins, J.M.}, \bibinfo{author}{Thuillier, S.}, \bibinfo{author}{Andrade-Campos, A.}, \bibinfo{author}{Coppieters, S.}, \bibinfo{year}{2022}.
\newblock \bibinfo{title}{Testing methodologies for the calibration of advanced plasticity models for sheet metals: A review}.
\newblock \bibinfo{journal}{Strain} \bibinfo{volume}{58}, \bibinfo{pages}{e12426}.
\newblock \DOIprefix\doi{https://doi.org/10.1111/str.12426}.
\bibitem[{Rossi et~al.(2015)Rossi, Lava, Pierron, Debruyne and Sasso}]{Rossi2015}
\bibinfo{author}{Rossi, M.}, \bibinfo{author}{Lava, P.}, \bibinfo{author}{Pierron, F.}, \bibinfo{author}{Debruyne, D.}, \bibinfo{author}{Sasso, M.}, \bibinfo{year}{2015}.
\newblock \bibinfo{title}{Effect of dic spatial resolution, noise and interpolation error on identification results with the vfm}.
\newblock \bibinfo{journal}{Strain} \bibinfo{volume}{51}, \bibinfo{pages}{206--222}.
\newblock \DOIprefix\doi{https://doi.org/10.1111/str.12134}.
\bibitem[{Rossi and Pierron(2012)}]{Rossi2012}
\bibinfo{author}{Rossi, M.}, \bibinfo{author}{Pierron, F.}, \bibinfo{year}{2012}.
\newblock \bibinfo{title}{On the use of simulated experiments in designing tests for material characterization from full-field measurements}.
\newblock \bibinfo{journal}{International Journal of Solids and Structures} \bibinfo{volume}{49}, \bibinfo{pages}{420--435}.
\newblock \DOIprefix\doi{https://doi.org/10.1016/j.ijsolstr.2011.09.025}.
\bibitem[{Roth and Mohr(2018)}]{Roth2018}
\bibinfo{author}{Roth, C.}, \bibinfo{author}{Mohr, D.}, \bibinfo{year}{2018}.
\newblock \bibinfo{title}{Determining the strain to fracture for simple shear for a wide range of sheet metals}.
\newblock \bibinfo{journal}{International Journal of Mechanical Sciences} \bibinfo{volume}{149}, \bibinfo{pages}{224--240}.
\newblock \DOIprefix\doi{https://doi.org/10.1016/j.ijmecsci.2018.10.007}.
\bibitem[{Roth and Mohr(2016)}]{ROTH2016}
\bibinfo{author}{Roth, C.C.}, \bibinfo{author}{Mohr, D.}, \bibinfo{year}{2016}.
\newblock \bibinfo{title}{Ductile fracture experiments with locally proportional loading histories}.
\newblock \bibinfo{journal}{International Journal of Plasticity} \bibinfo{volume}{79}, \bibinfo{pages}{328--354}.
\newblock \DOIprefix\doi{https://doi.org/10.1016/j.ijplas.2015.08.004}.
\bibitem[{Réthoré et~al.(2018)Réthoré, Leygue, Coret, Stainier and Verron}]{Rethore2018}
\bibinfo{author}{Réthoré, J.}, \bibinfo{author}{Leygue, A.}, \bibinfo{author}{Coret, M.}, \bibinfo{author}{Stainier, L.}, \bibinfo{author}{Verron, E.}, \bibinfo{year}{2018}.
\newblock \bibinfo{title}{Computational measurements of stress fields from digital images}.
\newblock \bibinfo{journal}{International Journal for Numerical Methods in Engineering} \bibinfo{volume}{113}, \bibinfo{pages}{1810--1826}.
\newblock \DOIprefix\doi{10.1002/nme.5690}.
\bibitem[{Segal(2002)}]{SEGAL2002331}
\bibinfo{author}{Segal, V.}, \bibinfo{year}{2002}.
\newblock \bibinfo{title}{Severe plastic deformation: simple shear versus pure shear}.
\newblock \bibinfo{journal}{Materials Science and Engineering: A} \bibinfo{volume}{338}, \bibinfo{pages}{331--344}.
\newblock \DOIprefix\doi{10.1016/S0921-5093(02)00066-7}.
\bibitem[{Shahriari et~al.(2016)Shahriari, Swersky, Wang, Adams and De~Freitas}]{shahriari2015bayesian}
\bibinfo{author}{Shahriari, B.}, \bibinfo{author}{Swersky, K.}, \bibinfo{author}{Wang, Z.}, \bibinfo{author}{Adams, R.P.}, \bibinfo{author}{De~Freitas, N.}, \bibinfo{year}{2016}.
\newblock \bibinfo{title}{Taking the human out of the loop: A review of bayesian optimization}.
\newblock \bibinfo{journal}{Proceedings of the IEEE} \bibinfo{volume}{104}, \bibinfo{pages}{148--175}.
\newblock \DOIprefix\doi{10.1109/JPROC.2015.2494218}.
\bibitem[{Shannon(1948)}]{Shannon1948mathematical}
\bibinfo{author}{Shannon, C.E.}, \bibinfo{year}{1948}.
\newblock \bibinfo{title}{A mathematical theory of communication}.
\newblock \bibinfo{journal}{The Bell System Technical Journal} \bibinfo{volume}{27}, \bibinfo{pages}{379--423}.
\newblock \DOIprefix\doi{10.1002/j.1538-7305.1948.tb01338.x}.
\bibitem[{Simo and Hughes(2006)}]{simo2006computational}
\bibinfo{author}{Simo, J.C.}, \bibinfo{author}{Hughes, T.J.}, \bibinfo{year}{2006}.
\newblock \bibinfo{title}{Computational inelasticity}. volume~\bibinfo{volume}{7}.
\newblock \bibinfo{publisher}{Springer Science \& Business Media}.
\bibitem[{Singer and Nelder(2009)}]{singer2009nelder}
\bibinfo{author}{Singer, S.}, \bibinfo{author}{Nelder, J.}, \bibinfo{year}{2009}.
\newblock \bibinfo{title}{Nelder--mead algorithm}.
\newblock \bibinfo{journal}{Scholarpedia} \bibinfo{volume}{4}, \bibinfo{pages}{2928}.
\bibitem[{Stainier et~al.(2019)Stainier, Leygue and Ortiz}]{stainier2019model}
\bibinfo{author}{Stainier, L.}, \bibinfo{author}{Leygue, A.}, \bibinfo{author}{Ortiz, M.}, \bibinfo{year}{2019}.
\newblock \bibinfo{title}{Model-free data-driven methods in mechanics: material data identification and solvers}.
\newblock \bibinfo{journal}{Computational Mechanics} \bibinfo{volume}{64}, \bibinfo{pages}{381--393}.
\newblock \DOIprefix\doi{https://doi.org/10.1007/s00466-019-01731-1}.
\bibitem[{Sutton et~al.(2009)Sutton, Orteu and Schreier}]{Sutton2009}
\bibinfo{author}{Sutton, M.A.}, \bibinfo{author}{Orteu, J.J.}, \bibinfo{author}{Schreier, H.}, \bibinfo{year}{2009}.
\newblock \bibinfo{title}{Image correlation for shape, motion and deformation measurements: basic concepts, theory and applications}.
\newblock \bibinfo{publisher}{Springer Science \& Business Media}.
\bibitem[{Tancogne-Dejean et~al.(2021)Tancogne-Dejean, Roth, Morgeneyer, Helfen and Mohr}]{tancogne2021ductile}
\bibinfo{author}{Tancogne-Dejean, T.}, \bibinfo{author}{Roth, C.C.}, \bibinfo{author}{Morgeneyer, T.F.}, \bibinfo{author}{Helfen, L.}, \bibinfo{author}{Mohr, D.}, \bibinfo{year}{2021}.
\newblock \bibinfo{title}{Ductile damage of aa2024-t3 under shear loading: Mechanism analysis through in-situ laminography}.
\newblock \bibinfo{journal}{Acta Materialia} \bibinfo{volume}{205}, \bibinfo{pages}{116556}.
\newblock \DOIprefix\doi{https://doi.org/10.1016/j.actamat.2020.116556}.
\bibitem[{Torrey and Shavlik(2010)}]{Torrey2010}
\bibinfo{author}{Torrey, L.}, \bibinfo{author}{Shavlik, J.}, \bibinfo{year}{2010}.
\newblock \bibinfo{title}{Transfer learning}, in: \bibinfo{booktitle}{Handbook of Research on Machine Learning Applications and Trends: Algorithms, Methods, and Techniques}. \bibinfo{publisher}{IGI Global}, pp. \bibinfo{pages}{242--264}.
\newblock \DOIprefix\doi{10.4018/978-1-60566-766-9.ch011}.
\bibitem[{Vilotic et~al.(2003)Vilotic, Planck, Grbic, Alexandrov and Chikanova}]{Vilotic2003}
\bibinfo{author}{Vilotic, D.}, \bibinfo{author}{Planck, M.}, \bibinfo{author}{Grbic, S.}, \bibinfo{author}{Alexandrov, S.}, \bibinfo{author}{Chikanova, N.}, \bibinfo{year}{2003}.
\newblock \bibinfo{title}{An approach to determining the workability diagram based on upsetting tests}.
\newblock \bibinfo{journal}{Fatigue \& Fracture of Engineering Materials \& Structures} \bibinfo{volume}{26}, \bibinfo{pages}{305--310}.
\newblock \DOIprefix\doi{10.1046/j.1460-2695.2003.00469.x}.
\bibitem[{Wierzbicki et~al.(2005)Wierzbicki, Bao, Lee and Bai}]{WIERZBICKI2005Seven}
\bibinfo{author}{Wierzbicki, T.}, \bibinfo{author}{Bao, Y.}, \bibinfo{author}{Lee, Y.W.}, \bibinfo{author}{Bai, Y.}, \bibinfo{year}{2005}.
\newblock \bibinfo{title}{Calibration and evaluation of seven fracture models}.
\newblock \bibinfo{journal}{International Journal of Mechanical Sciences} \bibinfo{volume}{47}, \bibinfo{pages}{719--743}.
\newblock \DOIprefix\doi{10.1016/j.ijmecsci.2005.03.003}. \bibinfo{note}{a Special Issue in Honour of Professor Stephen R. Reid's 60th Birthday}.
\bibitem[{Xu et~al.(2021)Xu, Huang and Darve}]{XU2021JComP}
\bibinfo{author}{Xu, K.}, \bibinfo{author}{Huang, D.Z.}, \bibinfo{author}{Darve, E.}, \bibinfo{year}{2021}.
\newblock \bibinfo{title}{Learning constitutive relations using symmetric positive definite neural networks}.
\newblock \bibinfo{journal}{Journal of Computational Physics} \bibinfo{volume}{428}, \bibinfo{pages}{110072}.
\newblock \DOIprefix\doi{10.1016/j.jcp.2020.110072}.
\bibitem[{Yamanaka et~al.(2020)Yamanaka, Kamijyo, Koenuma, Watanabe and Kuwabara}]{YAMANAKA2020}
\bibinfo{author}{Yamanaka, A.}, \bibinfo{author}{Kamijyo, R.}, \bibinfo{author}{Koenuma, K.}, \bibinfo{author}{Watanabe, I.}, \bibinfo{author}{Kuwabara, T.}, \bibinfo{year}{2020}.
\newblock \bibinfo{title}{Deep neural network approach to estimate biaxial stress-strain curves of sheet metals}.
\newblock \bibinfo{journal}{Materials \& Design} \bibinfo{volume}{195}, \bibinfo{pages}{108970}.
\newblock \DOIprefix\doi{10.1016/j.matdes.2020.108970}.
\bibitem[{Zhang et~al.(2017)Zhang, Min, Carsley, Stoughton and Lin}]{Zhang2017}
\bibinfo{author}{Zhang, L.}, \bibinfo{author}{Min, J.}, \bibinfo{author}{Carsley, J.}, \bibinfo{author}{Stoughton, T.}, \bibinfo{author}{Lin, J.}, \bibinfo{year}{2017}.
\newblock \bibinfo{title}{Experimental and theoretical investigation on the role of friction in nakazima testing}.
\newblock \bibinfo{journal}{International Journal of Mechanical Sciences} \bibinfo{volume}{133}, \bibinfo{pages}{217--226}.
\newblock \DOIprefix\doi{https://doi.org/10.1016/j.ijmecsci.2017.08.020}.
\bibitem[{Zhang et~al.(1999)Zhang, Hauge, Ødegård and Thaulow}]{ZHANG19993497}
\bibinfo{author}{Zhang, Z.}, \bibinfo{author}{Hauge, M.}, \bibinfo{author}{Ødegård, J.}, \bibinfo{author}{Thaulow, C.}, \bibinfo{year}{1999}.
\newblock \bibinfo{title}{Determining material true stress–strain curve from tensile specimens with rectangular cross-section}.
\newblock \bibinfo{journal}{International Journal of Solids and Structures} \bibinfo{volume}{36}, \bibinfo{pages}{3497--3516}.
\newblock \DOIprefix\doi{10.1016/S0020-7683(98)00153-X}.
\bibitem[{Zhao et~al.(2021)Zhao, Braatz and Bazant}]{ZHAO2021JComP}
\bibinfo{author}{Zhao, H.}, \bibinfo{author}{Braatz, R.D.}, \bibinfo{author}{Bazant, M.Z.}, \bibinfo{year}{2021}.
\newblock \bibinfo{title}{Image inversion and uncertainty quantification for constitutive laws of pattern formation}.
\newblock \bibinfo{journal}{Journal of Computational Physics} \bibinfo{volume}{436}, \bibinfo{pages}{110279}.
\newblock \DOIprefix\doi{10.1016/j.jcp.2021.110279}.
\bibitem[{Zhao et~al.(2023)Zhao, Deng, Cohen, Lim, Li, Fraggedakis, Jiang, Storey, Chueh, Braatz et~al.}]{zhao2023learning}
\bibinfo{author}{Zhao, H.}, \bibinfo{author}{Deng, H.D.}, \bibinfo{author}{Cohen, A.E.}, \bibinfo{author}{Lim, J.}, \bibinfo{author}{Li, Y.}, \bibinfo{author}{Fraggedakis, D.}, \bibinfo{author}{Jiang, B.}, \bibinfo{author}{Storey, B.D.}, \bibinfo{author}{Chueh, W.C.}, \bibinfo{author}{Braatz, R.D.}, et~al., \bibinfo{year}{2023}.
\newblock \bibinfo{title}{Learning heterogeneous reaction kinetics from x-ray videos pixel by pixel}.
\newblock \bibinfo{journal}{Nature} \bibinfo{volume}{621}, \bibinfo{pages}{289--294}.
\newblock \DOIprefix\doi{https://doi.org/10.1038/s41586-023-06393-x}.
\bibitem[{Zhao et~al.(2020)Zhao, Storey, Braatz and Bazant}]{ZHAO2020PRL}
\bibinfo{author}{Zhao, H.}, \bibinfo{author}{Storey, B.D.}, \bibinfo{author}{Braatz, R.D.}, \bibinfo{author}{Bazant, M.Z.}, \bibinfo{year}{2020}.
\newblock \bibinfo{title}{Learning the physics of pattern formation from images}.
\newblock \bibinfo{journal}{Phys. Rev. Lett.} \bibinfo{volume}{124}, \bibinfo{pages}{060201}.
\newblock \DOIprefix\doi{10.1103/PhysRevLett.124.060201}.
\bibitem[{Zhu et~al.(2014)Zhu, Xia, Luo, Gu and Zhou}]{zhu2014influence}
\bibinfo{author}{Zhu, J.}, \bibinfo{author}{Xia, Y.}, \bibinfo{author}{Luo, H.}, \bibinfo{author}{Gu, G.}, \bibinfo{author}{Zhou, Q.}, \bibinfo{year}{2014}.
\newblock \bibinfo{title}{Influence of flow rule and calibration approach on plasticity characterization of dp780 steel sheets using hill48 model}.
\newblock \bibinfo{journal}{International Journal of Mechanical Sciences} \bibinfo{volume}{89}, \bibinfo{pages}{148--157}.
\newblock \DOIprefix\doi{https://doi.org/10.1016/j.ijmecsci.2014.09.001}.
\bibitem[{Zhu et~al.(2018)Zhu, Zhang, Luo and Sahraei}]{ZHU2018APEN}
\bibinfo{author}{Zhu, J.}, \bibinfo{author}{Zhang, X.}, \bibinfo{author}{Luo, H.}, \bibinfo{author}{Sahraei, E.}, \bibinfo{year}{2018}.
\newblock \bibinfo{title}{Investigation of the deformation mechanisms of lithium-ion battery components using in-situ micro tests}.
\newblock \bibinfo{journal}{Applied Energy} \bibinfo{volume}{224}, \bibinfo{pages}{251--266}.
\newblock \DOIprefix\doi{10.1016/j.apenergy.2018.05.007}.

\end{thebibliography}





\end{document}